\documentclass[12pt, a4paper]{article}
\UseRawInputEncoding 
\usepackage{tikz-cd}
\usetikzlibrary{arrows,snakes,shapes.arrows,decorations.markings}
     \tikzset{>=triangle 90}
     \tikzstyle{bbc}=[draw,circle,fill=black,scale=.75]
     \tikzstyle{rc}=[circle,fill=red,scale=.6]
     \tikzstyle{wc}=[draw,circle,scale=.75]

\usepackage{color}
\newcommand{\hilight}[1]{\colorbox{yellow}{#1}}

\usepackage{latexsym}
\usepackage{amsmath, mathdots, amscd}
\usepackage{multirow}
\usepackage{amssymb}
\usepackage{graphicx}
\usepackage{euscript}
\usepackage{amsfonts}
\usepackage{verbatim}
\usepackage{epsfig}
\usepackage{hyperref}
\usepackage[margin=2cm,bottom=2.0cm]{geometry}
\usepackage{float}

\def\blue{\textcolor{blue}}

\def\be{\begin{eqnarray}}
\def\ee{\end{eqnarray}}
\def\0{\nonumber}
\def\d{\partial}
\newcommand\EE{\EuScript{E}}
\newcommand\EU{\EuScript{U}}
\newcommand\rem{$\clubsuit$}

\newcommand{\ttsmat}[4]{\big({ \textstyle {#1 \atop #3}{#2 \atop #4}}\big)}

\providecommand{\abs}[1]{\lvert#1\rvert}
\providecommand{\bigabs}[1]{\bigl\lvert#1\bigr\rvert}
\providecommand{\biggabs}[1]{\biggl\lvert#1\biggr\rvert}                                                                \providecommand{\Bigabs}[1]{\Bigl\lvert#1\Bigr\rvert}
\providecommand{\Biggabs}[1]{\Biggl\lvert#1\Biggr\rvert}
\providecommand{\norm}[1]{\lVert#1\rVert}
\providecommand{\bignorm}[1]{\bigl\lVert#1\bigr\rVert}
\providecommand{\biggnorm}[1]{\biggl\lVert#1\biggr\rVert}
\providecommand{\Bignorm}[1]{\Bigl\lVert#1\Bigr\rVert}
\providecommand{\Biggnorm}[1]{\Biggl\lVert#1\Biggr\rVert}

\providecommand{\supp}{\textnormal{supp}}
\providecommand{\grad}{\textnormal{grad}}
\providecommand{\diver}{\textnormal{div}}
\providecommand{\dx}{\textnormal{dx}}
\providecommand{\Exp}{\textnormal{Exp}}
\providecommand{\Spin}{\textnormal{Spin}}
\providecommand{\Cl}{\textnormal{Cl}}
\providecommand{\CCl}{\mathbb{C}\textnormal{l}}
\providecommand{\Diff}{\textnormal{Diff}}
\providecommand{\Tr}{\textnormal{Tr}}
\providecommand{\Ob}{\textnormal{Ob}}
\providecommand{\id}{\textnormal{id}}
\providecommand{\Id}{\textnormal{Id}}
\providecommand{\Funct}{\textnormal{Funct}}
\providecommand{\Spec}{\textnormal{Spec}}
\providecommand{\ker}{\textnormal{ker}}
\providecommand{\Ker}{\textnormal{Ker}}
\providecommand{\IIm}{\textnormal{Im}}
\providecommand{\RRe}{\textnormal{Re}}
\providecommand{\coker}{\textnormal{coker}}
\providecommand{\im}{\textnormal{im}}
\providecommand{\coim}{\textnormal{coim}}
\providecommand{\Kom}{\textnormal{Kom}}
\providecommand{\dom}{\textnormal{dom}}
\providecommand{\codom}{\textnormal{codom}}
\providecommand{\Cyl}{\textnormal{Cyl}}
\providecommand{\Hom}{\textnormal{Hom}}
\providecommand{\SpHom}{\textnormal{SpHom}}
\providecommand{\Iso}{\textnormal{Iso}}
\providecommand{\End}{\textnormal{End}}
\providecommand{\Aut}{\textnormal{Aut}}
\providecommand{\Ker}{\textnormal{Ker}}
\providecommand{\Sets}{\textnormal{Sets}}
\providecommand{\Vect}{\textnormal{Vect}}
\providecommand{\rk}{\textnormal{rk}}
\providecommand{\Int}{\textnormal{Int}}
\providecommand{\ch}{\textnormal{ch}}
\providecommand{\PD}{\textnormal{PD}}
\providecommand{\cpt}{\textnormal{cpt}}
\providecommand{\ev}{\textnormal{ev}}
\providecommand{\odd}{\textnormal{odd}}
\providecommand{\GL}{\textnormal{GL}}
\providecommand{\Hol}{\textnormal{Hol}}
\providecommand{\Tor}{\textnormal{Tor}}
\providecommand{\Free}{\textnormal{Free}}
\providecommand{\pfaff}{\textnormal{pfaff}}
\providecommand{\Maps}{\textnormal{Maps}}
\providecommand{\Pfaff}{\textnormal{Pfaff}}
\providecommand{\mod}{\textnormal{mod}}
\providecommand{\Sq}{\textnormal{Sq}}
\providecommand{\sq}{\textnormal{sq}}
\providecommand{\PP}{\mathbb{P}}
\providecommand{\ZZ}{\mathbb{Z}}

\newcommand{\ii}{\mathrm{i}}
\def\0{\nonumber}
\def\tz{\tilde z}
\def\tc{\tilde c}
\def\a{{\bf a}}
\def\b{{\bf b}}
\def\c{{\bf c}}
\def\cO{\mathcal{O}}
\def\cL{\mathcal{L}}

\begin{document}
\begin{titlepage}

\begin{flushright}

\end{flushright}
 
\vskip 1cm
\begin{center}
 
{\LARGE\bf \boldmath Probing 7-branes on Orbifolds} 
 
 \vskip 2cm
 
{\large Simone Giacomelli,$^{1,2}$ Marina Moleti,$^3$ and Raffaele Savelli$^{4}$}

 \vskip 0.9cm
 
 {\it  $^1$ Dipartimento di Fisica, Universit\`a di Milano-Bicocca,\\ Piazza della Scienza 3, I-20126 Milano, Italy \\[2mm]
 
$^2$ INFN, sezione di Milano-Bicocca,\\ Piazza della Scienza 3,  I-20126 Milano, Italy \\[2mm]
 
$^3$ International School for Advanced Studies (SISSA/ISAS),\\
Via Bonomea 265, I-34136 Trieste, Italy\\[2mm]
 
 $^4$ Dipartimento di Fisica \& INFN, Universit\`a di Roma ``Tor Vergata'', \\ Via della Ricerca
Scientifica 1, I-00133 Roma, Italy
 }
 \vskip 2cm
 
\abstract{\noindent D3 branes in the vicinity of an E$_6$, E$_7$, or E$_8$ stack of 7-branes in flat space are known to host at low energy a famous class of strongly-coupled $\mathcal{N}=2$ superconformal field theories featuring exceptional global symmetry. What if, instead, the 7-branes wrap an orbifold? We give a systematic characterization of such theories in the case of $\mathbb{C}^2/\mathbb{Z}_n$, and determine their main properties, like global symmetries, spectra of Coulomb-branch operators, and patterns of Higgs-branch flows. We put forward a set of rules to construct their magnetic quivers, which directly generalize what happens in the perturbative case, and later derive them using the duality with M5 branes probing M9 walls.}

\end{center}

\end{titlepage}

\tableofcontents

\section{Introduction}

The study of the low-energy physics of D-branes probing singular spaces in string theory is by now a classic subject in the field. It has its inception in the work of Douglas and Moore \cite{Douglas:1996sw}, where quiver diagrams were first introduced to provide a handy description of the effective field theories arising on the worldvolume of the probe. Since then, countless different setups of branes at singularities, oftentimes connected by dualities, have been scrutinized for just as many different purposes, from constructing explicit particle-physics models relevant for phenomenology to addressing more general questions about the strong-coupling regime of quantum field theories.

In recent years it has been discovered that D-branes probing non-perturbative backgrounds in Type IIB/F-theory lead to a whole new class of strongly-coupled field theories, including superconformal theories in four dimensions with $\mathcal{N}=3$ supersymmetry \cite{Garcia-Etxebarria:2015wns, Aharony:2016kai}, which were believed not to exist until not long ago. The $\mathcal{N}=3$ construction has been recently generalized to less supersymmetric setups \cite{Apruzzi:2020pmv, Giacomelli:2020jel, Giacomelli:2020gee, Bourget:2020mez, Heckman:2020svr, Kimura:2021sbt}, providing, among other things, an explicit realization of all the rank-one Coulomb-branch geometries classified in \cite{Argyres:2015ffa, Argyres:2015gha, Argyres:2016xua, Argyres:2016yzz}. This results clearly motivate an in-depth study of such geometric constructions in order to achieve a clearer understanding of the landscape of superconformal theories. The present work constitutes a step in this direction. 
 
As a particular class of such setups, one can consider D3 branes in Type IIB string theory as point-particles probing a singular space of the form $\mathbb{C}^2/\Gamma\times\mathbb{C}$, with $\Gamma$ a discrete subgroup of $SU(2)$, and spice it up by adding D7 branes and O7 planes extended along the first factor. The resulting four-dimensional (4d) field theories on the probes are $\mathcal{N}=2$  supersymmetric gauge theories admitting a perturbative formulation. Depending on the number of D3 and D7 branes in the game, some of them may feature conformal symmetry as well as a number of exactly marginal deformations. The well-known rules for drawing the corresponding quiver diagrams employ the technology of fractional branes to extract gauge groups and matter content of the effective theories (see e.g.~chapter 11 of \cite{Ibanez:2012zz} for an introduction to the case $\Gamma=\mathbb{Z}_n$, and also \cite{Aldazabal:2000sa, Cvetic:2001tj, Franco:2007ii, Bianchi:2013gka, Bianchi:2020fuk}).

If one is willing to leave the realm of weakly-coupled string theory, one may wonder what kind of 4d field theories one gets by allowing the presence of mutually non-perturbative 7-branes in the probed Type IIB background. Here, however, our knowledge disappears, since these theories are intrinsically strongly coupled, and the techniques developed in the early days to analyze them break down.

But nowadays, thanks to the recent impressive improvement in the understanding of strongly-coupled supersymmetric theories, we can count on a variety of new tools to investigate the low-energy physics of these more general string configurations. In particular, although a quiver of ``electric'' type is not available for such theories, due to the absence of a Langrangian description, a ``magnetic'' one \cite{Cabrera:2019izd} is. The latter is the quiver of the theory obtained by applying mirror symmetry \cite{Intriligator:1996ex} to the reduction of the original theory to three dimensions.\footnote{The so-called ``bad theories'' \cite{Gaiotto:2008ak} are known to admit a magnetic quiver, even though they do not have a conventional mirror dual \cite{Bourget:2021jwo}. This fact, however, will not be relevant for the present work.} It is a common feature of intrinsically strongly-coupled theories to admit a magnetic-quiver description (see e.g.~\cite{Benini:2010uu,Cabrera:2018jxt, Cabrera:2019dob, Cremonesi:2015lsa, Ferlito:2017xdq, Closset:2020scj, Closset:2020afy, Giacomelli:2020ryy}), and the Type IIB models mentioned above are no exception.

It is the aim of this paper to employ magnetic quivers to start shedding light on the 4d field theories arising from probing parallel 7-branes of general type wrapped on four-dimensional orbifolds, and to uncover some of their properties. For simplicity, we will focus on abelian orbifolds of $\mathbb{C}^2$, and pay special attention to superconformal field theories (SCFT). A \emph{necessary} condition to have conformal symmetry is to probe a background with (locally) constant axio-dilaton, or equivalently to (locally) cancel the background 7-brane charge. This condition is strong enough to restrict the possible 7-brane stacks to a handful of cases, that include the stacks carrying $SO(8)$, E$_6$, E$_7$, and E$_8$ symmetries.\footnote{There are other three cases with constant axio-dilaton, which, in the language of F-theory, correspond to the Kodaira singularities II, III, and IV. We will not discuss these cases, leaving them for future work.} Only the first case (a.k.a. the D$_4$ stack) is perturbative, and hence approachable with the standard rules of franctional branes. As for the other three cases, which in a smooth background yield the celebrated Minahan-Nemeschansky theories \cite{Minahan:1996cj}, the corresponding field theory on the probe has never been determined, to the best of our knowledge, even for the simplest orbifolds.

Here we fill in this gap by giving a systematic description of the superconformal field theories associated to the E$_6$, E$_7$, E$_8$ stacks wrapping $\mathbb{C}^2/\mathbb{Z}_n$ orbifolds. In particular, we determine some of their main properties, such as global symmetry, spectrum of Coulomb-branch (CB) operators and Hasse diagram of Higgs-branch flows. We also elucidate, by way of examples, the role of the choice of holonomy, at the origin and at infinity of the orbifold space, for the background 7-brane gauge field.

To this end, we will first reformulate the D$_4$ case in the language of magnetic quivers (see \cite{Tachikawa:2014qaa, Mekareeya:2015bla} for earlier work on this subject). This will allow us to guess the general rules for constructing the magnetic quivers associated to the non-perturbative cases. Similarly to the perturbative situation, depending on the order of the orbifold we will find either one or two inequivalent projections, leading respectively to a single or to a couple of distinct families of superconformal theories. Then, we will derive the results of our extrapolation using the framework of M-theory: Building on the results of \cite{Ganor:1996mu,Seiberg:1996vs, Ganor:1996pc}, it can be argued that D3 branes probing an E$_8$-type stack of 7-branes wrapped on $\mathbb{C}^2/\mathbb{Z}_n$ are dual to M5 branes on a torus probing a M9 wall wrapped on the same orbifold \cite{Giacomelli:2020gee}, in the sense that the low-energy 4d theories arising on the branes are the same.\footnote{See also \cite{Heckman:2022suy} for a closely-related discussion.} One of the goals of the present work is to further clarify this correspondence. In particular, a subtle point which we emphasize is that this Type IIB/M-theory correspondence is true provided that the theory arising from the M-theory setup is suitably mass-deformed to remove the $SU(n)$ global symmetry inherited from the six-dimensional (6d) SCFT. This is similar to the situation discussed in \cite{Ohmori:2015pia}.\footnote{In the case of $\mathcal{S}$-fold geometries this mass deformation is not needed since the $SU(n)$ global symmetry is already broken by the holonomies along the cycles of the torus.} 

The above-mentioned connection with 6d SCFTs is crucial for our analysis, since the magnetic quivers of the relevant 6d theories are already known \cite{Mekareeya:2017jgc}, and this will allow us to systematically derive those of the 4d theories we are interested in. All the SCFTs we will find are class $\mathcal{S}$ theories of type $A$ \cite{Gaiotto:2009we},\footnote{See \cite{Ohmori:2015pua, Zafrir:2015rga, DelZotto:2015rca, Baume:2021qho} for further discussions about the relation between class $\mathcal{S}$ theories and $T^2$ compactifications of $\mathcal{N}=(1,0)$ SCFTâs.} and their three-dimensional (3d) mirror is given by an affine Dynkin diagram to which we attach a collection of abelian $U(1)$ nodes in such a way that the resulting quiver is star-shaped.

At this stage, a comment is in order. As is well known, the Higgs branch of 4d theories on the worldvolume of D3 branes probing 7-branes of type $\mathfrak{g}$ is the $\mathfrak{g}$ instanton moduli space (in the case at hand on $\mathbb{C}^2/\mathbb{Z}_n$). These spaces have been studied extensively, starting from the mathematical work \cite{KN1990}, and the corresponding magnetic quivers have been determined in several special cases in \cite{Hanany:1999sj, Mekareeya:2015bla, Mekareeya:2017jgc}. These are given by flavored affine Dynkin diagrams, and coincide with ours only when there is a single attached $U(1)$ node. In other cases, the quivers coincide once the abelian nodes have been ungauged. It is therefore natural to associate the theories on probe D3 branes to the flavored quivers. Since, as we will see, the flavored quivers do not correspond to conformal theories, which are the main focus of the present work, we concentrate on the gauged version and refer to the corresponding SCFTs as theories on D3 branes by an abuse of language. We observe that theories featuring extra abelian global symmetries (and therefore multiple $U(1)$ nodes in the magnetic quiver) actually originate from partial mass deformations of the 6d theories leaving part of the $SU(n)$ symmetry unbroken. 

The paper is organized as follows. We start in Section \ref{Sec:PertTheories} by reviewing the 4d quiver gauge theories engineered by D3 brane probes of an $SO(8)$ stack of D7 branes wrapped on $\mathbb{C}^2/\mathbb{Z}_n$ orbifolds: We explicitly discuss all conformal theories arising for $n\leq4$, with particular emphasis on the Higgs mechanisms linking them, and on the two inequivalent projections appearing for even $n$. As a byproduct we derive a new S-duality between  $\mathcal{N}=2$ lagrangian theories. In Section \ref{Sec:Non-Pert} we tackle the exceptional 7-brane stacks: By rewriting the perturbative case in the language of magnetic quivers, we identify a general algorithm to construct the non-perturbative theories and to determine their properties. In Section \ref{Sec:M-theory} we give the M-theory derivation of the rules we proposed, focusing mainly on the case of the E$_8$ stack (even though we explain how to analyze the other cases), which is more directly connected to the M-theory setup. We finally draw our conclusions in Section \ref{Sec:Concl}. Appendices \ref{FIdef} and \ref{Sec:CBSpectrum} review some technical material needed to deal with magnetic quivers and Class $\mathcal{S}$ theories, whereas Appendix \ref{Sec:MoreEx} contains more examples with higher orbifold order which were too large to fit in the main text.

\section{Perturbative theories}\label{Sec:PertTheories}

In this introductory section we will describe the 4d theories obtained by probing with D3 branes a D$_4$ stack of 7-branes (4 D7/image-D7's on top of an O7$^-$ plane) wrapping the orbifolds $\mathbb{C}^2/\mathbb{Z}_n$. The aim is to highlight certain distinctive features of theirs, which we will then discover also in their non-perturbative counterparts.

The rules to derive the low-energy spectrum of such theories have long been known (see e.g.~\cite{Bianchi:2013gka}, where the present setting appears as a special case). The local complex coordinates of the internal threefold behave under the orbifold action as
\be
(z_1,z_2,z_3)\longrightarrow (e^{2\pi {\rm i}/n}z_1, e^{-2\pi {\rm i}/n}z_2,z_3)\,,
\ee
and the D7/O7's are located at $z_3=0$. Cutting a long story short, what happens is the following. The stack of D3-brane probes splits into $n$ fractional stacks, and so does the stack of D7 branes. If a stack of D3/D7 branes is its own orientifold image it carries a symplectic/orthogonal gauge/flavor symmetry, otherwise it carries an unitary gauge/flavor symmetry. On the one hand, fields in the 3-3 open-string sector describing movements of the probes along the $\{z_1,z_2\}$ plane form a number of hypermultiplets: If they connect two gauge groups that are the orientifold images of one another they transform in the antisymmetric representation of that gauge group in the unoriented theory, otherwise they remain bifundamentals after the orientifold quotient. On the other hand, fields in the 3-7 open-string sector form hypermultiplets transforming in the representations $\{(N_p,\bar{M}_p)\oplus(M_p,\bar{N}_p)\}$, where $N_p,M_p$ are the numbers of the $p$th fractional D3,D7 branes respectively. 

Let us discuss in detail what happens for $n\leq4$, focusing in particular on the superconformal theories.

\paragraph{$\bf n=1$} This is of course the well-known case of $N$ D3 branes probing a smooth D$_4$ stack of 7-branes \cite{Banks:1996nj}, leading to the quiver
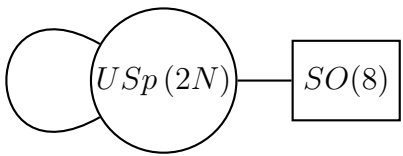
\begin{center}
\begin{tikzpicture}[-,thick, scale=0.8]
  \node[circle, draw, inner sep= 1pt](L1) at (10,0){$USp\left(2N\right)$};
  \node[draw, rectangle, minimum width=30pt, minimum height=30pt](L2) at (13,0){$SO(8)$};
 \path[every node/.style={font=\sffamily\small,
  		fill=white,inner sep=1pt}]
(L1) edge  (L2)
%(L2) edge [bend left] node[below=2mm] {$Q_i$}(L1)
(L1) edge [loop, in=210, out=150, looseness=5]  (L1)
;
\end{tikzpicture}\vspace{-1cm}
\end{center}\be {\label{n=1quivers}}\ee
Here $N$ stands for the number of D3 branes \emph{without} counting their orientifold images separately.\footnote{Such a convention will be more suited for the non-perturbative generalization.} The total number of D7 branes plus image-D7 branes is fixed to $8$ by conformal invariance (which agrees with the fact that a D$_4$ stack of 7-branes does not backreact on the axio-dilaton). Such a field theory has rank $N$ and its CB operators have conformal dimensions $2,4,6,\ldots,2N$. The segment between the two nodes is a bifundamental hypermultiplet, whereas the segment starting and ending on the gauge node is a hypermultiplet transforming in the antisymmetric representation of the gauge group. Geometrically, the latter field describes movements of the D3 stack along the (internal part of) the 7-brane worldvolume. Note that the antisymmetric representation of a symplectic algebra is \emph{not} irreducible, but can be decomposed into a symplectic trace (which is a singlet) and a traceless part. When the number of D3 branes is minimal (i.e.~$N=1$), the traceless part is absent and the antisymmetric field is just a free hypermultiplet (thus the corresponding loop-edge can be removed from the quiver). \\ 

Consider now the situation where the 7-brane stack is wrapped on an orbifold $\mathbb{C}^2/\mathbb{Z}_n$. When the order of the orbifold group is even, there are two inequivalent orientifold projections, corresponding to the two conjugacy classes of reflections in the dihedral group $\mathbb{D}_n$, the symmetry group of the Chan-Paton lattice: We indicate them with the subscripts $_{\bf v}$ and $_{\bf e}$, according to whether the involution is a reflection with respect to an axis passing through vertices and edges of the corresponding regular polygon respectively.

\paragraph{$\bf n=2_v$} In this case we have two vertices (the two square-roots of unity) and the orientifold involution fixes both of them. Applying the rules of fractional branes, we find the following quiver
\begin{center}
\begin{tikzpicture}[-,thick, scale=0.8]
  \node[circle, draw, inner sep= 1pt](L1) at (10,0){$USp\left(2N_1\right)$};
  \node[draw, rectangle, minimum width=30pt, minimum height=30pt](L2) at (13,0){$SO(2M_1)$};
 \node[circle, draw, inner sep= 1pt](L3) at (7,0){$USp\left(2N_0\right)$};
 \node[draw, rectangle, minimum width=30pt, minimum height=30pt](L4) at (4,0){$SO(2M_0)$};
 \path[every node/.style={font=\sffamily\small,
  		fill=white,inner sep=1pt}]
(L1) edge  (L2)
(L1) edge  (L3)
(L3) edge  (L4);
\end{tikzpicture}\vspace{-1cm}
\end{center}\be {}\ee
where $N_0/M_0,N_1/M_1$ are the number of fractional D3/D7 branes in the two stacks (without counting the orientifold images). Conformal invariance is imposed by two equations, corresponding to the vanishing of the beta functions of the two gauge nodes. We can write them as 
\begin{eqnarray}\label{Conf_n=2}
M_0+M_1=4\,,\nonumber\\
M_0-M_1=8(N_0-N_1)\,,
\end{eqnarray}
where the first equation says that we must be probing a 7-brane stack of type D$_4$. There are only three solutions to this system. The ``symmetric''\footnote{We will give this property a more fundamental meaning, when looking at magnetic quivers in Section \ref{Sec:D4rev}.} one is $M_0=M_1=4$ and $N_0=N_1\equiv N$ corresponding to the quiver
\begin{center}
\begin{tikzpicture}[-,thick, scale=0.8]
  \node[circle, draw, inner sep= 1pt](L1) at (10,0){$USp\left(2N\right)$};
  \node[draw, rectangle, minimum width=30pt, minimum height=30pt](L2) at (13,0){$SO(4)$};
 \node[circle, draw, inner sep= 1pt](L3) at (7,0){$USp\left(2N\right)$};
 \node[draw, rectangle, minimum width=30pt, minimum height=30pt](L4) at (4,0){$SO(4)$};
 \path[every node/.style={font=\sffamily\small,
  		fill=white,inner sep=1pt}]
(L1) edge  (L2)
(L1) edge  (L3)
(L3) edge  (L4);
\end{tikzpicture}\vspace{-1cm}
\end{center}\be {\label{SymmetricNZ2}}\ee
The above describes a rank-$2N$ theory, with pairs of CB operators of dimensions $2,4,6,\ldots,2N$.

Let us see what happens if we give non-trivial vev to the left-most fundamental flavors (analogous statements hold for the right-most flavors). The moment map here is a rank-$2$ matrix in the adjoint of $SO(4)$. Suppose we give the latter a vev in the principal nilpotent orbit (i.e.~the $[3,1]$). Then, the $SO(4)$ flavor symmetry is completely broken, together with a $USp(2)$ subgroup of the left-most gauge group. Consequently, the right-most gauge group gains two new fundamental flavors, such that the resulting quiver is 
\begin{center}
\begin{tikzpicture}[-,thick, scale=0.8]
  \node[circle, draw, inner sep= 1pt](L1) at (10,0){$USp\left(2N\right)$};
  \node[draw, rectangle, minimum width=30pt, minimum height=30pt](L2) at (14,0){$SO(8)$};
 \node[circle, draw, inner sep= 1pt](L3) at (6,0){$USp\left( 2N-2\right)$};
% \node[draw, rectangle, minimum width=30pt, minimum height=30pt](L4) at (4,0){$SO(4)$};
 \path[every node/.style={font=\sffamily\small,
  		fill=white,inner sep=1pt}]
(L1) edge  (L2)
(L1) edge  (L3);
%(L3) edge  (L4);
\end{tikzpicture}\vspace{-1cm}
\end{center}\be {\label{asymmetricQuiverSpSO}}\ee
corresponding to the rank-$(2N-2)$ superconformal point reached by the RG flow. This theory could have as well been obtained by making a different choice of Chan-Paton embedding of the orbifold group, that is $N_0=N$, $N_1=N-1$, $M_0=4$, $M_1=0$. This is one of the two other solutions of the conformality constraints \eqref{Conf_n=2} (the last solution is obviously obtained by $0\leftrightarrow1$). Turning on a nilpotent vev for $SO(8)$ in \eqref{asymmetricQuiverSpSO} leads us back to the quiver structure \eqref{SymmetricNZ2}, with $N$ lowered by one unit. These operations yield a cascade of SCFTs, oscillating between the quiver structures \eqref{SymmetricNZ2} and \eqref{asymmetricQuiverSpSO}, and ending when all the gauge groups disappear and only a pair of free hypers is left over.

Let us briefly discuss what happens if we give vev to the bifundamental hypermultiplet in the middle of the quiver \eqref{SymmetricNZ2}. Given its gauge quantum numbers, its condensation will break the gauge group to the diagonal $USp(2N)$ subgroup. This diagonal subgroup will have 4 fundamental hypers (the sum of the two hypers on each side), and a hyper in the antisymmetric representation (which corresponds to the $4N^2-N(2N+1)$ massless degrees of freedom of the bifundamental hyper left over after condensation).\footnote{Only for $N=1$, this operation is equivalent to the one discussed before of giving vev to the fundamental hypers on one side of the quiver.} The resulting quiver is \eqref{n=1quivers}. Geometrically, the operation of giving vev to the bifulndamental field corresponds to binding the two fractional D3 stacks together. The ensuing bound state then behaves as an integral stack of $N$ D3 branes, and can thus move off the orbifold singularity, as described by vevs for the antisymmetric field.

\paragraph{$\bf n=2_e$} In this case the orientifold involution exchanges the two fractional D3-brane stacks. Therefore, the resulting quiver is
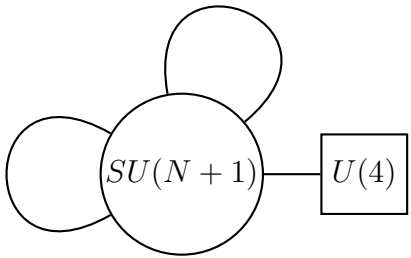
\begin{center}
\begin{tikzpicture}[-,thick, scale=0.8]
  \node[circle, draw, inner sep= 1pt](L1) at (10,0){$SU(N+1)$};
  \node[draw, rectangle, minimum width=30pt, minimum height=30pt](L2) at (13,0){$U(4)$};
 \path[every node/.style={font=\sffamily\small,
  		fill=white,inner sep=1pt}]
(L1) edge  (L2)
%(L2) edge [bend left] node[below=2mm] {$Q_i$}(L1)
(L1) edge [loop, in=210, out=150, looseness=5]  (L1)
(L1) edge [loop, in=40, out=100, looseness=5]  (L1);
\end{tikzpicture}\vspace{-1cm}
\end{center}\be {\label{n=2eQuivers}}\ee
where the loop-edges denote two hypermultiplets both transforming in the antisymmetric representation of the gauge group. Because of these antisymmetric hypers, the flavor symmetry gains an $U(2)$ factor. In this case, conformal invariance leaves $N$ unconstrained, but fixes to $4$ the number of D7 branes (as expected from local tadpole cancelation).

Strictly speaking, the gauge group of theory in the ultraviolet is $U(N)$. This is because, while the diagonal $U(1)$ between the D3 stack and its orientifold image is projected out, the relative one survives. As a consequence, the global symmetry of the above quiver looses a $U(1)$ and reaches the same rank of \eqref{SymmetricNZ2}. This extra abelian gauge group is crucial for matching the Higgs branch of such theories with $SO(8)$ instanton moduli spaces on $\mathbb{C}^2/\mathbb{Z}_2$. However, since here we are focusing on superconformal theories emerging in the infrared,\footnote{In particular, the class of theories \eqref{n=2eQuivers} has first appeared in \cite{Gukov:1998kt}.} the abelian gauge symmetries are irrelevant, and therefore will all be ignored throughout the paper.

The minimal choice $N=1$ is somewhat special, in that the two antisymmetric hypers are singlets and thus decouple, and the flavor symmetry of the interacting sector is enhanced to $SO(8)$. Therefore the interacting theory is identical to the one without orbifold, given by the quiver \eqref{n=1quivers} with $N=1$.

The $N=2$ case is also a bit special, because the two antisymmetric fields provide two extra fundamental hypermultiplets, which, together with the four fundamental hypers, gives $SU(3)$ SQCD with six fundamental flavors, here (unusually) engineered using Type IIB string theory.

The two antisymmetric fields originate from open strings connecting the fractional D3 stack to its orientifold image. This implies that, when some of their degrees of freedom condense, stack and image-stack bind together creating an integral D3-brane stack. The latter bound state, being now orientifold invariant, must carry a symplectic gauge group. The massless degrees of freedom left over, then, form a hypermultiplet transforming in the antisymmetric representation of the unbroken gauge group,\footnote{There is an extra singlet left over only when $N$ is odd.} corresponding to the fact that the bound state is now free to move along the 7-brane worldvolume. This precisely reproduces the family of theories \eqref{n=1quivers}. We summarize this process as follows
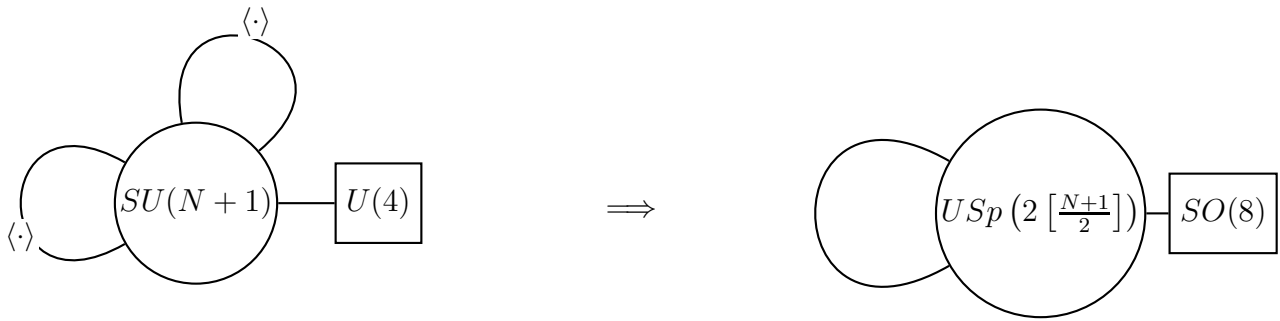
\begin{flushleft}
\begin{tikzpicture}[-,thick, scale=0.8]
  \node[circle, draw, inner sep= 1pt](L1) at (10,0){$SU(N+1)$};
  \node[draw, rectangle, minimum width=30pt, minimum height=30pt](L2) at (13,0){$U(4)$};
 \path[every node/.style={font=\sffamily\small,
  		fill=white,inner sep=1pt}]
(L1) edge  (L2)
%(L2) edge [bend left] node[below=2mm] {$Q_i$}(L1)
(L1) edge [loop, in=210, out=150, looseness=5] node[below=2mm] {$\langle\cdot\rangle$} (L1)
(L1) edge [loop, in=40, out=100, looseness=5] node[above=.01mm] {$\langle\cdot\rangle$}  (L1);
\end{tikzpicture}
\end{flushleft} 
\begin{center}\vspace{-2.4cm}$\Longrightarrow$\end{center}
\begin{flushright}\vspace{-2.6cm}
\begin{tikzpicture}[-,thick, scale=0.8]
  \node[circle, draw, inner sep= 1pt](L1) at (10,0){$USp\left(2\left[\frac{N+1}{2}\right]\right)$};
  \node[draw, rectangle, minimum width=30pt, minimum height=30pt](L2) at (13,0){$SO(8)$};
 \path[every node/.style={font=\sffamily\small,
  		fill=white,inner sep=1pt}]
(L1) edge  (L2)
%(L2) edge [bend left] node[below=2mm] {$Q_i$}(L1)
(L1) edge [loop, in=210, out=150, looseness=5] (L1);
\end{tikzpicture}
\end{flushright}

\paragraph{$\bf n=3$} Let us now discuss the $\mathbb{Z}_3$ orbifold. Here, as for all the odd cases, only one orientifold projection is possible, up to equivalences. There are three fractional D3-brane stacks associated to the three roots of unity: The $0$th root is orientifold invariant, whereas the other two roots are the orientifold image of one another. The resulting quiver is
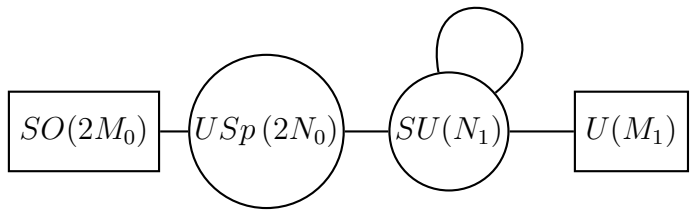
\begin{center}
\begin{tikzpicture}[-,thick, scale=0.8]
  \node[circle, draw, inner sep= 1pt](L1) at (10,0){$SU(N_1)$};
  \node[draw, rectangle, minimum width=30pt, minimum height=30pt](L2) at (13,0){$U(M_1)$};
 \node[circle, draw, inner sep= 1pt](L3) at (7,0){$USp\left(2N_0\right)$};
 \node[draw, rectangle, minimum width=30pt, minimum height=30pt](L4) at (4,0){$SO(2M_0)$};
 \path[every node/.style={font=\sffamily\small,
  		fill=white,inner sep=1pt}]
(L1) edge  (L2)
(L1) edge  (L3)
(L3) edge  (L4)
(L1) edge [loop, in=40, out=100, looseness=5] (L1);
\end{tikzpicture}\be {\label{n=3orbifold}}\ee
\end{center}
where the loop-edge indicates a hypermultiplet transforming in the antisymmetric representation of the unitary gauge group. Of course, there are flavor-symmetry enhancements in special cases: $U(M_1)$ becomes $SO(2M_1)$ when $N_1=2$ and $U(M_1+1)$ when $N_1=3$.  We can formulate the conditions for superconformal invariance as follows
\begin{eqnarray}
M_0+M_1&=&4\,,\nonumber\\
M_0-M_1&=&2(2N_0-N_1)\,,
\end{eqnarray}
where the first condition expresses, as usual, local 7-brane tadpole cancelation. There are five different families of solutions to the above equations, which we label by the number $N$:\footnote{Again the minimal choice $N=1$ falls back (modulo a free hyper) on a theory associated to a lower-order orbifold (in this case $\mathbb{Z}_2$). In Section \ref{Sec:D4rev}, using magnetic quivers, we will show exactly when the minimal choice for $N$ leads to such a degeneration.}

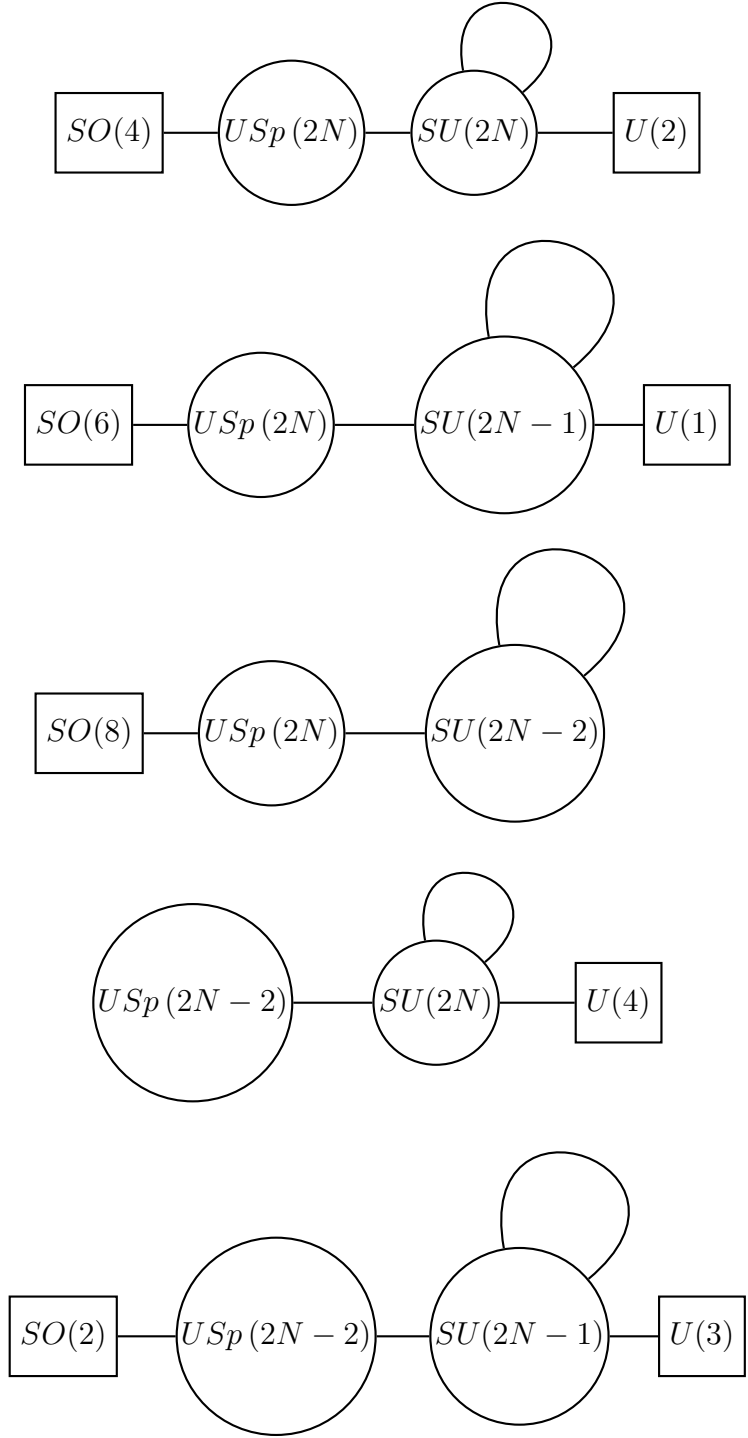
\begin{center}\vspace{-.8cm}
\begin{tikzpicture}[-,thick, scale=0.8]
  \node[circle, draw, inner sep= 1pt](L1) at (10,0){$SU(2N)$};
  \node[draw, rectangle, minimum width=30pt, minimum height=30pt](L2) at (13,0){$U(2)$};
 \node[circle, draw, inner sep= 1pt](L3) at (7,0){$USp\left(2N\right)$};
 \node[draw, rectangle, minimum width=30pt, minimum height=30pt](L4) at (4,0){$SO(4)$};
 \path[every node/.style={font=\sffamily\small,
  		fill=white,inner sep=1pt}]
(L1) edge  (L2)
(L1) edge  (L3)
(L3) edge  (L4)
(L1) edge [loop, in=40, out=100, looseness=5] (L1);
\end{tikzpicture}
\end{center}

\begin{center}\vspace{-1cm}
\begin{tikzpicture}[-,thick, scale=0.8]
  \node[circle, draw, inner sep= 1pt](L1) at (11,0){$SU(2N-1)$};
  \node[draw, rectangle, minimum width=30pt, minimum height=30pt](L2) at (14,0){$U(1)$};
 \node[circle, draw, inner sep= 1pt](L3) at (7,0){$USp\left(2N\right)$};
 \node[draw, rectangle, minimum width=30pt, minimum height=30pt](L4) at (4,0){$SO(6)$};
 \path[every node/.style={font=\sffamily\small,
  		fill=white,inner sep=1pt}]
(L1) edge  (L2)
(L1) edge  (L3)
(L3) edge  (L4)
(L1) edge [loop, in=40, out=100, looseness=5] (L1);
\end{tikzpicture}
\end{center}

\begin{center}\vspace{-1cm}
\begin{tikzpicture}[-,thick, scale=0.8]
  \node[circle, draw, inner sep= 1pt](L1) at (11,0){$SU(2N-2)$};
  %\node[draw, rectangle, minimum width=30pt, minimum height=30pt](L2) at (13,0){$U(M_1)$};
 \node[circle, draw, inner sep= 1pt](L3) at (7,0){$USp\left(2N\right)$};
 \node[draw, rectangle, minimum width=30pt, minimum height=30pt](L4) at (4,0){$SO(8)$};
 \path[every node/.style={font=\sffamily\small,
  		fill=white,inner sep=1pt}]
%(L1) edge  (L2)
(L1) edge  (L3)
(L3) edge  (L4)
(L1) edge [loop, in=40, out=100, looseness=5] (L1);
\end{tikzpicture}
\end{center}

\begin{center}\vspace{-.5cm}
\begin{tikzpicture}[-,thick, scale=0.8]
  \node[circle, draw, inner sep= 1pt](L1) at (10,0){$SU(2N)$};
  \node[draw, rectangle, minimum width=30pt, minimum height=30pt](L2) at (13,0){$U(4)$};
 \node[circle, draw, inner sep= 1pt](L3) at (6,0){$USp\left(2N-2\right)$};
% \node[draw, rectangle, minimum width=30pt, minimum height=30pt](L4) at (4,0){$SO(M_0)$};
 \path[every node/.style={font=\sffamily\small,
  		fill=white,inner sep=1pt}]
(L1) edge  (L2)
(L1) edge  (L3)
%(L3) edge  (L4)
(L1) edge [loop, in=40, out=100, looseness=5] (L1);
\end{tikzpicture}
\end{center}

\begin{center}\vspace{-.8cm}
\begin{tikzpicture}[-,thick, scale=0.8]
  \node[circle, draw, inner sep= 1pt](L1) at (11,0){$SU(2N-1)$};
  \node[draw, rectangle, minimum width=30pt, minimum height=30pt](L2) at (14,0){$U(3)$};
 \node[circle, draw, inner sep= 1pt](L3) at (7,0){$USp\left(2N-2\right)$};
 \node[draw, rectangle, minimum width=30pt, minimum height=30pt](L4) at (3.5,0){$SO(2)$};
 \path[every node/.style={font=\sffamily\small,
  		fill=white,inner sep=1pt}]
(L1) edge  (L2)
(L1) edge  (L3)
(L3) edge  (L4)
(L1) edge [loop, in=40, out=100, looseness=5] (L1);
\end{tikzpicture}\vspace{-1.3cm}
\end{center}\be \label{list_n=3}\ee
\\
The above theories are all interconnected by higgsing fundamental flavors. As we have already seen for the $\mathbb{Z}_2$ orbifold, this procedure triggers a cascade of theories ending eventually on a bunch of free hypers, after all gauge groups have been broken. For instance, one can go from the first to the second quiver by giving vev to one fundamental hypermultiplet on the r.h.s., which breaks the unitary gauge group to $SU(2N-1)$, and makes the symplectic gauge group gain one fundamental flavor. The complete breaking of the unitary flavor symmetry leads to the third quiver, whereas the complete breaking of the orthogonal flavor symmetry leads to the fourth quiver. From the latter, one can reach the fifth quiver by higgsing $SU(2N)$ to $SU(2N-1)$. A further higgsing to $SU(2N-2)$ yields back the first quiver with $N$ lowered by one unit. And so on, so forth.

The antisymmetric field comes from open strings connecting the (unitary) fractional D3-brane stack to its orientifold image. Giving it a vev binds together the two stacks, creating an orientifold-invariant one, carrying a symplectic gauge group. This operation connects $n=3$ quivers (like the first one or the third and the fourth ones in the above list) to $n=2$ quivers (like \eqref{SymmetricNZ2} or \eqref{asymmetricQuiverSpSO}). This is not so surprising, because only the diagonal vector fields of the fractional stack/image-stack remain massless, thus preventing the probe D3 branes to fully ``see'' the singularity. It is also possible to reach \eqref{SymmetricNZ2} and \eqref{asymmetricQuiverSpSO} respectively from the fifth and the second quiver above, by giving vev to the antisymmetric hyper as well as to one fundamental hyper.\footnote{What if instead we give vev to the bifundamental hypermultiplet? We land directly on the $n=1$ quivers \eqref{n=1quivers}, because the invariant fractional stack binds to the non-invariant one (and consequently to its image too), thus making the probe D3 branes unable to see the singularity at all.}

The superconformal theory represented e.g.~by the first quiver in the above list, has rank $3N-1$, with pairs of CB operators of even dimensions from $2$ to $N$ and a single CB operator for each odd dimension from $3$ to $2N-1$.

\paragraph{$\bf n=4_v$} We now come to the $\mathbb{Z}_4$ orbifold, with the choice of orientifold involution that fixes the $0$th and the $2$nd fourth roots of unity, and exchanges the other two roots. With generic numbers of D3/D7 branes, we get the following quiver

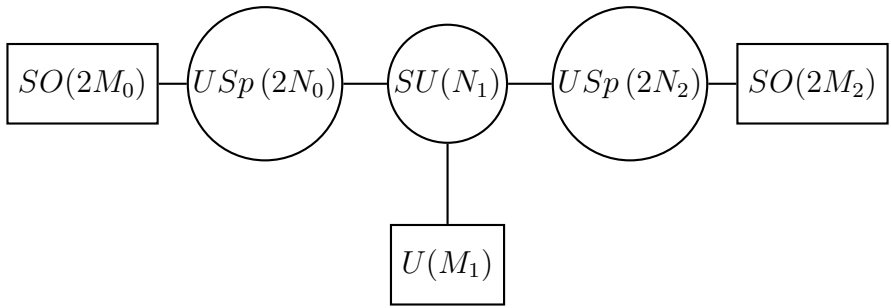
\begin{center}
\begin{tikzpicture}[-,thick, scale=0.8]
  \node[circle, draw, inner sep= 1pt](L1) at (10,0){$SU(N_1)$};
  \node[circle, draw, inner sep= 1pt](L2) at (13,0){$USp\left(2N_2\right)$};
 \node[circle, draw, inner sep= 1pt](L3) at (7,0){$USp\left(2N_0\right)$};
\node[draw, rectangle, minimum width=30pt, minimum height=30pt](L5) at (16,0){$SO(2M_2)$};
 \node[draw, rectangle, minimum width=30pt, minimum height=30pt](L4) at (4,0){$SO(2M_0)$};
 \node[draw, rectangle, minimum width=30pt, minimum height=30pt](L6) at (10,-3){$U(M_1)$};
 \path[every node/.style={font=\sffamily\small,
  		fill=white,inner sep=1pt}]
(L1) edge  (L2)
(L1) edge  (L3)
(L5) edge  (L2)
(L1) edge  (L6)
(L3) edge  (L4);
\end{tikzpicture}\vspace{-1cm}
\end{center}\be {\label{n=4orbifold}}\ee
where the $U(M_1)$ flavor symmetry gets enhanced to $SO(2M_1)$ when $N_1=2$. We find the following conditions for superconformal invariance
\begin{eqnarray}
M_0+M_1+M_2&=&4\,,\nonumber\\
M_0-M_2&=&2N_0-N_1\,,\nonumber\\
M_1&=&2(N_1-N_0-N_2)\,,
\end{eqnarray}
where we recognize in the first equation the requirement to cancel the D7 charge of the O7-plane. There are several solutions to this system, which produce families of quivers that are all interconnected by a pattern of higgsing for the three sets of fundamental flavors. We limit ourselves to writing the most ``symmetric'' family:
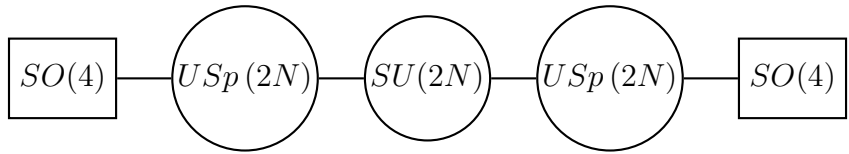
\begin{center}
\begin{tikzpicture}[-,thick, scale=0.8]
  \node[circle, draw, inner sep= 1pt](L1) at (10,0){$SU(2N)$};
  \node[circle, draw, inner sep= 1pt](L2) at (13,0){$USp\left(2N\right)$};
 \node[circle, draw, inner sep= 1pt](L3) at (7,0){$USp\left(2N\right)$};
\node[draw, rectangle, minimum width=30pt, minimum height=30pt](L5) at (16,0){$SO(4)$};
 \node[draw, rectangle, minimum width=30pt, minimum height=30pt](L4) at (4,0){$SO(4)$};
% \node[draw, rectangle, minimum width=30pt, minimum height=30pt](L6) at (10,-3){$U(2)$};
 \path[every node/.style={font=\sffamily\small,
  		fill=white,inner sep=1pt}]
(L1) edge  (L2)
(L1) edge  (L3)
(L5) edge  (L2)
%(L1) edge  (L6)
(L3) edge  (L4);
\end{tikzpicture}\vspace{-1cm}
\end{center}\be {\label{n=4orbifoldS}}\ee
representing a rank-$(4N-1)$ theory. As already described earlier, giving vevs to bifundamental flavors decreases the order of the orbifold seen by the probe D3 branes, because fractional stacks form bound states and are no longer able to fully resolve the singularity.

\paragraph{$\bf n=4_e$} The $\mathbb{Z}_4$ orbifold allows for another orientifold projection, under which, contrary to the previous one, no fractional stack is invariant. Therefore, the resulting quiver is
\begin{center}
\begin{tikzpicture}[-,thick, scale=0.8]
  \node[circle, draw, inner sep= 1pt](L1) at (10,0){$SU(N_3)$};
  \node[draw, rectangle, minimum width=30pt, minimum height=30pt](L2) at (13,0){$U(M_3)$};
 \node[circle, draw, inner sep= 1pt](L3) at (7,0){$SU(N_0)$};
 \node[draw, rectangle, minimum width=30pt, minimum height=30pt](L4) at (4,0){$U(M_0)$};
 \path[every node/.style={font=\sffamily\small,
  		fill=white,inner sep=1pt}]
(L1) edge  (L2)
(L1) edge  (L3)
(L3) edge  (L4)
(L3) edge [loop, in=80, out=140, looseness=5] (L3)
(L1) edge [loop, in=40, out=100, looseness=5] (L1);
\end{tikzpicture}\vspace{-1cm}
\end{center}\be{}\ee
where the loop-edges are hypermultiplets transforming in the antisymmetric representation of the corresponding gauge group. There are flavor-symmetry enhancements from $U(M_{0/3})$ to $SO(2M_{0/3})$ or $U(M_{0/3}+1)$ when $N_{0,3}=2$ or $N_{0,3}=3$ respectively.  The conditions for superconformal invariance are
\begin{eqnarray}
M_0+M_3&=&4\,,\nonumber\\
M_0-M_3&=&2(N_0-N_3)\,,
\end{eqnarray}
where the first equation indeed fixes the total number of D7 branes (without counting orientifold images) to $4$, as required by the absence of backreaction on the axio-dilaton. There are three solutions to this system (ignoring the mirror ones, obtained by exchanging $0\leftrightarrow3$):\footnote{Again, the minimal choice $N=1$ leads back to the $\mathbb{Z}_2$ orbifold.}
\begin{center}\vspace{-.8cm}
\begin{tikzpicture}[-,thick, scale=0.8]
  \node[circle, draw, inner sep= 1pt](L1) at (10.5,0){$SU(N+1)$};
  \node[draw, rectangle, minimum width=30pt, minimum height=30pt](L2) at (13.5,0){$U(2)$};
 \node[circle, draw, inner sep= 1pt](L3) at (7,0){$SU(N+1)$};
 \node[draw, rectangle, minimum width=30pt, minimum height=30pt](L4) at (4,0){$U(2)$};
 \path[every node/.style={font=\sffamily\small,
  		fill=white,inner sep=1pt}]
(L1) edge  (L2)
(L1) edge  (L3)
(L3) edge  (L4)
(L3) edge [loop, in=80, out=140, looseness=5] (L3)
(L1) edge [loop, in=40, out=100, looseness=5] (L1);
\end{tikzpicture}
\end{center}

\begin{center}\vspace{-1cm}
\begin{tikzpicture}[-,thick, scale=0.8]
  \node[circle, draw, inner sep= 1pt](L1) at (10,0){$SU(N+1)$};
  \node[draw, rectangle, minimum width=30pt, minimum height=30pt](L2) at (13,0){$U(3)$};
 \node[circle, draw, inner sep= 1pt](L3) at (7,0){$SU(N)$};
 \node[draw, rectangle, minimum width=30pt, minimum height=30pt](L4) at (4,0){$U(1)$};
 \path[every node/.style={font=\sffamily\small,
  		fill=white,inner sep=1pt}]
(L1) edge  (L2)
(L1) edge  (L3)
(L3) edge  (L4)
(L3) edge [loop, in=80, out=140, looseness=5] (L3)
(L1) edge [loop, in=40, out=100, looseness=5] (L1);
\end{tikzpicture}
\end{center}

\begin{center}\vspace{-1cm}
\begin{tikzpicture}[-,thick, scale=0.8]
  \node[circle, draw, inner sep= 1pt](L1) at (10,0){$SU(N+1)$};
  \node[draw, rectangle, minimum width=30pt, minimum height=30pt](L2) at (13,0){$U(4)$};
 \node[circle, draw, inner sep= 1pt](L3) at (6.5,0){$SU(N-1)$};
 %\node[draw, rectangle, minimum width=30pt, minimum height=30pt](L4) at (4,0){$U(M_0)$};
 \path[every node/.style={font=\sffamily\small,
  		fill=white,inner sep=1pt}]
(L1) edge  (L2)
(L1) edge  (L3)
%(L3) edge  (L4)
(L3) edge [loop, in=80, out=140, looseness=5] (L3)
(L1) edge [loop, in=40, out=100, looseness=5] (L1);
\end{tikzpicture}
\end{center}\vspace{-.9cm}
\be\label{ElQuiverZ4e}
\ee
One goes down in this list by successively higgsing the fundamental flavors on the l.h.s.~of the quiver. Vevs for any of the antisymmetric fields lead to $n=3$ quivers, whereas vevs for the bifundamental field yields $n=2_{\rm e}$ quivers. This is all compatible with the fact that such vevs create bound states between different pairs of fractional stacks, thus making the D3 branes no longer able to probe the full $\mathbb{Z}_4$ singularity.

\section{Non-perturbative superconformal theories}\label{Sec:Non-Pert}

In this section we will determine the superconformal field theories living on D3 branes probing a non-perturbative 7-brane which wraps a $\mathbb{C}^2/\mathbb{Z}_n$ singular space. We will use the results of the previous section concerning the 7-brane of type D$_4$ as a guideline to identify the right structure underlying this type of theories. After discussing in Section \ref{Sec:GenProp} some general properties we expect from such theories, and revisiting the perturbative case in Section \ref{Sec:D4rev}, we present the general rules to treat the non-perturbative cases in Section \ref{Sec:GenDisc}, and describe the properties of the resulting theories. To exemplify the rules, we discuss in detail the SCFTs arising from a E$_8$, E$_7$ and E$_6$ stack wrapping the orbifold $\mathbb{C}^2/\mathbb{Z}_2$ in Sections \ref{Sec:E8}, \ref{Sec:E7}, \ref{Sec:E6} respectively. We defer a detailed description of higher-order orbifolds to Appendix \ref{Sec:MoreEx}.

\subsection{General properties}\label{Sec:GenProp}

Let us start by discussing the features we should expect from the geometric realization of the theory. 

The field theories resulting from probing with $N$ D3 branes a stack of 7-branes wrapped on an orbifold $\mathbb{C}^2/\mathbb{Z}_n$ are expected to have rank at most $nN$. We start by pointing out that we can compute the $a$ and $c$ central charges of the field theory using holography \cite{Aharony:2007dj, Apruzzi:2020pmv}: 
\be\label{formulacc}\begin{array}{ll} 
8a-4c & = (n\Delta_7)N^2+(2n\Delta_7\epsilon+\Delta_7-1)N+\alpha N^0,\\
24(c-a) & = 6N(\Delta_7-1)+\beta N^0.
\end{array}\ee
In the above formula $\Delta_7$ denotes the deficit angle associated with the chosen 7-brane and the quantity $\epsilon$ is the D3 charge of the background, which receives both a topological contribution proportional to the Euler character of the background and a contribution which depends on the choice of holonomy at infinity and at the origin of $\mathbb{C}^2/\mathbb{Z}_n$ for the gauge field supported on the 7-brane. The available choices for a 7-brane of type $\mathfrak{g}$ are classified by all possible subsets of nodes of the affine $\mathfrak{g}$ Dynkin diagram such that the sum of their comarks (counted with multiplicity) is $n$.\footnote{Here we are assuming that the holonomies at the origin and at infinity are inner-automorphisms of $\mathfrak{g}$ of order $n$.} 

Apart from the central charges, we can make a prediction about the global symmetry of the theory and part of its Higgs branch flows:\footnote{These are the properties valid generically. Sporadic exceptions are possible.} The global symmetry is the product  of the isometry of the background ($SU(2)$ for $n=2$ and $U(1)$ for $n>2$) and the symmetry coming from the 7-brane, which depends on the choice of holonomies for the gauge field supported on the 7-brane. Regarding the set of Higgs-branch flows, we know this should include at least a flow to the corresponding $N$-instanton theory in flat space (a.k.a.~Minahan-Nemeschansky theory) for the following reason: By moving the D3 branes along the transverse $\mathbb{C}^2/\mathbb{Z}_n$, away from the singular point, the probes only see a 7-brane in flat space and consequently their worldvolume theory reduces to the $N$-instanton model. From this consideration we also conclude that (for $N=1$) the transverse slice is $\mathbb{C}^2/\mathbb{Z}_n$. 

Throughout this section, we will describe the field theories by means of their magnetic quivers \cite{Cabrera:2019izd}, i.e.~the mirror duals of their reduction to three dimensions. Such quivers will allow us to read off a number of properties of the SCFTs, including flavor symmetry, dimensions of CB operators and pattern of Higgs-branch flows.

\subsection{The 7-brane of type D$_4$ revisited}\label{Sec:D4rev}

Here we will reinterpret the perturbative case discussed in Section \ref{Sec:PertTheories} in a language that will turn out more appropriate for the non-perturbative generalizations. Looking at the $\mathbb{Z}_2$, $\mathbb{Z}_3$, and $\mathbb{Z}_4$ orbifolds will be enough to visualize the general behavior. The relevant magnetic quivers (or 3d mirrors) are discussed in \cite{Hanany:1999sj} and can also be derived using the class $\mathcal{S}$ technology developed in \cite{Benini:2010uu, Chacaltana:2010ks}.

\subsubsection*{$\bf\mathbb{C}^2/\mathbb{Z}_2$}

The global symmetry commuting with the $\mathbb{Z}_2$ holonomy can be $SO(8)$, $SU(4)\times U(1)$ or $SU(2)^4$. The case $SU(2)^4$ leads to the family of models $$\boxed{2}-USp(2N)-USp(2N)-\boxed{2}$$ corresponding to the family of electric quivers \eqref{SymmetricNZ2}. The CB spectrum includes two sets of operators of dimension $2,4,6,\dots ,2N-2,2N$ and the central charges are $$8a-4c=4N^2+2N\,;\quad 24(c-a)=6N\,,$$ which fits with \eqref{formulacc} if we set $\epsilon=1/8$ and $\alpha=\beta=0$. The corresponding magnetic quiver is 
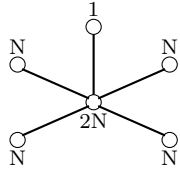
\be\label{D4NMQ}
\begin{tikzpicture}
\filldraw[fill= white] (0,0.5) circle [radius=0.1] node[below] {\scriptsize N};
\filldraw[fill= white] (1,1) circle [radius=0.1] node[below] {\scriptsize 2N};
\filldraw[fill= white] (2,0.5) circle [radius=0.1] node[below] {\scriptsize N};
\filldraw[fill= white] (0,1.5) circle [radius=0.1] node[above] {\scriptsize N};
\filldraw[fill= white] (2,1.5) circle [radius=0.1] node[above] {\scriptsize N};
\filldraw[fill= white] (1,2) circle [radius=0.1] node[above] {\scriptsize 1};
\draw [thick] (0.06, 0.56) -- (0.94,0.95) ;
\draw [thick] (1.94, 0.56) -- (1.06,0.95) ;
\draw [thick] (1, 1.1) -- (1,1.9) ;
\draw [thick] (0.06, 1.44) -- (0.94,1.05) ;
\draw [thick] (1.94, 1.44) -- (1.06,1.05) ;
\end{tikzpicture}\ee
We note that this is the affine Dynkin diagram of D$_4$ with an extra node of label $1$ attached to the central node. The extra abelian node is essentially associated to the bifundamental field linking the two copies of the gauge group. The quickest way to see why this is the magnetic quiver corresponding to \eqref{SymmetricNZ2} is to compare the global symmetries. One can read off the global symmetry associated to a magnetic quiver as follows: All the balanced subquivers contribute a simple factor of the symmetry according to its Dynkin-diagram shape; each unbalanced node contribute a $U(1)$; one Abelian node decouples and hence does not contribute to the global symmetry. For example, in the quiver \eqref{D4NMQ} the middle node is unbalanced while, the four nodes of label $N$, which instead are all balanced, form the Dynkin diagram of $SU(2)^4$. As a consequence, for generic values of $N$, the flavor symmetry is $SU(2)^4\times U(1)$, as expected from \eqref{SymmetricNZ2}.

If instead we choose an $SO(8)$-preserving holonomy we find the family \eqref{asymmetricQuiverSpSO}, i.e. $$USp(2N-2)-USp(2N)-\boxed{4}$$ whose magnetic quiver is
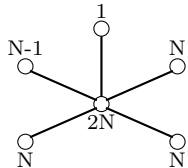
\be\label{SONMQ}
\begin{tikzpicture}
\filldraw[fill= white] (0,0.5) circle [radius=0.1] node[below] {\scriptsize N};
\filldraw[fill= white] (1,1) circle [radius=0.1] node[below] {\scriptsize 2N};
\filldraw[fill= white] (2,0.5) circle [radius=0.1] node[below] {\scriptsize N};
\filldraw[fill= white] (0,1.5) circle [radius=0.1] node[above] {\scriptsize N-1};
\filldraw[fill= white] (2,1.5) circle [radius=0.1] node[above] {\scriptsize N};
\filldraw[fill= white] (1,2) circle [radius=0.1] node[above] {\scriptsize 1};
\draw [thick] (0.06, 0.56) -- (0.94,0.95) ;
\draw [thick] (1.94, 0.56) -- (1.06,0.95) ;
\draw [thick] (1, 1.1) -- (1,1.9) ;
\draw [thick] (0.06, 1.44) -- (0.94,1.05) ;
\draw [thick] (1.94, 1.44) -- (1.06,1.05) ;
\end{tikzpicture}\ee
Here we have a balanced subquiver with the shape of the D$_4$ Dynkin diagram, and therefore the flavor symmetry is generically $SO(8)\times U(1)$, as in \eqref{asymmetricQuiverSpSO}.
The quiver \eqref{SONMQ} can be obtained from \eqref{D4NMQ} by higgsing one of the four $SU(2)$ symmetries. At the level of magnetic quivers, this process is implemented through the technique of ``quiver subtraction'', which we will explain in detail in Section \ref{Sec:E8}. The CB spectrum of the above family is almost as in the $SU(2)^4$ family, except for the fact that one of the dimension $2N$ operators is missing. In this case we should set $\epsilon=-3/8$, $\alpha=1$ and $\beta=-1$ in \eqref{formulacc} to recover the central charges.

From reiterated higgsing processes (or, as we will see, magnetic quiver subtractions), we find a sequence of RG flows connecting these two families of theories, summarized by the following Hasse diagram: 
\be\label{HDD41} 
\begin{tikzpicture} 
\node[] at (-0.8,0) {$\cdots$};
\node[] (A) at (0,0) {\scriptsize $SU(2)^4$}; 
\node[] at (0,0.3) {\scriptsize (N)};
\node[] (B) at (3,0) {\scriptsize $SO(8)$}; 
\node[] at (3,0.3) {\scriptsize (N)};
\node[] (C) at (6,0) {\scriptsize $SU(2)^4$}; 
\node[] at (6,0.3) {\scriptsize (N-1)}; 
\node[] (D) at (9,0) {\scriptsize $SO(8)$}; 
\node[] at (9,0.3) {\scriptsize (N-1)};
\node[] at (1.5,0.2) {\scriptsize $\mathfrak{a}_1$};
\node[] at (4.5,0.2) {\scriptsize $\mathfrak{d}_4$};
\node[] at (7.5,0.2) {\scriptsize $\mathfrak{a}_1$};
\node[] at (10.1,0) {$\cdots$};
\draw[->] (A) -- (B); 
\draw[->] (B) -- (C); 
\draw[->] (C) -- (D); 
\end{tikzpicture}
\ee 
where the symbols $\mathfrak{a}_1,\mathfrak{d}_4$ indicate that the higgsing involves the flavor symmetry $SU(2),SO(8)$ respectively.
This behavior is exactly the one we observed in Section \ref{Sec:PertTheories}, for the probe theories arising from the D$_4$ stack on the $\mathbb{C}^2/\mathbb{Z}_2$ orbifold of type $2_{\rm v}$.

The case of holonomy preserving a $SU(4)\times U(1)$ subgroup, which corresponds to the family of quivers \eqref{n=2eQuivers}, splits into two similar sequences of theories, according to whether the number of colors is even or odd.  One sequence is given by a $SU(2N)$ gauge theory with two antisymmetric hypermultiplets and four fundamentals, whose CB spectrum includes operators of dimension $2,3,4,\dots ,2N-1,2N$, and we need to set $\epsilon=-1/8$, $\alpha=-1$ and $\beta=1$ in \eqref{formulacc} to recover the central charges. The corresponding magnetic quiver is 
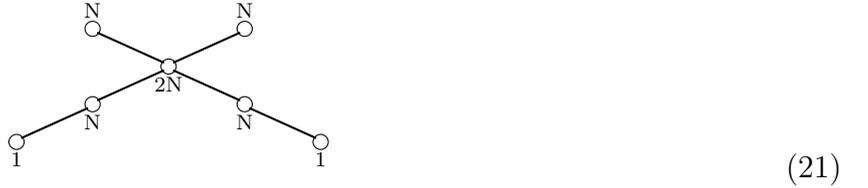
\be\label{SU4NMQ}
\begin{tikzpicture}
\filldraw[fill= white] (0,0.5) circle [radius=0.1] node[below] {\scriptsize N};
\filldraw[fill= white] (1,1) circle [radius=0.1] node[below] {\scriptsize 2N};
\filldraw[fill= white] (2,0.5) circle [radius=0.1] node[below] {\scriptsize N};
\filldraw[fill= white] (0,1.5) circle [radius=0.1] node[above] {\scriptsize N};
\filldraw[fill= white] (2,1.5) circle [radius=0.1] node[above] {\scriptsize N};
\filldraw[fill= white] (-1,0) circle [radius=0.1] node[below] {\scriptsize 1};
\filldraw[fill= white] (3,0) circle [radius=0.1] node[below] {\scriptsize 1};
\draw [thick] (0.06, 0.55) -- (0.94,0.95) ;
\draw [thick] (1.94, 0.55) -- (1.06,0.95) ;
\draw [thick] (-0.94, 0.05) -- (-0.06,0.45) ;
\draw [thick] (2.94, 0.05) -- (2.06,0.45) ;
\draw [thick] (0.06, 1.45) -- (0.94,1.05) ;
\draw [thick] (1.94, 1.45) -- (1.06,1.05) ;
\end{tikzpicture}\ee
This corresponds to attaching an extra node of label $1$ to \emph{two} of the four tails of the extended Dynkin diagram of D$_4$. Note that, for this operation to be non-trivial, we must consider more than one D3 probe, i.e.~$N\geq2$. For the minimal choice $N=1$, indeed, this family degenerates to the theory living on a single D3-brane probing the D$_4$ stack in flat space, i.e.~$SU(2)$ SQCD with four fundamental flavors. This is exactly what we noticed for the quiver \eqref{n=2eQuivers}. We would also like to point out that the global symmetry of the family \eqref{SU4NMQ} contains an extra $U(1)$ compared to \eqref{D4NMQ} and to the ultraviolet theory expected on the D3-brane worldvolume. Nevertheless, as we already stressed in Section \ref{Sec:PertTheories}, our focus is on the interacting SCFT emerging in the infrared, whose global symmetry includes the additional $U(1)$.

The other series of theories is given by a $SU(2N-1)$ gauge theory with two antisymmetric hypermultiplets and four fundamentals, whose magnetic quiver is 
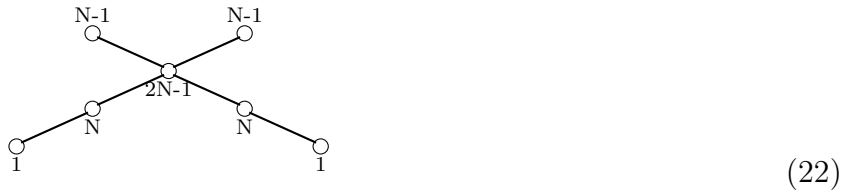
\be\label{SU42NMQ}
\begin{tikzpicture}
\filldraw[fill= white] (0,0.5) circle [radius=0.1] node[below] {\scriptsize N};
\filldraw[fill= white] (1,1) circle [radius=0.1] node[below] {\scriptsize 2N-1};
\filldraw[fill= white] (2,0.5) circle [radius=0.1] node[below] {\scriptsize N};
\filldraw[fill= white] (0,1.5) circle [radius=0.1] node[above] {\scriptsize N-1};
\filldraw[fill= white] (2,1.5) circle [radius=0.1] node[above] {\scriptsize N-1};
\filldraw[fill= white] (-1,0) circle [radius=0.1] node[below] {\scriptsize 1};
\filldraw[fill= white] (3,0) circle [radius=0.1] node[below] {\scriptsize 1};
\draw [thick] (0.06, 0.55) -- (0.94,0.95) ;
\draw [thick] (1.94, 0.55) -- (1.06,0.95) ;
\draw [thick] (-0.94, 0.05) -- (-0.06,0.45) ;
\draw [thick] (2.94, 0.05) -- (2.06,0.45) ;
\draw [thick] (0.06, 1.45) -- (0.94,1.05) ;
\draw [thick] (1.94, 1.45) -- (1.06,1.05) ;
\end{tikzpicture}\ee 
which is exactly the quiver we obtain by higgsing the $SU(4)$ symmetry of \eqref{SU4NMQ}. The CB spectrum of this second family includes operators of all possible integer dimensions from $2$ to $2N-1$, and in this case we should set $\epsilon=-5/8$ and $\alpha=\beta=-2$ in \eqref{formulacc} to recover the central charges. Again, from reiterated higgsing processes, we find a sequence of RG flows connecting these two families of theories, summarized by the following Hasse diagram (for large enough $N$): \be\label{HDD42} 
\begin{tikzpicture} 
\node[] at (-1.2,0) {$\cdots$};
\node[] (A) at (0,0) {\scriptsize $SU(4)\times U(1)$}; 
\node[] at (0,0.3) {\scriptsize (N)};
\node[] (B) at (3,0) {\scriptsize $SU(4)\times U(1)$}; 
\node[] at (3,0.3) {\scriptsize (N)};
\node[] (C) at (6,0) {\scriptsize $SU(4)\times U(1)$}; 
\node[] at (6,0.3) {\scriptsize (N-1)}; 
\node[] (D) at (9,0) {\scriptsize $SU(4)\times U(1)$}; 
\node[] at (9,0.3) {\scriptsize (N-1)};
\node[] at (1.5,0.2) {\scriptsize $\mathfrak{a}_3$};
\node[] at (4.5,0.2) {\scriptsize $\mathfrak{a}_3$};
\node[] at (7.5,0.2) {\scriptsize $\mathfrak{a}_3$};
\node[] at (10.3,0) {$\cdots$};
\draw[->] (A) -- (B); 
\draw[->] (B) -- (C); 
\draw[->] (C) -- (D); 
\end{tikzpicture}
\ee
where we see that the higgsing always involves the flavor symmetry $SU(4)$.

\subsubsection*{$\bf\mathbb{C}^2/\mathbb{Z}_3$}
In Section \ref{Sec:PertTheories} we saw that all of the O7$^-$ orientifold projections of the $\mathbb{C}^2/\mathbb{Z}_3$ orbifold along the 7-brane worldvolume are equivalent to each other. Therefore we have a single sequence of RG flows connecting the five families of superconformal theories described by the quivers in \eqref{list_n=3}. The family with flavor symmetry containing $SO(4)\times U(2)$, which is associated to the first of those quivers, has a magnetic quiver (or 3d mirror) of the form
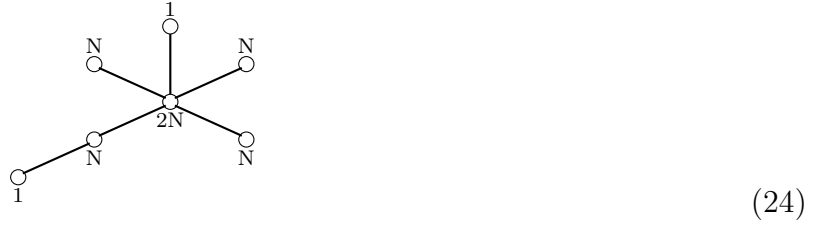
\be\label{SO4U2NMQ}
\begin{tikzpicture}
\filldraw[fill= white] (0,0.5) circle [radius=0.1] node[below] {\scriptsize N};
\filldraw[fill= white] (1,1) circle [radius=0.1] node[below] {\scriptsize 2N};
\filldraw[fill= white] (2,0.5) circle [radius=0.1] node[below] {\scriptsize N};
\filldraw[fill= white] (0,1.5) circle [radius=0.1] node[above] {\scriptsize N};
\filldraw[fill= white] (2,1.5) circle [radius=0.1] node[above] {\scriptsize N};
\filldraw[fill= white] (-1,0) circle [radius=0.1] node[below] {\scriptsize 1};
\filldraw[fill= white] (1,2) circle [radius=0.1] node[above] {\scriptsize 1};
\draw [thick] (0.06, 0.55) -- (0.94,0.95) ;
\draw [thick] (1.94, 0.55) -- (1.06,0.95) ;
\draw [thick] (-0.94, 0.05) -- (-0.06,0.45) ;
\draw [thick] (1, 1.1) -- (1,1.9) ;
\draw [thick] (0.06, 1.45) -- (0.94,1.05) ;
\draw [thick] (1.94, 1.45) -- (1.06,1.05) ;
\end{tikzpicture}\ee
We recognize this as the extended Dynkin diagram of D$_4$ with an extra node of label $1$ attached to both the central node and to one of the four tails. Also in this case we must consider at least two probe D3 branes, otherwise we reduce the order of the orbifold and get back \eqref{D4NMQ} with $N=1$. The magnetic quiver for the second and fourth theories in \eqref{list_n=3}, associated with a $SO(6)\times U(1)$-preserving holonomy, reads 
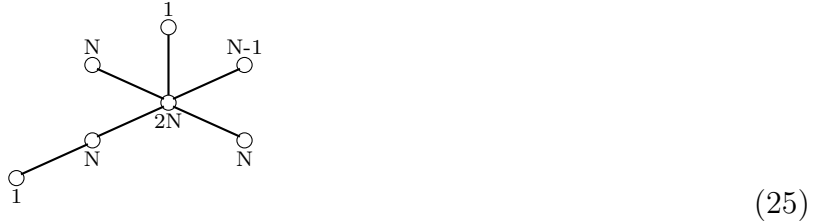
\be\label{SO6U1NMQ}
\begin{tikzpicture}
\filldraw[fill= white] (0,0.5) circle [radius=0.1] node[below] {\scriptsize N};
\filldraw[fill= white] (1,1) circle [radius=0.1] node[below] {\scriptsize 2N};
\filldraw[fill= white] (2,0.5) circle [radius=0.1] node[below] {\scriptsize N};
\filldraw[fill= white] (0,1.5) circle [radius=0.1] node[above] {\scriptsize N};
\filldraw[fill= white] (2,1.5) circle [radius=0.1] node[above] {\scriptsize N-1};
\filldraw[fill= white] (-1,0) circle [radius=0.1] node[below] {\scriptsize 1};
\filldraw[fill= white] (1,2) circle [radius=0.1] node[above] {\scriptsize 1};
\draw [thick] (0.06, 0.55) -- (0.94,0.95) ;
\draw [thick] (1.94, 0.55) -- (1.06,0.95) ;
\draw [thick] (-0.94, 0.05) -- (-0.06,0.45) ;
\draw [thick] (1, 1.1) -- (1,1.9) ;
\draw [thick] (0.06, 1.45) -- (0.94,1.05) ;
\draw [thick] (1.94, 1.45) -- (1.06,1.05) ;
\end{tikzpicture}\ee
The magnetic quiver for the third model  in \eqref{list_n=3}, which has trivial $SO(8)$-preserving holonomy, is 
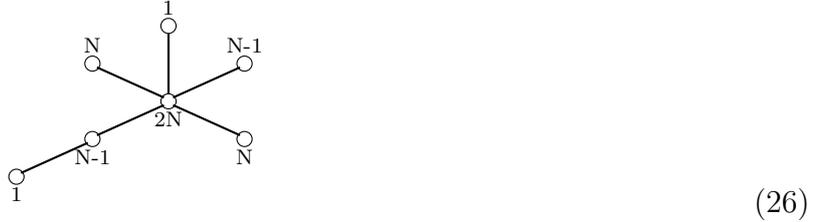
\be\label{SO8NMQ}
\begin{tikzpicture}
\filldraw[fill= white] (0,0.5) circle [radius=0.1] node[below] {\scriptsize N-1};
\filldraw[fill= white] (1,1) circle [radius=0.1] node[below] {\scriptsize 2N};
\filldraw[fill= white] (2,0.5) circle [radius=0.1] node[below] {\scriptsize N};
\filldraw[fill= white] (0,1.5) circle [radius=0.1] node[above] {\scriptsize N};
\filldraw[fill= white] (2,1.5) circle [radius=0.1] node[above] {\scriptsize N-1};
\filldraw[fill= white] (-1,0) circle [radius=0.1] node[below] {\scriptsize 1};
\filldraw[fill= white] (1,2) circle [radius=0.1] node[above] {\scriptsize 1};
\draw [thick] (0.06, 0.55) -- (0.94,0.95) ;
\draw [thick] (1.94, 0.55) -- (1.06,0.95) ;
\draw [thick] (-0.94, 0.05) -- (-0.06,0.45) ;
\draw [thick] (1, 1.1) -- (1,1.9) ;
\draw [thick] (0.06, 1.45) -- (0.94,1.05) ;
\draw [thick] (1.94, 1.45) -- (1.06,1.05) ;
\end{tikzpicture}\ee
Finally, the last theory  in \eqref{list_n=3} associated with $U(3)\times U(1)$-preserving holonomy is described by the magnetic quiver 
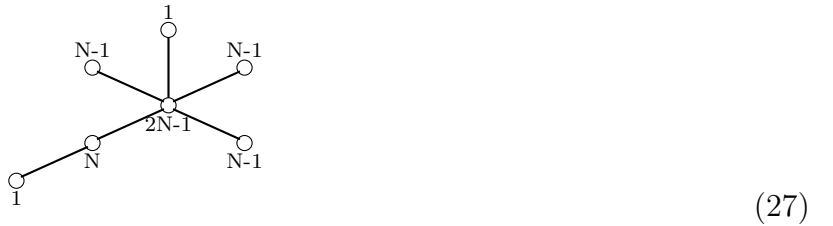
\be\label{SU3NMQ}
\begin{tikzpicture}
\filldraw[fill= white] (0,0.5) circle [radius=0.1] node[below] {\scriptsize N};
\filldraw[fill= white] (1,1) circle [radius=0.1] node[below] {\scriptsize 2N-1};
\filldraw[fill= white] (2,0.5) circle [radius=0.1] node[below] {\scriptsize N-1};
\filldraw[fill= white] (0,1.5) circle [radius=0.1] node[above] {\scriptsize N-1};
\filldraw[fill= white] (2,1.5) circle [radius=0.1] node[above] {\scriptsize N-1};
\filldraw[fill= white] (-1,0) circle [radius=0.1] node[below] {\scriptsize 1};
\filldraw[fill= white] (1,2) circle [radius=0.1] node[above] {\scriptsize 1};
\draw [thick] (0.06, 0.55) -- (0.94,0.95) ;
\draw [thick] (1.94, 0.55) -- (1.06,0.95) ;
\draw [thick] (-0.94, 0.05) -- (-0.06,0.45) ;
\draw [thick] (1, 1.1) -- (1,1.9) ;
\draw [thick] (0.06, 1.45) -- (0.94,1.05) ;
\draw [thick] (1.94, 1.45) -- (1.06,1.05) ;
\end{tikzpicture}\ee

By reiterated higgsing processes, we find a sequence of RG flows connecting the families of theories displayed in \eqref{list_n=3}, summarized by the following Hasse diagram (for large enough $N$):
\be
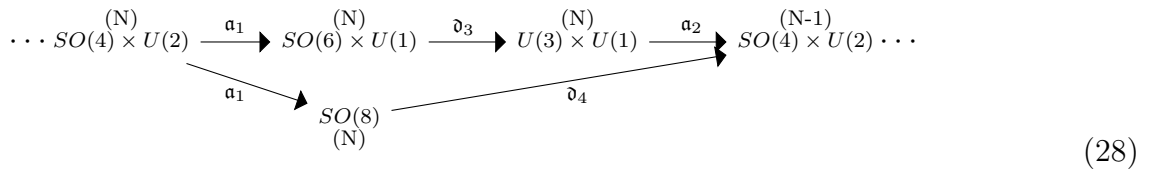
\label{HDD43} 
\begin{tikzpicture} 
\node[] at (-1.2,0) {$\cdots$};
\node[] (A) at (0,0) {\scriptsize $SO(4)\times U(2)$}; 
\node[] at (0,0.3) {\scriptsize (N)};
\node[] (B) at (3,0) {\scriptsize $SO(6)\times U(1)$}; 
\node[] at (3,0.3) {\scriptsize (N)};
\node[] (E) at (3,-1) {\scriptsize $SO(8)$}; 
\node[] at (3,-1.3) {\scriptsize (N)};
\node[] (C) at (6,0) {\scriptsize $U(3)\times U(1)$}; 
\node[] at (6,0.3) {\scriptsize (N)}; 
\node[] (D) at (9,0) {\scriptsize $SO(4)\times U(2)$}; 
\node[] at (9,0.3) {\scriptsize (N-1)};
\node[] at (1.5,0.2) {\scriptsize $\mathfrak{a}_1$};
\node[] at (1.5,-0.7) {\scriptsize $\mathfrak{a}_1$};
\node[] at (4.5,0.2) {\scriptsize $\mathfrak{d}_3$};
\node[] at (6,-0.7) {\scriptsize $\mathfrak{d}_4$};
\node[] at (7.5,0.2) {\scriptsize $\mathfrak{a}_2$};
\node[] at (10.25,0) {$\cdots$};
\draw[->] (A) -- (B); 
\draw[->] (B) -- (C); 
\draw[->] (C) -- (D); 
\draw[->] (A) -- (E);
\draw[->] (E) -- (D);
\end{tikzpicture}
\ee 
This exactly reproduces the process of higgsing fundamental flavors described in Section \ref{Sec:PertTheories}. From the $SO(4)\times U(2)$ theory we find the $SO(6)\times U(1)$ model by activating a vev for the $USp(2N)$ fundamentals, whereas we get the $SO(8)$ model by turning on a vev for the $SU(2N)$ fundamentals. 

Notice that the second and fourth quivers in \eqref{list_n=3} have the same magnetic quiver. These models also share the same flavor symmetry, â 't Hooft anomalies and CB spectrum, suggesting that they actually coincide, even though they arise from different orientifold projections. Indeed the statement is true for $N=1$, since they both reduce to $SU(2)$ SQCD with four flavors. For $N=2$ they are respectively 
\be\label{ccc}\boxed{3}-USp(4)-SU(3)-\boxed{2}\ee 
and \be\label{ddd} SU(2)-SU(4)-\boxed{4}\ee 
where the $SU(4)$ group also has a hypermultiplet in the antisymmetric representation. In order to relate the two theories, we exploit the Argyres-Seiberg duality \cite{Argyres:2007cn} for the $USp(4)$ gauge group on the left of \eqref{ccc}: This theory is known to be equivalent to a $SU(2)$ gauging of the rank-one E$_7$ Minahan-Nemeschansky theory. We can therefore rewrite \eqref{ccc} as 
$$SU(2)-E_7-SU(3)-\boxed{2}$$ 
where we have a $SU(2)\times SU(3)$ gauging of the E$_7$ Minahan-Nemeschansky theory. This is equivalent to \eqref{ddd} provided that a $SU(3)$ vector multiplet coupled to the $E_7$ theory and to two fundamentals is equivalent to the $SU(4)$ theory with 6 flavors and one antisymmetric. Remarkably, this duality was proposed by Argyres and Wittig in \cite{Argyres:2007tq} (see the eight entry in Table 2 of \cite{Argyres:2007tq}). We therefore find a peculiar duality between two different $\mathcal{N}=2$ superconformal lagrangian theories.

When we will come to non-perturbative theories in the next subsections, lacking the electric-quiver description, we will have to rely solely on magnetic quivers. And indeed oftentimes we will not manage to display all possible holonomy choices allowed by a given orbifold. Among all possible theories a distinguished role is played by the most ``symmetric'' family from which all other theories originate via a Higgs-branch flow. An example is given by \eqref{SO4U2NMQ} which describes the $SO(4)\times U(2)$ series we have discussed before. This family, for a 7-brane of type $\mathfrak{g}$, is characterized at the level of magnetic quivers by the fact that the non-abelian gauge nodes all have rank $Na_i^{\vee}$, where $a_i^{\vee}$ denote the comarks of the corresponding nodes in the Dynkin diagram of type $\mathfrak{g}$. This restriction is equivalent to requiring the  $\mathfrak{g}$ holonomies at the origin and at infinity of the orbifold singularity to coincide. In the perturbative $\mathfrak{g}=\mathfrak{d}_4$ case this corresponds to an homogeneous distribution of fractional D3 branes, or equivalently to the absence of ``unpaired'' fractional D3 charges. From now on, we will call this class of theories the ``canonical family''.

\subsubsection*{$\bf\mathbb{C}^2/\mathbb{Z}_4$}
Now we look at the $\mathbb{Z}_4$ orbifold of the D$_4$ stack, which, as we saw in Section \ref{Sec:PertTheories}, comes in two inequivalent versions.

The first version, i.e.~$4_{\rm v}$, features several families of theories, the canonical one being associated to the electric quiver \eqref{n=4orbifoldS}. Its magnetic dual is the snowflake quiver
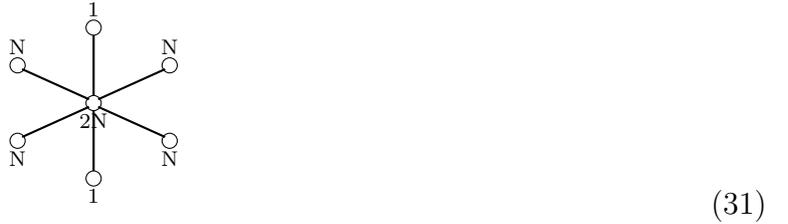
\be\label{SU24NMQ4v}
\begin{tikzpicture}
\filldraw[fill= white] (0,0.5) circle [radius=0.1] node[below] {\scriptsize N};
\filldraw[fill= white] (1,1) circle [radius=0.1] node[below] {\scriptsize 2N};
\filldraw[fill= white] (2,0.5) circle [radius=0.1] node[below] {\scriptsize N};
\filldraw[fill= white] (0,1.5) circle [radius=0.1] node[above] {\scriptsize N};
\filldraw[fill= white] (2,1.5) circle [radius=0.1] node[above] {\scriptsize N};
\filldraw[fill= white] (1,0) circle [radius=0.1] node[below] {\scriptsize 1};
\filldraw[fill= white] (1,2) circle [radius=0.1] node[above] {\scriptsize 1};
\draw [thick] (0.06, 0.55) -- (0.94,0.95) ;
\draw [thick] (1.94, 0.55) -- (1.06,0.95) ;
\draw [thick] (1, 0.9) -- (1,0.1) ;
\draw [thick] (1, 1.1) -- (1,1.9) ;
\draw [thick] (0.06, 1.45) -- (0.94,1.05) ;
\draw [thick] (1.94, 1.45) -- (1.06,1.05) ;
\end{tikzpicture}\ee
This corresponds to attaching two nodes of label $1$ to the central node of the affine Dynkin diagram of D$_4$. This family starts with a single probe brane, which gives rise to a rank-$3$ superconformal theory. Again, from reiterated higgsing processes, we find a sequence of RG flows connecting various families of theories associated to different holonomy choices. We summarize it by the following Hasse diagram:
\be\label{HDD44v} 
\begin{tikzpicture} 
\node[] at (-.8,0) {$\cdots$};
\node[] (A) at (0,0) {\scriptsize $SO(4)^2$}; 
\node[] at (0,0.3) {\scriptsize (N)};
\node[] (B) at (3,0) {\scriptsize $SU(2)^3$}; 
\node[] at (3,0.3) {\scriptsize (N)};
\node[] (C) at (6,0) {\scriptsize $SU(4)$}; 
\node[] at (6,0.3) {\scriptsize (N)}; 
\node[] (D) at (9,0) {\scriptsize $SU(2)$}; 
\node[] at (9,0.3) {\scriptsize (N)};
\node[] (E) at (12,0) {\scriptsize $SO(4)^2$}; 
\node[] at (12,0.3) {\scriptsize (N-1)};
\node[] at (1.5,0.2) {\scriptsize $\mathfrak{a}_1$};
\node[] at (4.5,0.2) {\scriptsize $\mathfrak{a}_1$};
\node[] at (7.5,0.2) {\scriptsize $\mathfrak{a}_3$};
\node[] at (10.5,0.2) {\scriptsize $\mathfrak{a}_1$};
\node[] at (12.8,0) {$\cdots$};
\draw[->] (A) -- (B); 
\draw[->] (B) -- (C); 
\draw[->] (C) -- (D); 
\draw[->] (D) -- (E); 
\end{tikzpicture}
\ee 
where we have displayed only the non-abelian part of the flavor symmetry.

Coming to the other $\mathbb{Z}_4$ orbi-orientifold, i.e.~$4_{\rm e}$, we have that the canonical family of superconformal theories is the one given by the first quiver in \eqref{ElQuiverZ4e}. The corresponding magnetic quiver is ($N\geq2$)
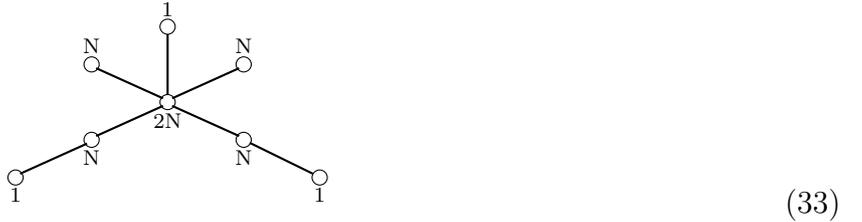
\be\label{U22NMQ4e}
\begin{tikzpicture}
\filldraw[fill= white] (0,0.5) circle [radius=0.1] node[below] {\scriptsize N};
\filldraw[fill= white] (1,1) circle [radius=0.1] node[below] {\scriptsize 2N};
\filldraw[fill= white] (2,0.5) circle [radius=0.1] node[below] {\scriptsize N};
\filldraw[fill= white] (0,1.5) circle [radius=0.1] node[above] {\scriptsize N};
\filldraw[fill= white] (2,1.5) circle [radius=0.1] node[above] {\scriptsize N};
\filldraw[fill= white] (-1,0) circle [radius=0.1] node[below] {\scriptsize 1};
\filldraw[fill= white] (1,2) circle [radius=0.1] node[above] {\scriptsize 1};
\filldraw[fill= white] (3,0) circle [radius=0.1] node[below] {\scriptsize 1};
\draw [thick] (0.06, 0.55) -- (0.94,0.95) ;
\draw [thick] (1.94, 0.55) -- (1.06,0.95) ;
\draw [thick] (-0.94, 0.05) -- (-0.06,0.45) ;
\draw [thick] (1, 1.1) -- (1,1.9) ;
\draw [thick] (0.06, 1.45) -- (0.94,1.05) ;
\draw [thick] (1.94, 1.45) -- (1.06,1.05) ;
\draw [thick] (2.91, 0.05) -- (2.08,0.46) ;
\end{tikzpicture}\ee
which is obtained by linking a node of label $1$ to the central node and to two tails of the affine Dynkin diagram of D$_4$. Consecutive higgsing processes allow us to find the following Hasse diagram (for large enough $N$):
\be\label{HDD44e} 
\begin{tikzpicture} 
\node[] at (-.8,0) {$\cdots$};
\node[] (A) at (0,0) {\scriptsize $SU(2)^2$}; 
\node[] at (0,0.3) {\scriptsize (N)};
\node[] (B) at (3,0) {\scriptsize $SU(3)$}; 
\node[] at (3,0.3) {\scriptsize (N)};
\node[] (C) at (6,0) {\scriptsize $SU(2)^2$}; 
\node[] at (6,0.3) {\scriptsize (N)}; 
\node[] (D) at (9,0) {\scriptsize $SU(3)$}; 
\node[] at (9,0.3) {\scriptsize (N)};
\node[] (E) at (12,0) {\scriptsize $SU(2)^2$}; 
\node[] at (12,0.3) {\scriptsize (N-1)};
\node[] at (1.5,0.2) {\scriptsize $\mathfrak{a}_1$};
\node[] at (4.5,0.2) {\scriptsize $\mathfrak{a}_2$};
\node[] at (7.5,0.2) {\scriptsize $\mathfrak{a}_1$};
\node[] at (10.5,0.2) {\scriptsize $\mathfrak{a}_2$};
\node[] at (12.8,0) {$\cdots$};
\draw[->] (A) -- (B); 
\draw[->] (B) -- (C); 
\draw[->] (C) -- (D); 
\draw[->] (D) -- (E); 
\end{tikzpicture}
\ee 
where we have displayed only the non-abelian part of the flavor symmetry.

\subsection{Rules of the game}\label{Sec:GenDisc}

By inspecting the perturbative examples we have just discussed, we can readily extract the following general rules.  
\begin{itemize}
\item The magnetic quiver of superconformal field theories derived from a stack of D3 branes probing a $\mathfrak{g}$-type 7-brane wrapped on the orbifold $\mathbb{C}^2/\mathbb{Z}_n$ can be obtained as follows: Take the affine Dynkin diagram of type $\mathfrak{g}$ and attach abelian nodes to the nodes of the Dynkin diagram with comarks $a_i^{\vee}$, in such a way that:
\begin{enumerate}
\item The resulting quiver is star-shaped.
\item Each node is attached with a single-valence bond.
\item $\sum_i a_i^{\vee}=n$ (counted with multiplicity).
\item None of the nodes are underbalanced.
\end{enumerate}
\item There are as many different (families of) theories as inequivalent ways of attaching nodes under the above conditions.
\item The \emph{canonical} theories are particularly simple to describe. In this case, as we have mentioned, the rank of the non-abelian nodes in the magnetic quiver is $Na_i^{\vee}$ where $N$ is the number of D3 branes. The total rank of the theory is $r=nN+1-K$ where $K$ is the total number of abelian nodes that have been attached. The dimensions of the CB operators can be arranged in ascending order into $N$ blocks of $n$ entries each, spaced out by $\Delta_7$ units, with the last block truncated by removing the $K-1$ highest entries.
\item All superconformal field theories belonging to a family can be obtained from the canonical theory by consecutive Higgs-branch RG flows, implemented by quiver subtraction.
\end{itemize}
All the SCFTs we can construct in this way are class $\mathcal{S}$ theories on a sphere. We review in Appendix \ref{Sec:CBSpectrum} the rules to derive the CB spectrum and 't Hooft anomalies of the theory. The requirements 1.~and 2.~above come from conformal invariance. While the first one is well known \cite{Benini:2010uu}, we motivate the second at the end of this subsection.

There is an important point to observe, concerning the case $\mathfrak{g}=\mathfrak{d}_4$. The perturbative theories we have constructed only involved at most two tails of the Dynkin diagram in the node-attaching process. It is easy to verify that, by involving either three or four tails, one necessarily ends up with theories that cannot be constructed using orientifolds and perturbative strings, and are intrinsically strongly coupled. Taking into account the above observation, it immediately follows that there are two inequivalent O7$^-$ projections when $n$ is even, and just a single one for $n$ odd. 

Remarkably, we find that for non-perturbative 7-branes as well we have either one or two families for given $n$. For instance, for the 7-brane of type E$_6$ there is a single $\mathbb{Z}_2$ orbifold, but there are two inequivalent $\mathbb{Z}_3$ ones! This fact has a nice and clear explanation in terms of fractional D3 branes and their ``images''. In the case of the orientifold of the $\mathbb{Z}_2$ orbifold, as we have seen in Section \ref{Sec:PertTheories}, we could either take both the fractional D3 branes to be orientifold-invariant stacks, leading to symplectic gauge theories, or take the fractional D3 branes to be the orientifold image of one another, leading to unitary gauge theories. By the same token, each D3-brane probing a 7-brane of type E$_6$ splits into three components, rotated by the same $\mathbb{Z}_3$ group acting on the F-theory torus. Hence, when we put the 7-branes on the orbifold $\mathbb{C}^2/\mathbb{Z}_3$, we can either choose all of the three fractional D3 branes to be $\mathbb{Z}_3$-invariant stacks, or ``identify'' the $\mathbb{Z}_3$ action intrinsic to E$_6$ (acting on the plane transverse to the 7-branes) with the $\mathbb{Z}_3$ orbifold action (acting on the 7-brane worldvolume). The first option corresponds to attaching a node of label $1$ to the central node of label $3N$ of the affine Dynkin diagram of E$_6$, whereas the second option, which requires $N>1$, corresponds to attaching a node of label $1$ to the end of each tail in the diagram. The exact same reasoning can be made for the other two non-perturbative 7-branes, leading to two different (families of) SCFTs when probing the E$_7$ stack on the orbifold $\mathbb{C}^2/\mathbb{Z}_4$ and the E$_8$ stack on the orbifold $\mathbb{C}^2/\mathbb{Z}_6$.

The perturbative case suggests an elegant mathematical structure underlying all of these facts. When probing the D$_4$ stack on the orbifold $\mathbb{C}^2/\mathbb{Z}_n$, the integral D3 branes split into a lattice of fractional D3 branes. The presence of the orientifold breaks the symmetry group of this lattice from the permutation group $S_n$ to the dihedral group $\mathbb{D}_n$. It is, however, more useful for us to see the latter as the semidirect product $\mathbb{D}_n\simeq\mathbb{Z}_n\rtimes\mathbb{Z}_2$, i.e.~the split extension of the ``orientifold group'' $\mathbb{Z}_2$ by the orbifold one $\mathbb{Z}_n$, with $\mathbb{Z}_2$ acting on $\mathbb{Z}_n$ by inversion. The inequivalent orientifold projections are then in one-to-one correspondence with the different conjugacy classes of embeddings of $\mathbb{Z}_2$ into $\mathbb{D}_n$. There is indeed a single such conjugacy class for odd $n$, and two for even $n$.

It is now very tempting to argue that the symmetry group $\mathfrak{g}^{(\mathbb{Z}_n)}$ of the lattice of fractional D3 branes originating from probing a $\mathfrak{g}$-type 7-brane wrapped on the orbifold $\mathbb{C}^2/\mathbb{Z}_n$ is a specific extension\footnote{Presumably an analogous conclusion can be made for the D- and E-type non-abelian orbifolds of $\mathbb{C}^2$, by plugging the orbifold group in the first place of the exact sequence.}
\be
\mathbb{Z}_n\longrightarrow\, \mathfrak{g}^{(\mathbb{Z}_n)}\longrightarrow\,\mathbb{Z}_{\Delta_7}
\ee
by $\mathbb{Z}_n$ of $\mathbb{Z}_{\Delta_7}$, which is precisely the group acting on the corresponding F-theory torus. We have just seen that $\mathfrak{d}_4^{(\mathbb{Z}_n)}=\mathbb{D}_n$. Inequivalent realizations of the orbifold theory should be related to conjugacy classes of embeddings of $\mathbb{Z}_{\Delta_7}$ into $\mathfrak{g}^{(\mathbb{Z}_n)}$. For instance, $\mathfrak{e}_6^{(\mathbb{Z}_2)}=\mathbb{Z}_6\equiv\mathbb{Z}_2\times\mathbb{Z}_3$ and $\mathfrak{e}_7^{(\mathbb{Z}_2)}=\mathbb{Z}_2\times\mathbb{Z}_4$ are both the trivial extension, but while the former only admits a single $\mathbb{Z}_3$ subgroup, the latter admits two non-conjugate $\mathbb{Z}_4$ subgroups. This implies the existence of two different families of canonical superconformal field theories for the E$_7$-type 7-brane on $\mathbb{C}^2/\mathbb{Z}_2$, which can be easily verified following the rules outlined above to derive the corresponding magnetic quivers.

It would be nice to explore this further, especially to see whether the above mathematical framework, which is rooted into the stringy origin of these field theories, is able to explain why e.g.~their CB operators have the dimensions that we are going to find using magnetic quivers. In what follows, indeed, we will concentrate on concretely applying the above rules to derive magnetic quivers (and from there a number of important properties of the field theories) for the easiest case of non-perturbative 7-branes on the $\mathbb{C}^2/\mathbb{Z}_2$ orbifold. We leave a detailed discussion of higher-order orbifolds to Appendix \ref{Sec:MoreEx}.

Let us conclude this subsection by motivating why we neglect multi-valence links for the abelian node(s) in the magnetic quivers. The reason is that this always leads to non conformal theories. Let us illustrate this with one example. Consider a D$_4$ Dynkin diagram with an abelian node connected with valence two at the central node: 
\be\label{doubleval}
\begin{tikzpicture}
\filldraw[fill= white] (0,0.5) circle [radius=0.1] node[below] {\scriptsize N};
\filldraw[fill= white] (1,1) circle [radius=0.1] node[below] {\scriptsize 2N};
\filldraw[fill= white] (2,0.5) circle [radius=0.1] node[below] {\scriptsize N};
\filldraw[fill= white] (0,1.5) circle [radius=0.1] node[above] {\scriptsize N};
\filldraw[fill= white] (2,1.5) circle [radius=0.1] node[above] {\scriptsize N};
\filldraw[fill= white] (1,2) circle [radius=0.1] node[above] {\scriptsize 1};
\draw [thick] (0.06, 0.55) -- (0.94,0.95) ;
\draw [thick] (1.94, 0.55) -- (1.06,0.95) ;
\draw [thick] (1.03, 1.1) -- (1.03,1.9) ;
\draw [thick] (0.97, 1.1) -- (0.97,1.9) ;
\draw [thick] (0.06, 1.45) -- (0.94,1.05) ;
\draw [thick] (1.94, 1.45) -- (1.06,1.05) ;
\end{tikzpicture}\ee 
As we have seen, if we replace the valence-two node with a pair of abelian nodes with valence one, we find the magnetic quiver \eqref{SU24NMQ4v}, associated with the four-dimensional SCFT \eqref{n=4orbifoldS}: 
\be\label{quivnew1}\boxed{2}-USp(2N)-SU(2N)-USp(2N)-\boxed{2}\ee 
The quiver \eqref{doubleval} is instead the mirror dual of  
\be\label{quivnew}\boxed{2}-USp(2N)-U(2N)-USp(2N)-\boxed{2}\ee 
which is infrared free in four dimensions due to the positive beta function for the $U(1)$ gauge factor. The duality between \eqref{quivnew} (seen as a 3d gauge theory) and \eqref{doubleval} can be easily understood by noticing that \eqref{quivnew} can be obtained starting from \eqref{quivnew1} by gauging a $U(1)$ baryonic symmetry (which corresponds on the mirror side to ungauging a $U(1)$). Since the resulting unitary gauge group is balanced, we expect to gain a $SU(2)$ topological symmetry, and therefore a $SU(2)$ flavor symmetry in the mirror theory. The only way to meet these requirements is to ungauge the abelian nodes in \eqref{SU24NMQ4v}, leading to 
\be\label{doubleval2}
\begin{tikzpicture}
\filldraw[fill= white] (0,0.5) circle [radius=0.1] node[below] {\scriptsize N};
\filldraw[fill= white] (1,1) circle [radius=0.1] node[below] {\scriptsize 2N};
\filldraw[fill= white] (2,0.5) circle [radius=0.1] node[below] {\scriptsize N};
\filldraw[fill= white] (0,1.5) circle [radius=0.1] node[above] {\scriptsize N};
\filldraw[fill= white] (2,1.5) circle [radius=0.1] node[above] {\scriptsize N};
\filldraw[fill= white] (1,0) circle [radius=0.1] node[below] {\scriptsize 1};
\filldraw[fill= white] (1,2) circle [radius=0.1] node[above] {\scriptsize 1};
\draw [thick] (0.06, 0.55) -- (0.94,0.95) ;
\draw [thick] (1.94, 0.55) -- (1.06,0.95) ;
\draw [thick] (1, 0.9) -- (1,0.1) ;
\draw [thick] (1, 1.1) -- (1,1.9) ;
\draw [thick] (0.06, 1.45) -- (0.94,1.05) ;
\draw [thick] (1.94, 1.45) -- (1.06,1.05) ;

\draw[->] (2.5,1) -- (4.5,1);
\node[] at (3.4,1.3) {\small Ungauging};

\filldraw[fill= white] (5,0.5) circle [radius=0.1] node[below] {\scriptsize N};
\filldraw[fill= white] (6,1) circle [radius=0.1] node[below] {\scriptsize 2N};
\filldraw[fill= white] (7,0.5) circle [radius=0.1] node[below] {\scriptsize N};
\filldraw[fill= white] (5,1.5) circle [radius=0.1] node[above] {\scriptsize N};
\filldraw[fill= white] (7,1.5) circle [radius=0.1] node[above] {\scriptsize N};
\node[draw, rectangle, minimum width=2pt, minimum height=2pt] at (6,2){\scriptsize 2};
\draw [thick] (5.06, 0.55) -- (5.94,0.95) ;
\draw [thick] (6.94, 0.55) -- (6.06,0.95) ;
\draw [thick] (6, 1.1) -- (6,1.78) ;
\draw [thick] (5.06, 1.45) -- (5.94,1.05) ;
\draw [thick] (6.94, 1.45) -- (6.06,1.05) ;
\end{tikzpicture}\ee   
Remembering that we can always ungauge a $U(1)$ factor in a flavorless quiver, since it acts trivially on all matter fields, we immediately see that the quiver on the right of \eqref{doubleval2} is equivalent to \eqref{doubleval}. Many similar examples are discussed in detail in \cite{Hanany:1999sj, Mekareeya:2015bla} and always correspond to non conformal gauge theories. When the source of non-conformality is due to the presence of unitary groups which are infrared free in four dimensions, we can make these theories conformal by removing the $U(1)$ factors, which in the magnetic quiver corresponds to replacing flavor nodes with abelian gauge nodes as in \eqref{doubleval2}. A different example discussed in \cite{Mekareeya:2015bla} is 
\be
\begin{tikzpicture}
\filldraw[fill= white] (0,0.5) circle [radius=0.1] node[below] {\scriptsize N};
\filldraw[fill= white] (1,1) circle [radius=0.1] node[below] {\scriptsize 2N};
\filldraw[fill= white] (2,0.5) circle [radius=0.1] node[below] {\scriptsize N};
\filldraw[fill= white] (0,1.5) circle [radius=0.1] node[above] {\scriptsize N};
\filldraw[fill= white] (2,1.5) circle [radius=0.1] node[above] {\scriptsize N};
\filldraw[fill= white] (-1,0.5) circle [radius=0.1] node[left] {\scriptsize 1};
\draw [thick] (0.06, 0.55) -- (0.94,0.95) ;
\draw [thick] (1.94, 0.55) -- (1.06,0.95) ;
\draw [thick] (-0.9, 0.47) -- (-0.1,0.47) ;
\draw [thick] (-0.9, 0.53) -- (-0.1,0.53) ;
\draw [thick] (0.06, 1.45) -- (0.94,1.05) ;
\draw [thick] (1.94, 1.45) -- (1.06,1.05) ;
\end{tikzpicture}\ee 
which is the magnetic quiver of the gauge theory 
$$USp(2N)-USp(2N)-\boxed{4}$$ 
Notice that in 4d the gauge group on the left is asymptotically free whereas that on the right is infrared free.

\subsection{The 7-brane of type E$_8$}\label{Sec:E8}

Let us start by discussing 7-branes of type E$_8$. Following the rules of Section \ref{Sec:GenDisc}, here we find a single family of superconformal theories when the 7-branes wrap the orbifold $\mathbb{C}^2/\mathbb{Z}_2$.

The allowed holonomies for $n=2$ break $E_8$ to $E_8$, $E_7\times SU(2)$ and $SO(16)$. For $N=1$ and $SO(16)$-preserving holonomy we can immediately identify the worldvolume theory on the D3 brane with the rank two $SO(16)\times SU(2)$ theory, which has precisely the expected global symmetry and is known to flow to the rank-one E$_8$ Minahan-Nemeschansky theory by turning on a nilpotent vev for the $SU(2)$ moment map \cite{Martone:2021ixp}. Said differently, the theory flows to the worldvolume theory of a single D3-brane probing a 7-brane of type E$_8$ in flat space with transverse slice $\mathbb{C}^2/\mathbb{Z}_2$. This theory therefore satisfies all the constraints we expect from the geometric picture. 
The magnetic quiver of its higher-rank generalizations reads
\be\label{D8NMQ}
\begin{tikzpicture}
\filldraw[fill= white] (0,0) circle [radius=0.1] node[below] {\scriptsize N};
\filldraw[fill= white] (1,0) circle [radius=0.1] node[below] {\scriptsize 2N};
\filldraw[fill= white] (2,0) circle [radius=0.1] node[below] {\scriptsize 3N};
\filldraw[fill= white] (3,0) circle [radius=0.1] node[below] {\scriptsize 4N};
\filldraw[fill= white] (4,0) circle [radius=0.1] node[below] {\scriptsize 5N};
\filldraw[fill= white] (5,0) circle [radius=0.1] node[below] {\scriptsize 6N};
\filldraw[fill= white] (6,0) circle [radius=0.1] node[below] {\scriptsize 4N};
\filldraw[fill= white] (5,1) circle [radius=0.1] node[above] {\scriptsize 3N};
\filldraw[fill= white] (7,0) circle [radius=0.1] node[below] {\scriptsize 2N};
\filldraw[fill= white] (8,0) circle [radius=0.1] node[below] {\scriptsize 1};
\draw [thick] (0.1, 0) -- (0.9,0) ;
\draw [thick] (1.1, 0) -- (1.9,0) ;
\draw [thick] (2.1, 0) -- (2.9,0) ;
\draw [thick] (3.1, 0) -- (3.9,0) ;
\draw [thick] (4.1, 0) -- (4.9,0) ;
\draw [thick] (5.1, 0) -- (5.9,0) ;
\draw [thick] (5, 0.1) -- (5,0.9) ;
\draw [thick] (6.1, 0) -- (6.9,0) ;
\draw [thick] (7.1, 0) -- (7.9,0) ;
\end{tikzpicture} \ee
which manifestly has a $SO(16)$ non-abelian global symmetry. The theory has rank $2N$ as expected and the CB spectrum includes operators of dimension 
$$(4,6)\,;\,(10,12)\,;\,\dots\,;\,(6N-2,6N)\,,$$ 
as can be verified using the rule reviewed in Appendix \ref{Sec:CBSpectrum}. The central charges are $$8a-4c=12N^2+6N\,;\qquad 24(c-a)=30N\,,$$ which is consistent with \eqref{formulacc} if we set $\alpha=\beta=0$ and $\epsilon=1/24$.

What happens if we turn on an expectation value for the $SO(16)$ moment map? We can argue, using the quiver subtraction technique \cite{Cabrera:2018ann, Bourget:2019aer}, that by turning on a minimal nilpotent vev we find the sequence of theories described by the magnetic quiver ($N>1$)
\be\label{E7NMQ}
\begin{tikzpicture}
\filldraw[fill= white] (0,0) circle [radius=0.1] node[below] {\scriptsize N-1};
\filldraw[fill= white] (1,0) circle [radius=0.1] node[below] {\scriptsize 2N-2};
\filldraw[fill= white] (2,0) circle [radius=0.1] node[below] {\scriptsize 3N-2};
\filldraw[fill= white] (3,0) circle [radius=0.1] node[below] {\scriptsize 4N-2};
\filldraw[fill= white] (4,0) circle [radius=0.1] node[below] {\scriptsize 5N-2};
\filldraw[fill= white] (5,0) circle [radius=0.1] node[below] {\scriptsize 6N-2};
\filldraw[fill= white] (6,0) circle [radius=0.1] node[below] {\scriptsize 4N-1};
\filldraw[fill= white] (5,1) circle [radius=0.1] node[above] {\scriptsize 3N-1};
\filldraw[fill= white] (7,0) circle [radius=0.1] node[below] {\scriptsize 2N};
\filldraw[fill= white] (8,0) circle [radius=0.1] node[below] {\scriptsize 1};
\draw [thick] (0.1, 0) -- (0.9,0) ;
\draw [thick] (1.1, 0) -- (1.9,0) ;
\draw [thick] (2.1, 0) -- (2.9,0) ;
\draw [thick] (3.1, 0) -- (3.9,0) ;
\draw [thick] (4.1, 0) -- (4.9,0) ;
\draw [thick] (5.1, 0) -- (5.9,0) ;
\draw [thick] (5, 0.1) -- (5,0.9) ;
\draw [thick] (6.1, 0) -- (6.9,0) ;
\draw [thick] (7.1, 0) -- (7.9,0) ;
\end{tikzpicture} \ee 
Indeed we can easily check that 
$$
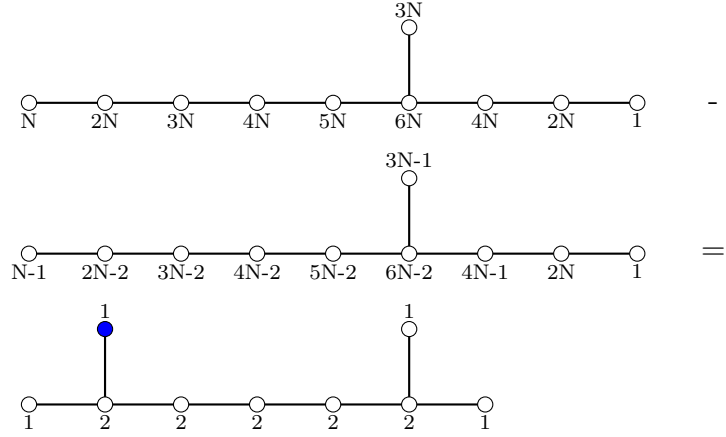
\begin{tikzpicture}
\filldraw[fill= white] (0,0) circle [radius=0.1] node[below] {\scriptsize N};
\filldraw[fill= white] (1,0) circle [radius=0.1] node[below] {\scriptsize 2N};
\filldraw[fill= white] (2,0) circle [radius=0.1] node[below] {\scriptsize 3N};
\filldraw[fill= white] (3,0) circle [radius=0.1] node[below] {\scriptsize 4N};
\filldraw[fill= white] (4,0) circle [radius=0.1] node[below] {\scriptsize 5N};
\filldraw[fill= white] (5,0) circle [radius=0.1] node[below] {\scriptsize 6N};
\filldraw[fill= white] (6,0) circle [radius=0.1] node[below] {\scriptsize 4N};
\filldraw[fill= white] (5,1) circle [radius=0.1] node[above] {\scriptsize 3N};
\filldraw[fill= white] (7,0) circle [radius=0.1] node[below] {\scriptsize 2N};
\filldraw[fill= white] (8,0) circle [radius=0.1] node[below] {\scriptsize 1}; 
\node[] at (9,0) {-};
\draw [thick] (0.1, 0) -- (0.9,0) ;
\draw [thick] (1.1, 0) -- (1.9,0) ;
\draw [thick] (2.1, 0) -- (2.9,0) ;
\draw [thick] (3.1, 0) -- (3.9,0) ;
\draw [thick] (4.1, 0) -- (4.9,0) ;
\draw [thick] (5.1, 0) -- (5.9,0) ;
\draw [thick] (5, 0.1) -- (5,0.9) ;
\draw [thick] (6.1, 0) -- (6.9,0) ;
\draw [thick] (7.1, 0) -- (7.9,0) ; 
\filldraw[fill= white] (0,-2) circle [radius=0.1] node[below] {\scriptsize N-1};
\filldraw[fill= white] (1,-2) circle [radius=0.1] node[below] {\scriptsize 2N-2};
\filldraw[fill= white] (2,-2) circle [radius=0.1] node[below] {\scriptsize 3N-2};
\filldraw[fill= white] (3,-2) circle [radius=0.1] node[below] {\scriptsize 4N-2};
\filldraw[fill= white] (4,-2) circle [radius=0.1] node[below] {\scriptsize 5N-2};
\filldraw[fill= white] (5,-2) circle [radius=0.1] node[below] {\scriptsize 6N-2};
\filldraw[fill= white] (6,-2) circle [radius=0.1] node[below] {\scriptsize 4N-1};
\filldraw[fill= white] (5,-1) circle [radius=0.1] node[above] {\scriptsize 3N-1};
\filldraw[fill= white] (7,-2) circle [radius=0.1] node[below] {\scriptsize 2N};
\filldraw[fill= white] (8,-2) circle [radius=0.1] node[below] {\scriptsize 1};
\node[] at (9,-2) {=};
\draw [thick] (0.1, -2) -- (0.9,-2) ;
\draw [thick] (1.1, -2) -- (1.9,-2) ;
\draw [thick] (2.1, -2) -- (2.9,-2) ;
\draw [thick] (3.1, -2) -- (3.9,-2) ;
\draw [thick] (4.1, -2) -- (4.9,-2) ;
\draw [thick] (5.1, -2) -- (5.9,-2) ;
\draw [thick] (5, -1.9) -- (5,-1.1) ;
\draw [thick] (6.1, -2) -- (6.9,-2) ;
\draw [thick] (7.1, -2) -- (7.9,-2) ;
\filldraw[fill= white] (0,-4) circle [radius=0.1] node[below] {\scriptsize 1};
\filldraw[fill= white] (1,-4) circle [radius=0.1] node[below] {\scriptsize 2};
\filldraw[fill= white] (2,-4) circle [radius=0.1] node[below] {\scriptsize 2};
\filldraw[fill= white] (3,-4) circle [radius=0.1] node[below] {\scriptsize 2};
\filldraw[fill= white] (4,-4) circle [radius=0.1] node[below] {\scriptsize 2};
\filldraw[fill= white] (5,-4) circle [radius=0.1] node[below] {\scriptsize 2};
\filldraw[fill= white] (6,-4) circle [radius=0.1] node[below] {\scriptsize 1};
\filldraw[fill= white] (5,-3) circle [radius=0.1] node[above] {\scriptsize 1};
\filldraw[fill= blue] (1,-3) circle [radius=0.1] node[above] {\scriptsize 1};
\draw [thick] (0.1, -4) -- (0.9,-4) ;
\draw [thick] (1.1, -4) -- (1.9,-4) ;
\draw [thick] (2.1, -4) -- (2.9,-4) ;
\draw [thick] (3.1, -4) -- (3.9,-4) ;
\draw [thick] (4.1, -4) -- (4.9,-4) ;
\draw [thick] (5.1, -4) -- (5.9,-4) ;
\draw [thick] (5, -3.9) -- (5,-3.1) ;
\draw [thick] (1, -3.9) -- (1,-3.1) ;\end{tikzpicture}$$
where we can recognize the quiver describing the $\mathfrak{d}_8$ minimal nilpotent orbit and the blue node has been introduced to rebalance.

The family of theories \eqref{E7NMQ} has rank $2N-1$ and CB spectrum 
$$(4,6)\,;\,(10,12)\,;\,\dots\,;\,(6N-2)\,,$$ 
which reproduces the spectrum of the $SO(16)$ family, except for the fact that the operator of dimension $6N$ is missing. The central charges are 
$$ 8a-4c=12N^2-6N+1\,;\qquad 24(c-a)=30N-13\,.$$
This is again consistent with \eqref{formulacc} if we set $\alpha=1$, $\beta=-13$ and $\epsilon=-11/24$. Furthermore, from the quiver we conclude that the global symmetry includes a $E_7\times SU(2)$ subgroup and for this reason we claim that this set of theories corresponds to the $E_7\times SU(2)$ series. 

We can now observe that also the theories in the $E_7\times SU(2)$ series can flow to other SCFTâs upon turning on a minimal nilpotent vev for the moment map of its global symmetry. In the case of the $E_7$ factor they flow back to the $SO(16)$ family, more precisely to the model corresponding to $N-1$ D3 probes. If we instead turn on a nilpotent vev for the $SU(2)$ moment map, we find a new set of theories which we associate with the $E_8$ series. The corresponding magnetic quiver is (for $N\geq2$) 
 \be
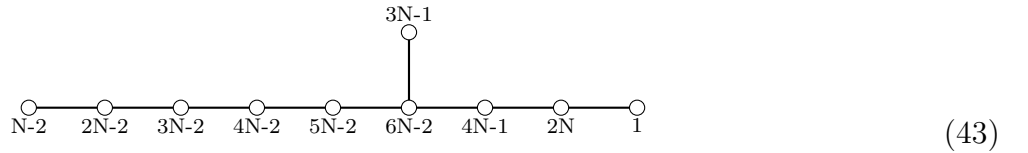
\label{E8NMQ}
\begin{tikzpicture}
\filldraw[fill= white] (0,0) circle [radius=0.1] node[below] {\scriptsize N-2};
\filldraw[fill= white] (1,0) circle [radius=0.1] node[below] {\scriptsize 2N-2};
\filldraw[fill= white] (2,0) circle [radius=0.1] node[below] {\scriptsize 3N-2};
\filldraw[fill= white] (3,0) circle [radius=0.1] node[below] {\scriptsize 4N-2};
\filldraw[fill= white] (4,0) circle [radius=0.1] node[below] {\scriptsize 5N-2};
\filldraw[fill= white] (5,0) circle [radius=0.1] node[below] {\scriptsize 6N-2};
\filldraw[fill= white] (6,0) circle [radius=0.1] node[below] {\scriptsize 4N-1};
\filldraw[fill= white] (5,1) circle [radius=0.1] node[above] {\scriptsize 3N-1};
\filldraw[fill= white] (7,0) circle [radius=0.1] node[below] {\scriptsize 2N};
\filldraw[fill= white] (8,0) circle [radius=0.1] node[below] {\scriptsize 1};
\draw [thick] (0.1, 0) -- (0.9,0) ;
\draw [thick] (1.1, 0) -- (1.9,0) ;
\draw [thick] (2.1, 0) -- (2.9,0) ;
\draw [thick] (3.1, 0) -- (3.9,0) ;
\draw [thick] (4.1, 0) -- (4.9,0) ;
\draw [thick] (5.1, 0) -- (5.9,0) ;
\draw [thick] (5, 0.1) -- (5,0.9) ;
\draw [thick] (6.1, 0) -- (6.9,0) ;
\draw [thick] (7.1, 0) -- (7.9,0) ;
\end{tikzpicture} \ee 
The CB spectrum of these theories includes operators of dimension 
$$(4,6)\,;\,(10,12)\,;\,\dots\,;\,(6N-8)\,;\,(6N-2)\,.$$ 
For $N=2$ this reduces to the $D_1^{20}(E_8)$ theory discovered in \cite{Giacomelli:2017ckh}.
We can summarize this web of RG flows with the following Hasse diagram ($N\geq2$):
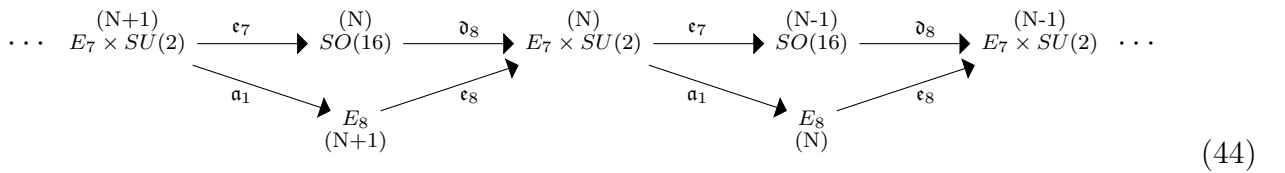
\be\label{HDE8} 
\begin{tikzpicture} 
\node[] at (-4.3,0) {$\cdots$};
\node[] (A) at (0,0) {\scriptsize $SO(16)$}; 
\node[] at (0,0.3) {\scriptsize (N)};
\node[] (B) at (3,0) {\scriptsize $E_7\times SU(2)$}; 
\node[] at (3,0.3) {\scriptsize (N)};
\node[] (C) at (0,-1) {\scriptsize $E_8$}; 
\node[] at (0,-1.3) {\scriptsize (N+1)}; 
\node[] (D) at (-3,0) {\scriptsize $E_7\times SU(2)$}; 
\node[] at (-3,0.3) {\scriptsize (N+1)};
\node[] at (1.5,0.2) {\scriptsize $\mathfrak{d}_8$};
\node[] at (1.5,-0.7) {\scriptsize $\mathfrak{e}_8$};
\node[] at (-1.5,0.2) {\scriptsize $\mathfrak{e}_7$};
\node[] at (-1.5,-0.7) {\scriptsize $\mathfrak{a}_1$};
\node[] (E) at (6,0) {\scriptsize $SO(16)$}; 
\node[] at (6,0.3) {\scriptsize (N-1)};
\node[] (F) at (9,0) {\scriptsize $E_7\times SU(2)$}; 
\node[] at (9,0.3) {\scriptsize (N-1)};
\node[] (G) at (6,-1) {\scriptsize $E_8$}; 
\node[] at (6,-1.3) {\scriptsize (N)}; 
\node[] at (7.5,0.2) {\scriptsize $\mathfrak{d}_8$};
\node[] at (7.5,-0.7) {\scriptsize $\mathfrak{e}_8$};
\node[] at (4.5,0.2) {\scriptsize $\mathfrak{e}_7$};
\node[] at (4.5,-0.7) {\scriptsize $\mathfrak{a}_1$}; 
\node[] at (10.3,0) {$\cdots$};
\draw[->] (A) -- (B); 
\draw[->] (C) -- (B); 
\draw[->] (D) -- (A); 
\draw[->] (D) -- (C); 
\draw[->] (E) -- (F); 
\draw[->] (G) -- (F); 
\draw[->] (B) -- (E); 
\draw[->] (B) -- (G);
\end{tikzpicture}
\ee

\subsection{The 7-brane of type E$_7$} \label{Sec:E7}
In the case of the E$_7$ stack of 7-branes we find two different families of superconformal theories. The allowed $\mathbb{Z}_2$ holonomies for $E_7$ preserve the following subgroups: $E_7$, $E_6\times U(1)$, $SU(8)$ and $SO(12)\times SU(2)$. According to the rules of Section \ref{Sec:GenDisc}, we expect to find \emph{canonical} theories for the cases $SU(8)$ and $E_6\times U(1)$. 

\paragraph{$\bullet$} Let us start from the former case. This is described by the magnetic quiver  
\be\label{SU8NMQ}
\begin{tikzpicture}
\filldraw[fill= white] (0,0) circle [radius=0.1] node[below] {\scriptsize N};
\filldraw[fill= white] (1,0) circle [radius=0.1] node[below] {\scriptsize 2N};
\filldraw[fill= white] (2,0) circle [radius=0.1] node[below] {\scriptsize 3N};
\filldraw[fill= white] (3,0) circle [radius=0.1] node[below] {\scriptsize 4N};
\filldraw[fill= white] (4,0) circle [radius=0.1] node[below] {\scriptsize 3N};
\filldraw[fill= white] (5,0) circle [radius=0.1] node[below] {\scriptsize 2N};
\filldraw[fill= white] (6,0) circle [radius=0.1] node[below] {\scriptsize N};
\filldraw[fill= white] (3,1) circle [radius=0.1] node[above] {\scriptsize 2N};
\filldraw[fill= white] (4,1) circle [radius=0.1] node[above] {\scriptsize 1}; 
\draw [thick] (0.1, 0) -- (0.9,0) ;
\draw [thick] (1.1, 0) -- (1.9,0) ;
\draw [thick] (2.1, 0) -- (2.9,0) ;
\draw [thick] (3.1, 0) -- (3.9,0) ;
\draw [thick] (4.1, 0) -- (4.9,0) ;
\draw [thick] (5.1, 0) -- (5.9,0) ;
\draw [thick] (3, 0.1) -- (3,0.9) ;
\draw [thick] (3.1, 1) -- (3.9,1) ;
\end{tikzpicture}\ee
The CB operators have dimension 
$$(3,4)\,;\,(7,8)\,;\,\dots\,;\, (4N-1,4N)\,,$$ 
and the central charges are $$8a-4c=8N^2+4N\,;\qquad 24(c-a)=18N\,,$$ which is consistent with the general formula \eqref{formulacc} if we set $\Delta_7=4$, $\alpha=\beta=0$ and $\epsilon=1/16$. For $N=1$ we recover the rank-two $SU(8)\times SU(2)$ theory which is known to flow to the E$_7$ Minahan-Nemeschansky theory upon turning on a nilpotent vev for the $SU(2)$ moment map \cite{Martone:2021ixp}, in agreement with our general expectations.

By higgsing the $SU(8)$ global symmetry, we find a family of theories with flavor symmetry containing $E_6\times U(1)$, with magnetic quiver 
\be\label{E6NMQ}
\begin{tikzpicture}
\filldraw[fill= white] (0,0) circle [radius=0.1] node[below] {\scriptsize N-1};
\filldraw[fill= white] (1,0) circle [radius=0.1] node[below] {\scriptsize 2N-1};
\filldraw[fill= white] (2,0) circle [radius=0.1] node[below] {\scriptsize 3N-1};
\filldraw[fill= white] (3,0) circle [radius=0.1] node[below] {\scriptsize 4N-1};
\filldraw[fill= white] (4,0) circle [radius=0.1] node[below] {\scriptsize 3N-1};
\filldraw[fill= white] (5,0) circle [radius=0.1] node[below] {\scriptsize 2N-1};
\filldraw[fill= white] (6,0) circle [radius=0.1] node[below] {\scriptsize N-1};
\filldraw[fill= white] (3,1) circle [radius=0.1] node[above] {\scriptsize 2N};
\filldraw[fill= white] (4,1) circle [radius=0.1] node[above] {\scriptsize 1}; 
\draw [thick] (0.1, 0) -- (0.9,0) ;
\draw [thick] (1.1, 0) -- (1.9,0) ;
\draw [thick] (2.1, 0) -- (2.9,0) ;
\draw [thick] (3.1, 0) -- (3.9,0) ;
\draw [thick] (4.1, 0) -- (4.9,0) ;
\draw [thick] (5.1, 0) -- (5.9,0) ;
\draw [thick] (3, 0.1) -- (3,0.9) ;
\draw [thick] (3.1, 1) -- (3.9,1) ;
\end{tikzpicture}\ee
In this case the CB spectrum includes operators of dimension 
$$(3,4)\,;\,(7,8)\,;\,\dots\,;\, (4N-1)\,.$$ 
This is the same as the spectrum of the previous series, apart from the fact that the operator of dimension $4N$ is missing. The central charges are reproduced by setting $\alpha=1$, $\beta=-7$ and $\epsilon=-7/16$. 

As a consistency check, we can observe using quiver subtraction that these two series are connected by a sequence of RG flows, as expected. This is summarized by the following Hasse diagram:
\be\label{HDE7} 
\begin{tikzpicture} 
\node[] at (-0.8,0) {$\cdots$};
\node[] (A) at (0,0) {\scriptsize $SU(8)$}; 
\node[] at (0,0.3) {\scriptsize (N)};
\node[] (B) at (3,0) {\scriptsize $E_6\times U(1)$}; 
\node[] at (3,0.3) {\scriptsize (N)};
\node[] (C) at (6,0) {\scriptsize $SU(8)$}; 
\node[] at (6,0.3) {\scriptsize (N-1)}; 
\node[] (D) at (9,0) {\scriptsize $E_6\times U(1)$}; 
\node[] at (9,0.3) {\scriptsize (N-1)};
\node[] at (1.5,0.2) {\scriptsize $\mathfrak{a}_7$};
\node[] at (4.5,0.2) {\scriptsize $\mathfrak{e}_6$};
\node[] at (7.5,0.2) {\scriptsize $\mathfrak{a}_7$};
\node[] at (10.1,0) {$\cdots$};
\draw[->] (A) -- (B); 
\draw[->] (B) -- (C); 
\draw[->] (C) -- (D); 
\end{tikzpicture}
\ee 

\paragraph{$\bullet$} The second, inequivalent option is the canonical family with global symmetry\footnote{There are two extra $U(1)$'s which enhance to $SO(4)$ for the minimal number of probe branes, $N=2$. By higgsing the $SO(4)$ one binds the two fractional D3-brane stacks together: The bound state (a pair of integral D3 branes) can now be pulled off the singularity, describing the rank-$2$ Minahan-Nemeschansky theory of type E$_7$.} $E_6\times U(1)$, with magnetic quiver ($N>1$):
\be\label{E6U1NMQ}
\begin{tikzpicture}
\filldraw[fill= white] (0,0) circle [radius=0.1] node[below] {\scriptsize N};
\filldraw[fill= white] (1,0) circle [radius=0.1] node[below] {\scriptsize 2N};
\filldraw[fill= white] (2,0) circle [radius=0.1] node[below] {\scriptsize 3N};
\filldraw[fill= white] (3,0) circle [radius=0.1] node[below] {\scriptsize 4N};
\filldraw[fill= white] (4,0) circle [radius=0.1] node[below] {\scriptsize 3N};
\filldraw[fill= white] (5,0) circle [radius=0.1] node[below] {\scriptsize 2N};
\filldraw[fill= white] (6,0) circle [radius=0.1] node[below] {\scriptsize N};
\filldraw[fill= white] (3,1) circle [radius=0.1] node[above] {\scriptsize 2N};
\filldraw[fill= white] (-1,0) circle [radius=0.1] node[below] {\scriptsize 1}; 
\filldraw[fill= white] (7,0) circle [radius=0.1] node[below] {\scriptsize 1}; 
\draw [thick] (0.1, 0) -- (0.9,0) ;
\draw [thick] (1.1, 0) -- (1.9,0) ;
\draw [thick] (2.1, 0) -- (2.9,0) ;
\draw [thick] (3.1, 0) -- (3.9,0) ;
\draw [thick] (4.1, 0) -- (4.9,0) ;
\draw [thick] (5.1, 0) -- (5.9,0) ;
\draw [thick] (3, 0.1) -- (3,0.9) ;
\draw [thick] (-0.9, 0) -- (-0.1,0) ;
\draw [thick] (6.1, 0) -- (6.9,0) ;
\end{tikzpicture}\ee
In this case the CB spectrum includes operators of dimension 
$$(4,5)\,;\,(8,9)\,;\,\dots\,;\, (4N)\,.$$ 
By higgsing the $E_6$ global symmetry, we find a family of theories with non-abelian flavor symmetry\footnote{For $N=2$ the non-abelian global symmetry enhances to $SU(10)$: This theory has two CB operators of dimensions $4$ and $5$. See also \cite{Martone:2021ixp, Martone:2021drm} for a detailed discussion about this theory.} $SU(8)$, with magnetic quiver ($N>1$)
\be\label{SU8bisNMQ}
\begin{tikzpicture}
\filldraw[fill= white] (0,0) circle [radius=0.1] node[below] {\scriptsize N};
\filldraw[fill= white] (1,0) circle [radius=0.1] node[below] {\scriptsize 2N-1};
\filldraw[fill= white] (2,0) circle [radius=0.1] node[below] {\scriptsize 3N-2};
\filldraw[fill= white] (3,0) circle [radius=0.1] node[below] {\scriptsize 4N-3};
\filldraw[fill= white] (4,0) circle [radius=0.1] node[below] {\scriptsize 3N-2};
\filldraw[fill= white] (5,0) circle [radius=0.1] node[below] {\scriptsize 2N-1};
\filldraw[fill= white] (6,0) circle [radius=0.1] node[below] {\scriptsize N};
\filldraw[fill= white] (3,1) circle [radius=0.1] node[above] {\scriptsize 2N-2};
\filldraw[fill= white] (-1,0) circle [radius=0.1] node[below] {\scriptsize 1}; 
\filldraw[fill= white] (7,0) circle [radius=0.1] node[below] {\scriptsize 1}; 
\draw [thick] (0.1, 0) -- (0.9,0) ;
\draw [thick] (1.1, 0) -- (1.9,0) ;
\draw [thick] (2.1, 0) -- (2.9,0) ;
\draw [thick] (3.1, 0) -- (3.9,0) ;
\draw [thick] (4.1, 0) -- (4.9,0) ;
\draw [thick] (5.1, 0) -- (5.9,0) ;
\draw [thick] (3, 0.1) -- (3,0.9) ;
\draw [thick] (-0.9, 0) -- (-0.1,0) ;
\draw [thick] (6.1, 0) -- (6.9,0) ;
\end{tikzpicture}\ee
The CB operators have dimension 
$$(4,5)\,;\,(8,9)\,;\,\dots\,;\, (4N-4,4N-3)\,,$$
which is the same as the spectrum of the previous series, apart from the fact that the operator of dimension $4N$ is missing.

We again observe, using quiver subtraction, that these two series are connected by a sequence of RG flows, as expected. This is summarized by the following Hasse diagram for large-enough $N$:
\be\label{HDE7bis} 
\begin{tikzpicture} 
\node[] at (-1,0) {$\cdots$};
\node[] (A) at (0,0) {\scriptsize $E_6\times U(1)$}; 
\node[] at (0,0.3) {\scriptsize (N)};
\node[] (B) at (3,0) {\scriptsize $SU(8)$}; 
\node[] at (3,0.3) {\scriptsize (N)};
\node[] (C) at (6,0) {\scriptsize $E_6\times U(1)$}; 
\node[] at (6,0.3) {\scriptsize (N-1)}; 
\node[] (D) at (9,0) {\scriptsize $SU(8)$}; 
\node[] at (9,0.3) {\scriptsize (N-1)};
\node[] at (1.5,0.2) {\scriptsize $\mathfrak{e}_6$};
\node[] at (4.5,0.2) {\scriptsize $\mathfrak{a}_7$};
\node[] at (7.5,0.2) {\scriptsize $\mathfrak{e}_6$};
\node[] at (9.8,0) {$\cdots$};
\draw[->] (A) -- (B); 
\draw[->] (B) -- (C); 
\draw[->] (C) -- (D); 
\end{tikzpicture}
\ee

\subsection{The 7-brane of type E$_6$}\label{Sec:E6}

Let us conclude this survey with the case of the 7-brane of type E$_6$. The allowed $\mathbb{Z}_2$ holonomies for $E_6$ preserve the following subgroups: $E_6$, $SO(10)\times U(1)$, and $SU(6)\times SU(2)$. All of them are realized as follows. 

As we have explained in Section \ref{Sec:GenDisc}, probing an E$_6$ stack of 7-branes wrapped on the orbifold $\mathbb{C}^2/\mathbb{Z}_2$ leads to a single \emph{canonical} set of superconformal theories. This has the following magnetic quiver ($N>1$):
\be\label{E6Z2NMQ}
\begin{tikzpicture}
\filldraw[fill= white] (0,0) circle [radius=0.1] node[below] {\scriptsize 1};
\filldraw[fill= white] (1,0) circle [radius=0.1] node[below] {\scriptsize N};
\filldraw[fill= white] (2,0) circle [radius=0.1] node[below] {\scriptsize 2N};
\filldraw[fill= white] (3,0) circle [radius=0.1] node[below] {\scriptsize 3N};
\filldraw[fill= white] (4,0) circle [radius=0.1] node[below] {\scriptsize 2N};
\filldraw[fill= white] (5,0) circle [radius=0.1] node[below] {\scriptsize N};
\filldraw[fill= white] (6,0) circle [radius=0.1] node[below] {\scriptsize 1};
\filldraw[fill= white] (3,1) circle [radius=0.1] node[above] {\scriptsize 2N};
\filldraw[fill= white] (4,1) circle [radius=0.1] node[above] {\scriptsize N}; 
\draw [thick] (0.1, 0) -- (0.9,0) ;
\draw [thick] (1.1, 0) -- (1.9,0) ;
\draw [thick] (2.1, 0) -- (2.9,0) ;
\draw [thick] (3.1, 0) -- (3.9,0) ;
\draw [thick] (4.1, 0) -- (4.9,0) ;
\draw [thick] (5.1, 0) -- (5.9,0) ;
\draw [thick] (3, 0.1) -- (3,0.9) ;
\draw [thick] (3.1, 1) -- (3.9,1) ;
\end{tikzpicture}\ee
The flavor symmetry\footnote{There are two extra $U(1)$'s which enhance to $SO(4)$ for the minimal number of probe branes, $N=2$. Higgsing the $SO(4)$ binds the fractional D3-brane stacks together to form a pair of integral D3 probes, which can move off the singularity. This yields the rank-$2$ Minahan-Nemeschansky theory of type E$_6$.} is $SO(10)\times U(1)$ and the spectrum of CB operators reads
$$(3,4)\,;\,(6,7)\,;\,\dots\,;\, (3N)\,.$$ 

By higgsing the $SO(10)$ symmetry, using the quiver-subtraction technique, we land on the quiver
\be\label{E6Z2NMQ1}
\begin{tikzpicture}
\filldraw[fill= white] (0,0) circle [radius=0.1] node[below] {\scriptsize 1};
\filldraw[fill= white] (1,0) circle [radius=0.1] node[below] {\scriptsize N};
\filldraw[fill= white] (2,0) circle [radius=0.1] node[below] {\scriptsize 2N-1};
\filldraw[fill= white] (3,0) circle [radius=0.1] node[below] {\scriptsize 3N-2};
\filldraw[fill= white] (4,0) circle [radius=0.1] node[below] {\scriptsize 2N-1};
\filldraw[fill= white] (5,0) circle [radius=0.1] node[below] {\scriptsize N};
\filldraw[fill= white] (6,0) circle [radius=0.1] node[below] {\scriptsize 1};
\filldraw[fill= white] (3,1) circle [radius=0.1] node[above] {\scriptsize 2N-2};
\filldraw[fill= white] (4,1) circle [radius=0.1] node[above] {\scriptsize N-1}; 
\draw [thick] (0.1, 0) -- (0.9,0) ;
\draw [thick] (1.1, 0) -- (1.9,0) ;
\draw [thick] (2.1, 0) -- (2.9,0) ;
\draw [thick] (3.1, 0) -- (3.9,0) ;
\draw [thick] (4.1, 0) -- (4.9,0) ;
\draw [thick] (5.1, 0) -- (5.9,0) ;
\draw [thick] (3, 0.1) -- (3,0.9) ;
\draw [thick] (3.1, 1) -- (3.9,1) ;
\end{tikzpicture}\ee
corresponding to a family of theories with non-abelian global symmetry $SU(6)\times SU(2)$ (enhancing to $SU(8)\times SU(2)$ when $N=2$) and CB spectrum
$$(3,4)\,;\,(6,7)\,;\,\dots\,;\, (3N-3,3N-2)\,.$$
We notice here that the rank-$2$ theory obtained for $N=2$ in \eqref{E6Z2NMQ1} is the same as the theory in \eqref{SU8NMQ} for $N=1$. The fact that the type of 7-brane changes should not bother us, since there may be multiple ultraviolet realizations of the same SCFT in the infrared.

At this point, we have two options. If we higgs the $SU(6)$, we get back to \eqref{E6Z2NMQ}, with $N\to N-1$ (for $N=2$, higgsing the $SU(8)$ gives us the rank-$1$ E$_6$ Minahan-Nemeschansky theory). If instead we Higgs the $SU(2)$, we end up with the quiver
\be\label{E6Z2NMQ2}
\begin{tikzpicture}
\filldraw[fill= white] (0,0) circle [radius=0.1] node[below] {\scriptsize 1};
\filldraw[fill= white] (1,0) circle [radius=0.1] node[below] {\scriptsize N};
\filldraw[fill= white] (2,0) circle [radius=0.1] node[below] {\scriptsize 2N-1};
\filldraw[fill= white] (3,0) circle [radius=0.1] node[below] {\scriptsize 3N-2};
\filldraw[fill= white] (4,0) circle [radius=0.1] node[below] {\scriptsize 2N-1};
\filldraw[fill= white] (5,0) circle [radius=0.1] node[below] {\scriptsize N};
\filldraw[fill= white] (6,0) circle [radius=0.1] node[below] {\scriptsize 1};
\filldraw[fill= white] (3,1) circle [radius=0.1] node[above] {\scriptsize 2N-2};
\filldraw[fill= white] (4,1) circle [radius=0.1] node[above] {\scriptsize N-2}; 
\draw [thick] (0.1, 0) -- (0.9,0) ;
\draw [thick] (1.1, 0) -- (1.9,0) ;
\draw [thick] (2.1, 0) -- (2.9,0) ;
\draw [thick] (3.1, 0) -- (3.9,0) ;
\draw [thick] (4.1, 0) -- (4.9,0) ;
\draw [thick] (5.1, 0) -- (5.9,0) ;
\draw [thick] (3, 0.1) -- (3,0.9) ;
\draw [thick] (3.1, 1) -- (3.9,1) ;
\end{tikzpicture}\ee
corresponding to a family of theories with non-abelian global symmetry\footnote{When $N=2$ this is equivalent to the rank-$1$ Minahan-Nemeschansky theory of type E$_7$!} $E_6$  and CB spectrum
$$(3,4)\,;\,(6,7)\,;\,\dots\,;\, (3N-2).$$

Finally, by higgsing the $E_6$ global symmetry of the above family we end up with the $SU(6)\times SU(2)$ family \eqref{E6Z2NMQ1} with $N\to N-1$. The pattern just described can be summarized with the following Hasse diagram for large-enough $N$:
\be\label{HDE6Z2} 
\begin{tikzpicture} 
\node[] at (-4.3,0) {$\cdots$};
\node[] (A) at (0,0) {\scriptsize $SO(10)\times U(1)$}; 
\node[] at (0,0.3) {\scriptsize (N)};
\node[] (B) at (3,0) {\scriptsize $SU(6)\times SU(2)$}; 
\node[] at (3,0.3) {\scriptsize (N)};
\node[] (C) at (0,-1) {\scriptsize $E_6$}; 
\node[] at (0,-1.3) {\scriptsize (N+1)}; 
\node[] (D) at (-3,0) {\scriptsize $SU(6)\times SU(2)$}; 
\node[] at (-3,0.3) {\scriptsize (N+1)};
\node[] at (1.5,0.2) {\scriptsize $\mathfrak{d}_5$};
\node[] at (1.5,-0.7) {\scriptsize $\mathfrak{e}_6$};
\node[] at (-1.5,0.2) {\scriptsize $\mathfrak{a}_5$};
\node[] at (-1.5,-0.7) {\scriptsize $\mathfrak{a}_1$};
\node[] (E) at (6,0) {\scriptsize $SO(10)\times U(1)$}; 
\node[] at (6,0.3) {\scriptsize (N-1)};
\node[] (F) at (9,0) {\scriptsize $SU(6)\times SU(2)$}; 
\node[] at (9,0.3) {\scriptsize (N-1)};
\node[] (G) at (6,-1) {\scriptsize $E_6$}; 
\node[] at (6,-1.3) {\scriptsize (N)}; 
\node[] at (7.5,0.2) {\scriptsize $\mathfrak{d}_5$};
\node[] at (7.5,-0.7) {\scriptsize $\mathfrak{e}_6$};
\node[] at (4.5,0.2) {\scriptsize $\mathfrak{a}_5$};
\node[] at (4.5,-0.7) {\scriptsize $\mathfrak{a}_1$}; 
\node[] at (10.3,0) {$\cdots$};
\draw[->] (A) -- (B); 
\draw[->] (C) -- (B); 
\draw[->] (D) -- (A); 
\draw[->] (D) -- (C); 
\draw[->] (E) -- (F); 
\draw[->] (G) -- (F); 
\draw[->] (B) -- (E); 
\draw[->] (B) -- (G);
\end{tikzpicture}
\ee

\section{M-theory derivation and 6d SCFTs}\label{Sec:M-theory}

All $\mathcal{N}=2$ theories discussed in \cite{Apruzzi:2020pmv, Giacomelli:2020jel, Giacomelli:2020gee} arise via mass deformation of the torus compactification of 6d $\mathcal{N}=(1,0)$ theories and, as we will now see, this holds also for the models we have discussed in the previous sections. Our task now is to exploit the connection with six-dimensional theories to achieve a uniform description of the theories we have discussed so far and their generalizations for other types of orbifold. The connection with 6d theories follows, as we have mentioned in the introduction, from the equivalence between M5 branes probing the M9 wall in M-theory and D3 branes probing exceptional 7-branes. We will begin by reviewing the properties we need of the relevant 6d $\mathcal{N}=(1,0)$ theories. 

\subsection{Orbi-instanton theories and their Higgs branch}   

The orbi-instanton models can be uniformly described in M-theory as the worldvolume theories on a stack of $N$ M5 branes inside a M9 wall \cite{Horava:1996ma} which wraps a ADE orbifold singularity \cite{DelZotto:2014hpa, Heckman:2018pqx}. As was argued in \cite{ Heckman:2018pqx}, these models constitute the basic building blocks from which all 6d SCFTs can be constructed. We focus on the case of abelian orbifolds  $\mathbb{C}^2/\mathbb{Z}_n$. The resulting 6d theories have $\mathcal{H}\times SU(n)$, where $\mathcal{H}$ is the subgroup of $E_8$ which commutes with the chosen holonomy for the $E_8$ gauge fields at infinity in   $\mathbb{C}^2/\mathbb{Z}_n$. The allowed holonomies are in one-to-one correspondence with the inequivalent ways of choosing nodes (with multiplicities) of the affine $E_8$ Dynkin diagram such that the sum of the corresponding comarks is equal to $n$. Here we recognize the combinatorics for a 7-brane of $E_8$ type wrapping the same orbifold singularity. As we will momentarily see, this is not a coincidence and there is a precise map between the two families of theories.

One way of describing the above family of 6d theories is via the effective lagrangian theory at a generic point of their tensor branch. This is a linear quiver with $N$ gauge groups of the form (see \cite{Mekareeya:2017jgc})\footnote{For a detailed discussion about 6d supersymmetric lagrangians and their ultraviolet completion see also \cite{Bhardwaj:2015xxa, Heckman:2015bfa, Bhardwaj:2015oru}.}
\be\label{6dlagr} G_1-SU(m_2)-\dots -SU(m_{N})-\boxed{n}\ee
where $m_i\leq m_{i+1}$ and on the right the quiver ends with $n$ fundamentals of $SU(m_{N})$. The first gauge group on the left $G_1$ is either: 
\begin{itemize}
\item $USp(m_1)$ with $m_1+8$ fundamentals; 
\item $SU(m_1)$ with an antisymmetric hypermultiplet and $m_1+8$ flavors; 
\item $SU(m_1=6)$ with 15 fundamentals and a half-hyper in the rank 3 antisymmetric.
\end{itemize}
The data specifying the quiver can be reconstructed from the choice of holonomy at infinity, as explained in detail in \cite{Mekareeya:2017jgc}.

When we put this theory on a torus we get a $\mathcal{N}=2$ SCFT in four dimensions with the same global symmetry. From the data of the 6d lagrangian we can reconstruct the CB spectrum of the 4d theory using the algorithm presented in \cite{Ohmori:2018ona}, which works as follows: We start from the last $SU(m_N)$ gauge group\footnote{Here we assume $N>n$ for ease of exposition. See \cite{Ohmori:2018ona} for the details.}. This produces a set of $m_N$ CB operators in 4d whose dimension is $6$ and $6+c_i$ where $c_i$ are the degrees of Casimir invariants of $SU(m_N)$. For example for $m_N=4$ the Casimirs have degree $2,3,4$ and therefore the corresponding CB operators have dimension $6,8,9,10$. Similarly, the $SU(m_{N-1})$ gauge group next to it gives rise to a second set of CB operators with dimension $12$ plus the Casimir degrees of $SU(m_{N-1})$. We keep going in this way, adding at each step CB operators whose dimension is dictated by the Casimirs of the corresponding gauge group. For example, if the last gauge group is $SU(3)$, it gives rise to a set of three operators with dimension $6N,6N+2$ and $6N+3$. Overall, the rank of the resulting $4d$ SCFT is equal to the rank of \eqref{6dlagr} plus $N$.

We can describe the Higgs branch of orbi-instanton theories (or their counterpart in lower dimensions obtained via  dimensional reduction) using magnetic quivers. These are always star-shaped with three tails and are explicitly constructed in \cite{Mekareeya:2017jgc}. Roughly speaking, we can derive them by starting from the 3d mirrors of the effective lagrangians \eqref{6dlagr} on the tensor branch and adding $N$ times the rank-$1$ $E_8$ quiver (see \cite{Mekareeya:2017jgc, Cabrera:2019izd} for more details). Since the quivers are known, we can exploit them to study the Hasse diagram of the 6d theory by performing quiver subtraction. We observe that for given $n$ the theories obtained by choosing different holonomies at infinity are all connected by a web of higgsings (see \cite{Heckman:2015ola, Heckman:2015axa, Heckman:2016ssk, Heckman:2022suy} for further studies about RG flows in 6d). Let us discuss explicitly the $n=2$ and $n=3$ cases for concreteness. 

\paragraph{$n=2$ orbi-instantons and their magnetic quivers} 
In the $n=2$ case we have three possible 6d theories, distinguished by the allowed $\mathbb{Z}_2$ holonomies. We label them using the unbroken subgroup $\mathcal{H}$, which can be $E_8$, $SO(16)$ and $E_7\times SU(2)$. The corresponding magnetic quivers are as follows: 
\be\label{k2quivers}\begin{tabular}{|c|c|}
\hline
Subgroup $\mathcal{H}$ & Magnetic Quivers\\
\hline
$SO(16)$& \begin{tikzpicture}
\filldraw[fill= white] (-2,0) circle [radius=0.1] node[below] {\scriptsize 1};
\filldraw[fill= white] (-1,0) circle [radius=0.1] node[below] {\scriptsize 2};
\filldraw[fill= white] (0,0) circle [radius=0.1] node[below] {\scriptsize N+2};
\filldraw[fill= white] (1,0) circle [radius=0.1] node[below] {\scriptsize 2N+2};
\filldraw[fill= white] (2,0) circle [radius=0.1] node[below] {\scriptsize 3N+2};
\filldraw[fill= white] (3,0) circle [radius=0.1] node[below] {\scriptsize 4N+2};
\filldraw[fill= white] (4,0) circle [radius=0.1] node[below] {\scriptsize 5N+2};
\filldraw[fill= white] (5,0) circle [radius=0.1] node[below] {\scriptsize 6N+2};
\filldraw[fill= white] (6,0) circle [radius=0.1] node[below] {\scriptsize 4N+1};
\filldraw[fill= white] (5,1) circle [radius=0.1] node[above] {\scriptsize 3N+1};
\filldraw[fill= white] (7,0) circle [radius=0.1] node[below] {\scriptsize 2N};

\draw [thick] (-0.1, 0) -- (-0.9,0) ;
\draw [thick] (-1.1, 0) -- (-1.9,0) ;
\draw [thick] (0.1, 0) -- (0.9,0) ;
\draw [thick] (1.1, 0) -- (1.9,0) ;
\draw [thick] (2.1, 0) -- (2.9,0) ;
\draw [thick] (3.1, 0) -- (3.9,0) ;
\draw [thick] (4.1, 0) -- (4.9,0) ;
\draw [thick] (5.1, 0) -- (5.9,0) ;
\draw [thick] (5, 0.1) -- (5,0.9) ;
\draw [thick] (6.1, 0) -- (6.9,0) ;
\end{tikzpicture}\\
\hline
$E_7\times SU(2)$& \begin{tikzpicture}
\filldraw[fill= white] (-2,0) circle [radius=0.1] node[below] {\scriptsize 1};
\filldraw[fill= white] (-1,0) circle [radius=0.1] node[below] {\scriptsize 2};
\filldraw[fill= white] (0,0) circle [radius=0.1] node[below] {\scriptsize N+1};
\filldraw[fill= white] (1,0) circle [radius=0.1] node[below] {\scriptsize 2N};
\filldraw[fill= white] (2,0) circle [radius=0.1] node[below] {\scriptsize 3N};
\filldraw[fill= white] (3,0) circle [radius=0.1] node[below] {\scriptsize 4N};
\filldraw[fill= white] (4,0) circle [radius=0.1] node[below] {\scriptsize 5N};
\filldraw[fill= white] (5,0) circle [radius=0.1] node[below] {\scriptsize 6N};
\filldraw[fill= white] (6,0) circle [radius=0.1] node[below] {\scriptsize 4N};
\filldraw[fill= white] (5,1) circle [radius=0.1] node[above] {\scriptsize 3N};
\filldraw[fill= white] (7,0) circle [radius=0.1] node[below] {\scriptsize 2N};

\draw [thick] (-0.1, 0) -- (-0.9,0) ;
\draw [thick] (-1.1, 0) -- (-1.9,0) ;
\draw [thick] (0.1, 0) -- (0.9,0) ;
\draw [thick] (1.1, 0) -- (1.9,0) ;
\draw [thick] (2.1, 0) -- (2.9,0) ;
\draw [thick] (3.1, 0) -- (3.9,0) ;
\draw [thick] (4.1, 0) -- (4.9,0) ;
\draw [thick] (5.1, 0) -- (5.9,0) ;
\draw [thick] (5, 0.1) -- (5,0.9) ;
\draw [thick] (6.1, 0) -- (6.9,0) ;
\end{tikzpicture}\\
\hline
$E_8$& \begin{tikzpicture}
\filldraw[fill= white] (-2,0) circle [radius=0.1] node[below] {\scriptsize 1};
\filldraw[fill= white] (-1,0) circle [radius=0.1] node[below] {\scriptsize 2};
\filldraw[fill= white] (0,0) circle [radius=0.1] node[below] {\scriptsize N};
\filldraw[fill= white] (1,0) circle [radius=0.1] node[below] {\scriptsize 2N};
\filldraw[fill= white] (2,0) circle [radius=0.1] node[below] {\scriptsize 3N};
\filldraw[fill= white] (3,0) circle [radius=0.1] node[below] {\scriptsize 4N};
\filldraw[fill= white] (4,0) circle [radius=0.1] node[below] {\scriptsize 5N};
\filldraw[fill= white] (5,0) circle [radius=0.1] node[below] {\scriptsize 6N};
\filldraw[fill= white] (6,0) circle [radius=0.1] node[below] {\scriptsize 4N};
\filldraw[fill= white] (5,1) circle [radius=0.1] node[above] {\scriptsize 3N};
\filldraw[fill= white] (7,0) circle [radius=0.1] node[below] {\scriptsize 2N};

\draw [thick] (-0.1, 0) -- (-0.9,0) ;
\draw [thick] (-1.1, 0) -- (-1.9,0) ;
\draw [thick] (0.1, 0) -- (0.9,0) ;
\draw [thick] (1.1, 0) -- (1.9,0) ;
\draw [thick] (2.1, 0) -- (2.9,0) ;
\draw [thick] (3.1, 0) -- (3.9,0) ;
\draw [thick] (4.1, 0) -- (4.9,0) ;
\draw [thick] (5.1, 0) -- (5.9,0) ;
\draw [thick] (5, 0.1) -- (5,0.9) ;
\draw [thick] (6.1, 0) -- (6.9,0) ;
\end{tikzpicture}\\
\hline
\end{tabular}
\ee
Using quiver subtraction we find the following Hasse diagram: 
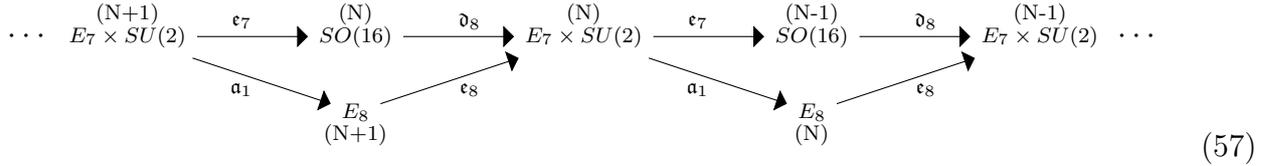
\be\label{HD6d2} 
\begin{tikzpicture} 
\node[] at (-4.3,0) {$\cdots$};
\node[] (A) at (0,0) {\scriptsize $SO(16)$}; 
\node[] at (0,0.3) {\scriptsize (N)};
\node[] (B) at (3,0) {\scriptsize $E_7\times SU(2)$}; 
\node[] at (3,0.3) {\scriptsize (N)};
\node[] (C) at (0,-1) {\scriptsize $E_8$}; 
\node[] at (0,-1.3) {\scriptsize (N+1)}; 
\node[] (D) at (-3,0) {\scriptsize $E_7\times SU(2)$}; 
\node[] at (-3,0.3) {\scriptsize (N+1)};
\node[] at (1.5,0.2) {\scriptsize $\mathfrak{d}_8$};
\node[] at (1.5,-0.7) {\scriptsize $\mathfrak{e}_8$};
\node[] at (-1.5,0.2) {\scriptsize $\mathfrak{e}_7$};
\node[] at (-1.5,-0.7) {\scriptsize $\mathfrak{a}_1$};
\node[] (E) at (6,0) {\scriptsize $SO(16)$}; 
\node[] at (6,0.3) {\scriptsize (N-1)};
\node[] (F) at (9,0) {\scriptsize $E_7\times SU(2)$}; 
\node[] at (9,0.3) {\scriptsize (N-1)};
\node[] (G) at (6,-1) {\scriptsize $E_8$}; 
\node[] at (6,-1.3) {\scriptsize (N)}; 
\node[] at (7.5,0.2) {\scriptsize $\mathfrak{d}_8$};
\node[] at (7.5,-0.7) {\scriptsize $\mathfrak{e}_8$};
\node[] at (4.5,0.2) {\scriptsize $\mathfrak{e}_7$};
\node[] at (4.5,-0.7) {\scriptsize $\mathfrak{a}_1$}; 
\node[] at (10.3,0) {$\cdots$};
\draw[->] (A) -- (B); 
\draw[->] (C) -- (B); 
\draw[->] (D) -- (A); 
\draw[->] (D) -- (C); 
\draw[->] (E) -- (F); 
\draw[->] (G) -- (F); 
\draw[->] (B) -- (E); 
\draw[->] (B) -- (G);
\end{tikzpicture}
\ee 
Notice that the RG flows interpolate between 6d theories labelled by a different number of M5 branes and the relevant higgsings only involve expectation values for the $\mathcal{H}$ moment maps, and therefore do not involve operators charged under the $SU(2)$ symmetry coming from the $\mathbb{Z}_2$ orbifold. In other words, this is not the full Hasse diagram of the theory but is enough for our purposes. Notice also that \eqref{HD6d2} and \eqref{HDE8} are identical, and we will momentarily interpret this fact.

\paragraph{$n=3$ orbi-instantons and their magnetic quivers} \label{k=3quiv}
We can repeat the previous analysis for $n=3$. The possible subgroups left unbroken by the holonomies are $E_8$, $E_7\times U(1)$, $E_6\times SU(3)$, $SO(14)\times U(1)$ and $SU(9)$. The corresponding magnetic quivers for the 6d SCFTs are as follows (see \cite{Cabrera:2019izd}): 
\be\label{Table:SU(3)}\begin{tabular}{|c|c|}
\hline
Subgroup $\mathcal{H}$ & Magnetic Quivers\\
\hline
$SU(9)$& \begin{tikzpicture}
\filldraw[fill= white] (-3,0) circle [radius=0.1] node[below] {\scriptsize 1};
\filldraw[fill= white] (-2,0) circle [radius=0.1] node[below] {\scriptsize 2};
\filldraw[fill= white] (-1,0) circle [radius=0.1] node[below] {\scriptsize 3};
\filldraw[fill= white] (0,0) circle [radius=0.1] node[below] {\scriptsize N+3};
\filldraw[fill= white] (1,0) circle [radius=0.1] node[below] {\scriptsize 2N+3};
\filldraw[fill= white] (2,0) circle [radius=0.1] node[below] {\scriptsize 3N+3};
\filldraw[fill= white] (3,0) circle [radius=0.1] node[below] {\scriptsize 4N+3};
\filldraw[fill= white] (4,0) circle [radius=0.1] node[below] {\scriptsize 5N+3};
\filldraw[fill= white] (5,0) circle [radius=0.1] node[below] {\scriptsize 6N+3};
\filldraw[fill= white] (6,0) circle [radius=0.1] node[below] {\scriptsize 4N+2};
\filldraw[fill= white] (5,1) circle [radius=0.1] node[above] {\scriptsize 3N+1};
\filldraw[fill= white] (7,0) circle [radius=0.1] node[below] {\scriptsize 2N+1};

\draw [thick] (-0.1, 0) -- (-0.9,0) ;
\draw [thick] (-1.1, 0) -- (-1.9,0) ;
\draw [thick] (-2.1, 0) -- (-2.9,0) ;
\draw [thick] (0.1, 0) -- (0.9,0) ;
\draw [thick] (1.1, 0) -- (1.9,0) ;
\draw [thick] (2.1, 0) -- (2.9,0) ;
\draw [thick] (3.1, 0) -- (3.9,0) ;
\draw [thick] (4.1, 0) -- (4.9,0) ;
\draw [thick] (5.1, 0) -- (5.9,0) ;
\draw [thick] (5, 0.1) -- (5,0.9) ;
\draw [thick] (6.1, 0) -- (6.9,0) ;
\end{tikzpicture}\\
\hline
$SO(14)\times U(1)$& \begin{tikzpicture}
\filldraw[fill= white] (-3,0) circle [radius=0.1] node[below] {\scriptsize 1};
\filldraw[fill= white] (-2,0) circle [radius=0.1] node[below] {\scriptsize 2};
\filldraw[fill= white] (-1,0) circle [radius=0.1] node[below] {\scriptsize 3};
\filldraw[fill= white] (0,0) circle [radius=0.1] node[below] {\scriptsize N+2};
\filldraw[fill= white] (1,0) circle [radius=0.1] node[below] {\scriptsize 2N+2};
\filldraw[fill= white] (2,0) circle [radius=0.1] node[below] {\scriptsize 3N+2};
\filldraw[fill= white] (3,0) circle [radius=0.1] node[below] {\scriptsize 4N+2};
\filldraw[fill= white] (4,0) circle [radius=0.1] node[below] {\scriptsize 5N+2};
\filldraw[fill= white] (5,0) circle [radius=0.1] node[below] {\scriptsize 6N+2};
\filldraw[fill= white] (6,0) circle [radius=0.1] node[below] {\scriptsize 4N+1};
\filldraw[fill= white] (5,1) circle [radius=0.1] node[above] {\scriptsize 3N+1};
\filldraw[fill= white] (7,0) circle [radius=0.1] node[below] {\scriptsize 2N};

\draw [thick] (-0.1, 0) -- (-0.9,0) ;
\draw [thick] (-1.1, 0) -- (-1.9,0) ;
\draw [thick] (-2.1, 0) -- (-2.9,0) ;
\draw [thick] (0.1, 0) -- (0.9,0) ;
\draw [thick] (1.1, 0) -- (1.9,0) ;
\draw [thick] (2.1, 0) -- (2.9,0) ;
\draw [thick] (3.1, 0) -- (3.9,0) ;
\draw [thick] (4.1, 0) -- (4.9,0) ;
\draw [thick] (5.1, 0) -- (5.9,0) ;
\draw [thick] (5, 0.1) -- (5,0.9) ;
\draw [thick] (6.1, 0) -- (6.9,0) ;
\end{tikzpicture}\\
\hline
$E_6\times SU(3)$& \begin{tikzpicture}
\filldraw[fill= white] (-3,0) circle [radius=0.1] node[below] {\scriptsize 1};
\filldraw[fill= white] (-2,0) circle [radius=0.1] node[below] {\scriptsize 2};
\filldraw[fill= white] (-1,0) circle [radius=0.1] node[below] {\scriptsize 3};
\filldraw[fill= white] (0,0) circle [radius=0.1] node[below] {\scriptsize N+2};
\filldraw[fill= white] (1,0) circle [radius=0.1] node[below] {\scriptsize 2N+1};
\filldraw[fill= white] (2,0) circle [radius=0.1] node[below] {\scriptsize 3N};
\filldraw[fill= white] (3,0) circle [radius=0.1] node[below] {\scriptsize 4N};
\filldraw[fill= white] (4,0) circle [radius=0.1] node[below] {\scriptsize 5N};
\filldraw[fill= white] (5,0) circle [radius=0.1] node[below] {\scriptsize 6N};
\filldraw[fill= white] (6,0) circle [radius=0.1] node[below] {\scriptsize 4N};
\filldraw[fill= white] (5,1) circle [radius=0.1] node[above] {\scriptsize 3N};
\filldraw[fill= white] (7,0) circle [radius=0.1] node[below] {\scriptsize 2N};

\draw [thick] (-0.1, 0) -- (-0.9,0) ;
\draw [thick] (-1.1, 0) -- (-1.9,0) ;
\draw [thick] (-2.1, 0) -- (-2.9,0) ;
\draw [thick] (0.1, 0) -- (0.9,0) ;
\draw [thick] (1.1, 0) -- (1.9,0) ;
\draw [thick] (2.1, 0) -- (2.9,0) ;
\draw [thick] (3.1, 0) -- (3.9,0) ;
\draw [thick] (4.1, 0) -- (4.9,0) ;
\draw [thick] (5.1, 0) -- (5.9,0) ;
\draw [thick] (5, 0.1) -- (5,0.9) ;
\draw [thick] (6.1, 0) -- (6.9,0) ;
\end{tikzpicture}\\
\hline
$E_7\times U(1)$& \begin{tikzpicture}
\filldraw[fill= white] (-3,0) circle [radius=0.1] node[below] {\scriptsize 1};
\filldraw[fill= white] (-2,0) circle [radius=0.1] node[below] {\scriptsize 2};
\filldraw[fill= white] (-1,0) circle [radius=0.1] node[below] {\scriptsize 3};
\filldraw[fill= white] (0,0) circle [radius=0.1] node[below] {\scriptsize N+1};
\filldraw[fill= white] (1,0) circle [radius=0.1] node[below] {\scriptsize 2N};
\filldraw[fill= white] (2,0) circle [radius=0.1] node[below] {\scriptsize 3N};
\filldraw[fill= white] (3,0) circle [radius=0.1] node[below] {\scriptsize 4N};
\filldraw[fill= white] (4,0) circle [radius=0.1] node[below] {\scriptsize 5N};
\filldraw[fill= white] (5,0) circle [radius=0.1] node[below] {\scriptsize 6N};
\filldraw[fill= white] (6,0) circle [radius=0.1] node[below] {\scriptsize 4N};
\filldraw[fill= white] (5,1) circle [radius=0.1] node[above] {\scriptsize 3N};
\filldraw[fill= white] (7,0) circle [radius=0.1] node[below] {\scriptsize 2N};

\draw [thick] (-0.1, 0) -- (-0.9,0) ;
\draw [thick] (-1.1, 0) -- (-1.9,0) ;
\draw [thick] (-2.1, 0) -- (-2.9,0) ;
\draw [thick] (0.1, 0) -- (0.9,0) ;
\draw [thick] (1.1, 0) -- (1.9,0) ;
\draw [thick] (2.1, 0) -- (2.9,0) ;
\draw [thick] (3.1, 0) -- (3.9,0) ;
\draw [thick] (4.1, 0) -- (4.9,0) ;
\draw [thick] (5.1, 0) -- (5.9,0) ;
\draw [thick] (5, 0.1) -- (5,0.9) ;
\draw [thick] (6.1, 0) -- (6.9,0) ;
\end{tikzpicture}\\
\hline
$E_8$& \begin{tikzpicture}
\filldraw[fill= white] (-3,0) circle [radius=0.1] node[below] {\scriptsize 1};
\filldraw[fill= white] (-2,0) circle [radius=0.1] node[below] {\scriptsize 2};
\filldraw[fill= white] (-1,0) circle [radius=0.1] node[below] {\scriptsize 3};
\filldraw[fill= white] (0,0) circle [radius=0.1] node[below] {\scriptsize N};
\filldraw[fill= white] (1,0) circle [radius=0.1] node[below] {\scriptsize 2N};
\filldraw[fill= white] (2,0) circle [radius=0.1] node[below] {\scriptsize 3N};
\filldraw[fill= white] (3,0) circle [radius=0.1] node[below] {\scriptsize 4N};
\filldraw[fill= white] (4,0) circle [radius=0.1] node[below] {\scriptsize 5N};
\filldraw[fill= white] (5,0) circle [radius=0.1] node[below] {\scriptsize 6N};
\filldraw[fill= white] (6,0) circle [radius=0.1] node[below] {\scriptsize 4N};
\filldraw[fill= white] (5,1) circle [radius=0.1] node[above] {\scriptsize 3N};
\filldraw[fill= white] (7,0) circle [radius=0.1] node[below] {\scriptsize 2N};

\draw [thick] (-0.1, 0) -- (-0.9,0) ;
\draw [thick] (-1.1, 0) -- (-1.9,0) ;
\draw [thick] (-2.1, 0) -- (-2.9,0) ;
\draw [thick] (0.1, 0) -- (0.9,0) ;
\draw [thick] (1.1, 0) -- (1.9,0) ;
\draw [thick] (2.1, 0) -- (2.9,0) ;
\draw [thick] (3.1, 0) -- (3.9,0) ;
\draw [thick] (4.1, 0) -- (4.9,0) ;
\draw [thick] (5.1, 0) -- (5.9,0) ;
\draw [thick] (5, 0.1) -- (5,0.9) ;
\draw [thick] (6.1, 0) -- (6.9,0) ;
\end{tikzpicture}\\
\hline
\end{tabular}
\ee
Using again quiver subtraction we find also in this case a Hasse diagram interpolating between all possible theories. 
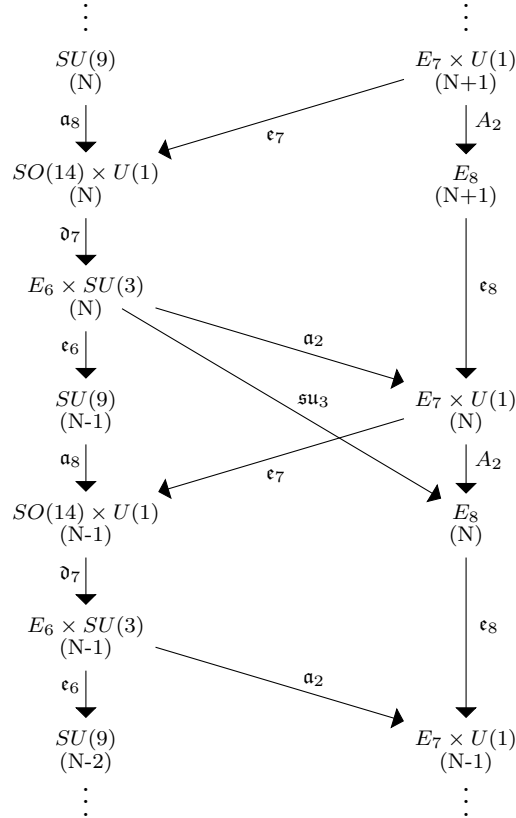
\be\label{HD6d3} 
\begin{tikzpicture} 
\node[] at (0,.7) {$\vdots$};
\node[] (A) at (0,0) {\scriptsize $SU(9)$}; 
\node[] (a) at (0,-0.3) {\scriptsize (N)};
\node[] (B) at (0,-1.5) {\scriptsize $SO(14)\times U(1)$}; 
\node[] (b) at (0,-1.8) {\scriptsize (N)};
\node[] (C) at (0,-3) {\scriptsize $E_6\times SU(3)$}; 
\node[] (c) at (0,-3.3) {\scriptsize (N)}; 
\node[] (D) at (0,-4.5) {\scriptsize $SU(9)$}; 
\node[] (d) at (0,-4.8) {\scriptsize (N-1)};
\node[] (E) at (0,-6) {\scriptsize $SO(14)\times U(1)$}; 
\node[] (e) at (0,-6.3) {\scriptsize (N-1)};
\node[] (F) at (0,-7.5) {\scriptsize $E_6\times SU(3)$}; 
\node[] (f) at (0,-7.8) {\scriptsize (N-1)};
\node[] (G) at (0,-9) {\scriptsize $SU(9)$}; 
\node[] (g) at (0,-9.3) {\scriptsize (N-2)}; 
\node[] at (0,-9.7) {$\vdots$};
\node[] (H) at (5,-4.5) {\scriptsize $E_7\times U(1)$}; 
\node[] (h) at (5,-4.8) {\scriptsize (N)};
\node[] (I) at (5,-6) {\scriptsize $E_8$}; 
\node[] (i) at (5,-6.3) {\scriptsize (N)};
\node[] (L) at (5,-9) {\scriptsize $E_7\times U(1)$}; 
\node[] (l) at (5,-9.3) {\scriptsize (N-1)};
\node[] (M) at (5,0) {\scriptsize $E_7\times U(1)$}; 
\node[] (m) at (5,-0.3) {\scriptsize (N+1)};
\node[] (N) at (5,-1.5) {\scriptsize $E_8$}; 
\node[] (n) at (5,-1.8) {\scriptsize (N+1)};
\node[] at (5,-9.7) {$\vdots$};
\node[] at (5,.7) {$\vdots$};
\node[] at (-0.2,-2.3) {\scriptsize $\mathfrak{d}_7$};
\node[] at (-0.2,-3.8) {\scriptsize $\mathfrak{e}_6$};
\node[] at (-0.2,-0.8) {\scriptsize $\mathfrak{a}_8$};
\node[] at (-0.2,-6.8) {\scriptsize $\mathfrak{d}_7$};
\node[] at (-0.2,-8.3) {\scriptsize $\mathfrak{e}_6$};
\node[] at (-0.2,-5.3) {\scriptsize $\mathfrak{a}_8$}; 
\node[] at (5.3,-3) {\scriptsize $\mathfrak{e}_8$};
\node[] at (5.3,-7.5) {\scriptsize $\mathfrak{e}_8$};
\node[] at (5.3,-0.8) {\scriptsize $A_2$};
\node[] at (5.3,-5.3) {\scriptsize $A_2$};
\node[] at (3,-3.7) {\scriptsize $\mathfrak{a}_2$};
\node[] at (3,-8.2) {\scriptsize $\mathfrak{a}_2$};
\node[] at (3,-4.5) {\scriptsize $\mathfrak{su}_3$};
\node[] at (2.5,-1) {\scriptsize $\mathfrak{e}_7$};
\node[] at (2.5,-5.5) {\scriptsize $\mathfrak{e}_7$};
\draw[->] (a) -- (B); 
\draw[->] (b) -- (C); 
\draw[->] (c) -- (D); 
\draw[->] (d) -- (E); 
\draw[->] (e) -- (F); 
\draw[->] (f) -- (G); 
\draw[->] (m) -- (N); 
\draw[->] (n) -- (H); 
\draw[->] (h) -- (I); 
\draw[->] (i) -- (L); 
\draw[->] (C) -- (H); 
\draw[->] (C) -- (I);
\draw[->] (F) -- (L);
\draw[->] (M) -- (B);
\draw[->] (H) -- (E);
\end{tikzpicture}
\ee 
In the Hasse diagram $\mathfrak{su}_3$ denotes the principal nilpotent orbit of $SU(3)$ and $A_2$ the Kleinian singularity $\mathbb{C}^2/\mathbb{Z}_3$.

\subsection{Mass deformations and Type IIB theories}\label{massdef}

At this stage the natural question is how the theories we have discussed in Section \ref{Sec:Non-Pert} are related to the 6d models we have just discussed, or more precisely their double dimensional reduction. 
The answer is that they are connected by a mass deformation for the $SU(n)$ symmetry. This is to be contrasted with the setup discussed in \cite{Giacomelli:2020gee}, where that symmetry was broken by the almost commuting holonomies which are not present in the case of interest for us. 

To understand why mass deformations play a role, it is useful to make an analogy with the case of D3 branes probing a $\mathbb{C}^2/\mathbb{Z}_n$ singularity, whose worldvolume theory is known to coincide with the Douglas-Moore circular quiver. This however is only true once we have turned on the B-field. As is discussed in detail in \cite{Ohmori:2015pia}, this brane setup is related after a T-duality (and a lift to M-theory) to the torus compactification of M5 branes probing the same orbifold and from this realization we clearly see that the resulting field theory has $SU(n)\times SU(n)$ global symmetry, a feature we do not expect from the Type IIB setup. The key point is that turning on the B-field in Type IIB is equivalent to activating mass terms which break $SU(n)\times SU(n)$ to $U(1)^{n-1}$. Only after this mass deformation the 4d theory reduces to the familiar circular quiver (see  \cite{Ohmori:2015pia}).
The theories we have discussed in Section  \ref{Sec:Non-Pert} are analogous to the Douglas-Moore models and appear only once we have mass deformed the naive dimensional reduction of the underlying 6d theory. We can readily verify this is indeed the case for $n=2$. Since the models at hand are strongly-coupled it is not obvious how to analyze the RG flow in 4d and therefore we find it more convenient to work in terms of the corresponding magnetic quivers and implement mass deformations using FI parameters. This was developed in detail in \cite{vanBeest:2021xyt} and is reviewed in Appendix \ref{FIdef}. In order to turn on a mass for the $SU(2)$ global symmetry of $n=2$ theories, we turn on FI parameters at the leftmost $U(1)$ and $U(2)$ nodes of the quivers \eqref{k2quivers}. This move clearly does not affect the global symmetry $\mathcal{H}$. It is straightforward to check, using the rules reviewed in Appendix \ref{FIdef}, that the $SO(16)$ quiver in  \eqref{k2quivers} reduces to \eqref{D8NMQ}, the $E_7\times SU(2)$ quiver reduces to \eqref{E7NMQ} and finally the $E_8$ quiver becomes  \eqref{E8NMQ}.

An interesting observation is that the chain of higgsings we have discussed before for 6d theories involves expectation values for operators charged under $\mathcal{H}$ only, and therefore is not affected by the mass deformation for the $SU(n)$ factor. As a result, the 6d Hasse diagram described before should apply to the models discussed in Section \ref{Sec:Non-Pert} as well. This explains why the Hasse diagram for $k=2$ is identical to that of Type IIB theories with a 7-brane of type E$_8$ wrapping $\mathbb{C}^2/\mathbb{Z}_2$. We would like to stress that with this approach we can entirely derive the properties of the E$_8$ 7-brane theories from known results about 6d theories in an algorithmic way, therefore bypassing the guesswork of Section  \ref{Sec:Non-Pert}. More general choices of 7-brane are now derived by activating further mass deformations which break the group $\mathcal{H}$. 

Let us discuss explicitly the $n=3$ case as well to see how this works for higher orbifold orders. We start from the $SU(9)$ quiver. In this case we should turn on two mass parameters, which can be done at the magnetic quiver level as follows: 
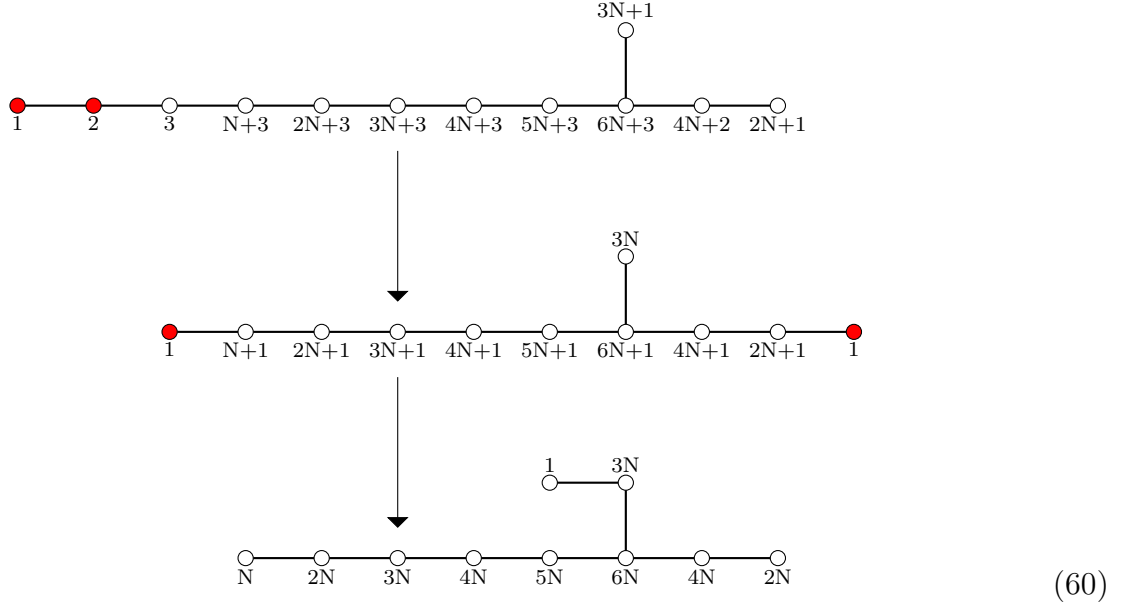
\be\label{SU9def}
\begin{tikzpicture}
\filldraw[fill= red] (-3,0) circle [radius=0.1] node[below] {\scriptsize 1};
\filldraw[fill= red] (-2,0) circle [radius=0.1] node[below] {\scriptsize 2};
\filldraw[fill= white] (-1,0) circle [radius=0.1] node[below] {\scriptsize 3};
\filldraw[fill= white] (0,0) circle [radius=0.1] node[below] {\scriptsize N+3};
\filldraw[fill= white] (1,0) circle [radius=0.1] node[below] {\scriptsize 2N+3};
\filldraw[fill= white] (2,0) circle [radius=0.1] node[below] {\scriptsize 3N+3};
\filldraw[fill= white] (3,0) circle [radius=0.1] node[below] {\scriptsize 4N+3};
\filldraw[fill= white] (4,0) circle [radius=0.1] node[below] {\scriptsize 5N+3};
\filldraw[fill= white] (5,0) circle [radius=0.1] node[below] {\scriptsize 6N+3};
\filldraw[fill= white] (6,0) circle [radius=0.1] node[below] {\scriptsize 4N+2};
\filldraw[fill= white] (5,1) circle [radius=0.1] node[above] {\scriptsize 3N+1};
\filldraw[fill= white] (7,0) circle [radius=0.1] node[below] {\scriptsize 2N+1};

\draw [thick] (-0.1, 0) -- (-0.9,0) ;
\draw [thick] (-1.1, 0) -- (-1.9,0) ;
\draw [thick] (-2.1, 0) -- (-2.9,0) ;
\draw [thick] (0.1, 0) -- (0.9,0) ;
\draw [thick] (1.1, 0) -- (1.9,0) ;
\draw [thick] (2.1, 0) -- (2.9,0) ;
\draw [thick] (3.1, 0) -- (3.9,0) ;
\draw [thick] (4.1, 0) -- (4.9,0) ;
\draw [thick] (5.1, 0) -- (5.9,0) ;
\draw [thick] (5, 0.1) -- (5,0.9) ;
\draw [thick] (6.1, 0) -- (6.9,0) ;

\draw [->] (2,-0.6) -- (2,-2.6) ;

\filldraw[fill= red] (-1,-3) circle [radius=0.1] node[below] {\scriptsize 1};
\filldraw[fill= white] (0,-3) circle [radius=0.1] node[below] {\scriptsize N+1};
\filldraw[fill= white] (1,-3) circle [radius=0.1] node[below] {\scriptsize 2N+1};
\filldraw[fill= white] (2,-3) circle [radius=0.1] node[below] {\scriptsize 3N+1};
\filldraw[fill= white] (3,-3) circle [radius=0.1] node[below] {\scriptsize 4N+1};
\filldraw[fill= white] (4,-3) circle [radius=0.1] node[below] {\scriptsize 5N+1};
\filldraw[fill= white] (5,-3) circle [radius=0.1] node[below] {\scriptsize 6N+1};
\filldraw[fill= white] (6,-3) circle [radius=0.1] node[below] {\scriptsize 4N+1};
\filldraw[fill= white] (5,-2) circle [radius=0.1] node[above] {\scriptsize 3N};
\filldraw[fill= white] (7,-3) circle [radius=0.1] node[below] {\scriptsize 2N+1};
\filldraw[fill= red] (8,-3) circle [radius=0.1] node[below] {\scriptsize 1};

\draw [thick] (-0.1, -3) -- (-0.9,-3) ;
\draw [thick] (7.1, -3) -- (7.9,-3) ;
\draw [thick] (0.1, -3) -- (0.9,-3) ;
\draw [thick] (1.1, -3) -- (1.9,-3) ;
\draw [thick] (2.1, -3) -- (2.9,-3) ;
\draw [thick] (3.1, -3) -- (3.9,-3) ;
\draw [thick] (4.1, -3) -- (4.9,-3) ;
\draw [thick] (5.1, -3) -- (5.9,-3) ;
\draw [thick] (5, -2.1) -- (5,-2.9) ;
\draw [thick] (6.1, -3) -- (6.9,-3) ;

\draw [->] (2,-3.6) -- (2,-5.6) ;

\filldraw[fill= white] (0,-6) circle [radius=0.1] node[below] {\scriptsize N};
\filldraw[fill= white] (1,-6) circle [radius=0.1] node[below] {\scriptsize 2N};
\filldraw[fill= white] (2,-6) circle [radius=0.1] node[below] {\scriptsize 3N};
\filldraw[fill= white] (3,-6) circle [radius=0.1] node[below] {\scriptsize 4N};
\filldraw[fill= white] (4,-6) circle [radius=0.1] node[below] {\scriptsize 5N};
\filldraw[fill= white] (5,-6) circle [radius=0.1] node[below] {\scriptsize 6N};
\filldraw[fill= white] (6,-6) circle [radius=0.1] node[below] {\scriptsize 4N};
\filldraw[fill= white] (5,-5) circle [radius=0.1] node[above] {\scriptsize 3N};
\filldraw[fill= white] (7,-6) circle [radius=0.1] node[below] {\scriptsize 2N};
\filldraw[fill= white] (4,-5) circle [radius=0.1] node[above] {\scriptsize 1};

\draw [thick] (4.1, -5) -- (4.9,-5) ;
\draw [thick] (0.1, -6) -- (0.9,-6) ;
\draw [thick] (1.1, -6) -- (1.9,-6) ;
\draw [thick] (2.1, -6) -- (2.9,-6) ;
\draw [thick] (3.1, -6) -- (3.9,-6) ;
\draw [thick] (4.1, -6) -- (4.9,-6) ;
\draw [thick] (5.1, -6) -- (5.9,-6) ;
\draw [thick] (5, -5.1) -- (5,-5.9) ;
\draw [thick] (6.1, -6) -- (6.9,-6) ;
\end{tikzpicture}
\ee 
After the mass deformation we land on a \emph{canonical} theory with rank-$3N$ whose CB spectrum includes $N$ triples of operators of dimension: 
$$ (3,5,6)\,;\,(9,11,12)\,;\,\dots\,;\,(6N-3,6N-1,6N)\,.$$
We can calculate the central charges $8a-4c$ and $24(c-a)$ respectively from the CB spectrum and the magnetic quiver. The result is consistent with the holographic formula in \eqref{formulacc} if we set $\alpha=\beta=0$. Moreover, we have $\epsilon=\frac{1}{18}$. \\
We can proceed with the analysis of FI deformations for all other holonomy choices in \eqref{Table:SU(3)}, or alternatively we can analyze the Hasse diagram of the theory to find the missing quivers. We follow the latter option.
The effect of turning on a vev along a $SU(9)$ nilpotent orbit is reproduced by quiver subtraction. The higgsing leads to the theory with $SO(14)\times U(1)$ global symmetry:
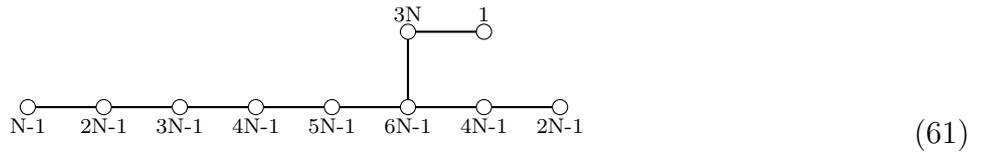
\be\label{E8k3Na8}
\begin{tikzpicture}
\filldraw[fill= white] (0,0) circle [radius=0.1] node[below] {\scriptsize N-1};
\filldraw[fill= white] (1,0) circle [radius=0.1] node[below] {\scriptsize 2N-1};
\filldraw[fill= white] (2,0) circle [radius=0.1] node[below] {\scriptsize 3N-1};
\filldraw[fill= white] (3,0) circle [radius=0.1] node[below] {\scriptsize 4N-1};
\filldraw[fill= white] (4,0) circle [radius=0.1] node[below] {\scriptsize 5N-1};
\filldraw[fill= white] (5,0) circle [radius=0.1] node[below] {\scriptsize 6N-1};
\filldraw[fill= white] (6,0) circle [radius=0.1] node[below] {\scriptsize 4N-1};
\filldraw[fill= white] (5,1) circle [radius=0.1] node[above] {\scriptsize 3N};
\filldraw[fill= white] (6,1) circle [radius=0.1] node[above] {\scriptsize 1};
\filldraw[fill= white] (7,0) circle [radius=0.1] node[below] {\scriptsize 2N-1};

\draw [thick] (0.1, 0) -- (0.9,0) ;
\draw [thick] (1.1, 0) -- (1.9,0) ;
\draw [thick] (2.1, 0) -- (2.9,0) ;
\draw [thick] (3.1, 0) -- (3.9,0) ;
\draw [thick] (4.1, 0) -- (4.9,0) ;
\draw [thick] (5.1, 0) -- (5.9,0) ;
\draw [thick] (5, 0.1) -- (5,0.9) ;
\draw [thick] (5.1, 1) -- (5.9,1) ;
\draw [thick] (6.1, 0) -- (6.9,0) ;
\end{tikzpicture} \ee 
The theory has rank $3N-1$ and the CB operators carry dimension:
$$(3,5,6)\,;\,(9,11,12)\,;\, \dots\,;\,(6N-3, 6N-1)$$ 
which is the same sequence as for \eqref{SU9def} except for the missing $6N$.
Consistency with \eqref{formulacc} requires $\alpha=1,\ \beta=-8, \epsilon=-\frac{5}{18}$.\\
Turning on a nilpotent vev for $SO(14)$ we land on the $E_6\times SU(3)$ theory:
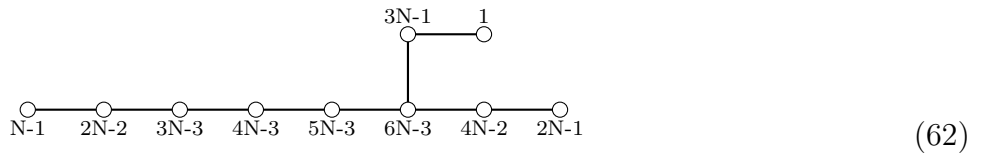
\be\label{E8k3Nd7}
\begin{tikzpicture}
\filldraw[fill= white] (0,0) circle [radius=0.1] node[below] {\scriptsize N-1};
\filldraw[fill= white] (1,0) circle [radius=0.1] node[below] {\scriptsize 2N-2};
\filldraw[fill= white] (2,0) circle [radius=0.1] node[below] {\scriptsize 3N-3};
\filldraw[fill= white] (3,0) circle [radius=0.1] node[below] {\scriptsize 4N-3};
\filldraw[fill= white] (4,0) circle [radius=0.1] node[below] {\scriptsize 5N-3};
\filldraw[fill= white] (5,0) circle [radius=0.1] node[below] {\scriptsize 6N-3};
\filldraw[fill= white] (6,0) circle [radius=0.1] node[below] {\scriptsize 4N-2};
\filldraw[fill= white] (5,1) circle [radius=0.1] node[above] {\scriptsize 3N-1};
\filldraw[fill= white] (6,1) circle [radius=0.1] node[above] {\scriptsize 1};
\filldraw[fill= white] (7,0) circle [radius=0.1] node[below] {\scriptsize 2N-1};

\draw [thick] (0.1, 0) -- (0.9,0) ;
\draw [thick] (1.1, 0) -- (1.9,0) ;
\draw [thick] (2.1, 0) -- (2.9,0) ;
\draw [thick] (3.1, 0) -- (3.9,0) ;
\draw [thick] (4.1, 0) -- (4.9,0) ;
\draw [thick] (5.1, 0) -- (5.9,0) ;
\draw [thick] (5, 0.1) -- (5,0.9) ;
\draw [thick] (5.1, 1) -- (5.9,1) ;
\draw [thick] (6.1, 0) -- (6.9,0) ;
\end{tikzpicture} \ee 
Now we have a rank $3N-2$ theory whose CB spectrum includes operators of dimension: 
$$ (3,5,6)\,;\,\dots\,;\,(6N-3)\,.$$ 
The sequence differs from the $SU(9)$ case only by the absence of operators of dimension $6N-1$ and $6N$. In this case \eqref{formulacc} is reproduced for $\alpha=4$, $\beta=-19$ and $\epsilon=-\frac{5}{18}$. 
Now we are allowed to perform two different higgsings. By activating a minimal nilpotent for the $SU(3)$ moment map we find: 
\be\label{E8k3Na2}
\begin{tikzpicture}
\filldraw[fill= white] (0,0) circle [radius=0.1] node[below] {\scriptsize N-2};
\filldraw[fill= white] (1,0) circle [radius=0.1] node[below] {\scriptsize 2N-3};
\filldraw[fill= white] (2,0) circle [radius=0.1] node[below] {\scriptsize 3N-3};
\filldraw[fill= white] (3,0) circle [radius=0.1] node[below] {\scriptsize 4N-3};
\filldraw[fill= white] (4,0) circle [radius=0.1] node[below] {\scriptsize 5N-3};
\filldraw[fill= white] (5,0) circle [radius=0.1] node[below] {\scriptsize 6N-3};
\filldraw[fill= white] (6,0) circle [radius=0.1] node[below] {\scriptsize 4N-2};
\filldraw[fill= white] (5,1) circle [radius=0.1] node[above] {\scriptsize 3N-1};
\filldraw[fill= white] (6,1) circle [radius=0.1] node[above] {\scriptsize 1};
\filldraw[fill= white] (7,0) circle [radius=0.1] node[below] {\scriptsize 2N-1};

\draw [thick] (0.1, 0) -- (0.9,0) ;
\draw [thick] (1.1, 0) -- (1.9,0) ;
\draw [thick] (2.1, 0) -- (2.9,0) ;
\draw [thick] (3.1, 0) -- (3.9,0) ;
\draw [thick] (4.1, 0) -- (4.9,0) ;
\draw [thick] (5.1, 0) -- (5.9,0) ;
\draw [thick] (5, 0.1) -- (5,0.9) ;
\draw [thick] (5.1, 1) -- (5.9,1) ;
\draw [thick] (6.1, 0) -- (6.9,0) ;
\end{tikzpicture} \ee 
The associated global symmetry is $E_7\times U(1)$. The theory has rank $3N-3$ and the CB spectrum reads:
$$ (3,5,6)\,;\,\dots\,;\,(6N-9,6N-7)\,;\,(6N-3).$$
Then we have $\alpha=17$, $\beta=-21$, $\epsilon= -\frac{17}{18}$.
One can check that higgsing the $E_7$ degrees of freedom leads to quiver \ref{E8k3Na8} with $N\rightarrow N-1$.
If instead we consider a principal nilpotent vev for $SU(3)$ we get the $E_8$ quiver
\be\label{E8k3Na3}
\begin{tikzpicture}
\filldraw[fill= white] (0,0) circle [radius=0.1] node[below] {\scriptsize N-3};
\filldraw[fill= white] (1,0) circle [radius=0.1] node[below] {\scriptsize 2N-3};
\filldraw[fill= white] (2,0) circle [radius=0.1] node[below] {\scriptsize 3N-3};
\filldraw[fill= white] (3,0) circle [radius=0.1] node[below] {\scriptsize 4N-3};
\filldraw[fill= white] (4,0) circle [radius=0.1] node[below] {\scriptsize 5N-3};
\filldraw[fill= white] (5,0) circle [radius=0.1] node[below] {\scriptsize 6N-3};
\filldraw[fill= white] (6,0) circle [radius=0.1] node[below] {\scriptsize 4N-2};
\filldraw[fill= white] (5,1) circle [radius=0.1] node[above] {\scriptsize 3N-1};
\filldraw[fill= white] (6,1) circle [radius=0.1] node[above] {\scriptsize 1};
\filldraw[fill= white] (7,0) circle [radius=0.1] node[below] {\scriptsize 2N-1};

\draw [thick] (0.1, 0) -- (0.9,0) ;
\draw [thick] (1.1, 0) -- (1.9,0) ;
\draw [thick] (2.1, 0) -- (2.9,0) ;
\draw [thick] (3.1, 0) -- (3.9,0) ;
\draw [thick] (4.1, 0) -- (4.9,0) ;
\draw [thick] (5.1, 0) -- (5.9,0) ;
\draw [thick] (5, 0.1) -- (5,0.9) ;
\draw [thick] (5.1, 1) -- (5.9,1) ;
\draw [thick] (6.1, 0) -- (6.9,0) ;
\end{tikzpicture} \ee 
The underlying 4d theory has rank $3N-5$ and CB operators of scaling dimension 
$$ (3,5,6)\,;\,\dots\,;\,(6N-15,6N-13)\,;(6N-9)\,;\,(6N-3)\,.$$ 
In this case we find $\alpha=57$, $\beta=-21$, $\epsilon= -\frac{29}{18}$.

If we instead assign a minimal nilpotent vev to the $E_6$ moment map we bounce back to a $SU(9)$ realization with $N\rightarrow N-1$.
The series of higgsings we have just described reproduces the 6d Hasse diagram as expected. The endpoint of the chain of higgsings is the rank-$1$ E$_6$ Minahan-Nemeschansky theory.

\be\label{E8k3hasse} 
\begin{tikzpicture} 
\node[] at (-12.8,0) {$\cdots$};
\node[] (H) at (-11.5,0) {\scriptsize $E_6\times SU(3)$}; 
\node[] at (-11.5,0.3) {\scriptsize (N+1)};
\node[] (A) at (0,0) {\scriptsize $SU(9)$}; 
\node[] at (0,0.3) {\scriptsize (N-1)};

\node[] (C) at (0,-1) {\scriptsize $E_7\times U(1)$}; 
\node[] at (0,-1.3) {\scriptsize (N)}; 
\node[] (D) at (-3,0) {\scriptsize $E_6\times SU(3)$}; 
\node[] at (-3,0.3) {\scriptsize (N)};
\node[] at (-6,0.3) {\scriptsize (N)};
\node[] (E) at (-6,0) {\scriptsize $SO(14)\times U(1)$}; 
\node[] (F) at (-9,0) {\scriptsize $SU(9)$}; 
\node[] at (-9,0.3) {\scriptsize (N)};

\node[] (I) at (-9,-1) {\scriptsize $E_7\times U(1)$}; 
\node[] at (-9,-1.3) {\scriptsize (N+1)};
\node[] (L) at (-6,-1) {\scriptsize $E_8$}; 
\node[] at (-6,-1.3) {\scriptsize (N+1)};

%\node[] at (1.3,0.2) {\scriptsize $\mathfrak{a}_8$};
%\node[] at (1.7,-0.6) {\scriptsize $\mathfrak{e}_7$};
\node[] at (-1.5,0.2) {\scriptsize $\mathfrak{a}_2$};
\node[] at (-1.5,-0.7) {\scriptsize $\mathfrak{e}_6$};
\node[] at (-10.2,0.2) {\scriptsize $\mathfrak{a}_2$};
\node[] at (-7.8,0.2) {\scriptsize $\mathfrak{a}_8$};
\node[] at (-7.5,-1.2) {\scriptsize $A_2$};
%\node[] at (1.5,-1.2) {\scriptsize $A_2$};
\node[] at (-3,-1.2) {\scriptsize $\mathfrak{e}_8$};
\node[] at (-8.4,-0.4) {\scriptsize $\mathfrak{su}_3$};
%\node[] at (0.6,-0.4) {\scriptsize $\mathfrak{su}_3$};
\node[] at (-7.3,-0.6) {\scriptsize $\mathfrak{e}_7$};
\node[] at (-4.5,0.2) {\scriptsize $\mathfrak{d}_7$};
\node[] at (-10.2,-0.7) {\scriptsize $\mathfrak{e}_6$};
\node[] at (1.5,0) {$\cdots$};
\draw[->] (D) -- (A); 
\draw[->] (D) -- (C); 
\draw[->] (F) -- (E); 
\draw[->] (E) -- (D); 
\draw[->] (H) -- (F);
\draw[->] (H) -- (I);  
\draw[->] (I) -- (E); 
\draw[->] (I) -- (L); 
\draw[->] (L) -- (C); 
\draw[->] (H) -- (L);
\end{tikzpicture}
\ee
\begin{eqnarray*}
\begin{tikzpicture} 
\node[] at (-4.5,0) {$\cdots$};
\node[] (A) at (0,0) {\scriptsize $SU(9)$}; 
\node[] at (0,0.3) {\scriptsize (2)};
\node[] (B) at (3,0) {\scriptsize $SO(14)\times U(1)$}; 
\node[] at (3,0.3) {\scriptsize (1)};
\node[] (C) at (0,-1) {\scriptsize $E_7\times U(1)$}; 
\node[] at (0,-1.3) {\scriptsize (2)}; 
\node[] (D) at (-3,0) {\scriptsize $E_6\times SU(3)$}; 
\node[] at (-3,0.3) {\scriptsize (2)};
\node[] (L) at (-4.5,-1) {\scriptsize $E_8$}; 
\node[] at (-4.5,-1.3) {\scriptsize (3)};

\node[] at (1.5,0.2) {\scriptsize $\mathfrak{a}_8$};
\node[] at (-2.5,-1.2) {\scriptsize $\mathfrak{e}_8$};
\node[] at (1.5,-0.7) {\scriptsize $\mathfrak{e}_7$};
\node[] at (-1.5,0.2) {\scriptsize $\mathfrak{a}_2$};
\node[] at (-1.5,-0.7) {\scriptsize $\mathfrak{e}_6$};
\node[] (G) at (6,0) {\scriptsize $E_6$}; 
\node[] at (6,0.3) {\scriptsize (1)};
\node[] at (5,0.2) {\scriptsize $\mathfrak{d}_7$};
\draw[->] (A) -- (B); 
\draw[->] (C) -- (B); 
\draw[->] (D) -- (A); 
\draw[->] (D) -- (C); 
\draw[->] (B) -- (G);
\draw[->] (L) -- (C);
\end{tikzpicture}
\end{eqnarray*}

The second canonical theory from a 7-brane of type E$_8$ on $\mathbb{C}^2/\mathbb{Z}_3$ arises via mass deformation of the $SO(14)\times U(1)$ theory, whose magnetic quiver is given in \eqref{Table:SU(3)}. We find 
\be\label{SO14def}
\begin{tikzpicture}
\filldraw[fill= red] (-3,0) circle [radius=0.1] node[below] {\scriptsize 1};
\filldraw[fill= red] (-2,0) circle [radius=0.1] node[below] {\scriptsize 2};
\filldraw[fill= white] (-1,0) circle [radius=0.1] node[below] {\scriptsize 3};
\filldraw[fill= white] (0,0) circle [radius=0.1] node[below] {\scriptsize N+2};
\filldraw[fill= white] (1,0) circle [radius=0.1] node[below] {\scriptsize 2N+2};
\filldraw[fill= white] (2,0) circle [radius=0.1] node[below] {\scriptsize 3N+2};
\filldraw[fill= white] (3,0) circle [radius=0.1] node[below] {\scriptsize 4N+2};
\filldraw[fill= white] (4,0) circle [radius=0.1] node[below] {\scriptsize 5N+2};
\filldraw[fill= white] (5,0) circle [radius=0.1] node[below] {\scriptsize 6N+2};
\filldraw[fill= white] (6,0) circle [radius=0.1] node[below] {\scriptsize 4N+1};
\filldraw[fill= white] (5,1) circle [radius=0.1] node[above] {\scriptsize 3N+1};
\filldraw[fill= white] (7,0) circle [radius=0.1] node[below] {\scriptsize 2N};

\draw [thick] (-0.1, 0) -- (-0.9,0) ;
\draw [thick] (-1.1, 0) -- (-1.9,0) ;
\draw [thick] (-2.1, 0) -- (-2.9,0) ;
\draw [thick] (0.1, 0) -- (0.9,0) ;
\draw [thick] (1.1, 0) -- (1.9,0) ;
\draw [thick] (2.1, 0) -- (2.9,0) ;
\draw [thick] (3.1, 0) -- (3.9,0) ;
\draw [thick] (4.1, 0) -- (4.9,0) ;
\draw [thick] (5.1, 0) -- (5.9,0) ;
\draw [thick] (5, 0.1) -- (5,0.9) ;
\draw [thick] (6.1, 0) -- (6.9,0) ;

\draw [->] (2,-0.6) -- (2,-2.6) ;

\filldraw[fill= red] (-1,-3) circle [radius=0.1] node[below] {\scriptsize 1};
\filldraw[fill= white] (0,-3) circle [radius=0.1] node[below] {\scriptsize N};
\filldraw[fill= white] (1,-3) circle [radius=0.1] node[below] {\scriptsize 2N};
\filldraw[fill= white] (2,-3) circle [radius=0.1] node[below] {\scriptsize 3N};
\filldraw[fill= white] (3,-3) circle [radius=0.1] node[below] {\scriptsize 4N};
\filldraw[fill= white] (4,-3) circle [radius=0.1] node[below] {\scriptsize 5N};
\filldraw[fill= white] (5,-3) circle [radius=0.1] node[below] {\scriptsize 6N};
\filldraw[fill= white] (6,-3) circle [radius=0.1] node[below] {\scriptsize 4N};
\filldraw[fill= white] (5,-2) circle [radius=0.1] node[above] {\scriptsize 3N};
\filldraw[fill= white] (7,-3) circle [radius=0.1] node[below] {\scriptsize 2N};
\filldraw[fill= red] (8,-3) circle [radius=0.1] node[below] {\scriptsize 1};

\draw [thick] (-0.1, -3) -- (-0.9,-3) ;
\draw [thick] (7.1, -3) -- (7.9,-3) ;
\draw [thick] (0.1, -3) -- (0.9,-3) ;
\draw [thick] (1.1, -3) -- (1.9,-3) ;
\draw [thick] (2.1, -3) -- (2.9,-3) ;
\draw [thick] (3.1, -3) -- (3.9,-3) ;
\draw [thick] (4.1, -3) -- (4.9,-3) ;
\draw [thick] (5.1, -3) -- (5.9,-3) ;
\draw [thick] (5, -2.1) -- (5,-2.9) ;
\draw [thick] (6.1, -3) -- (6.9,-3) ;

\draw [->] (2,-3.6) -- (2,-5.6) ;

\filldraw[fill= white] (0,-6) circle [radius=0.1] node[below] {\scriptsize N-1};
\filldraw[fill= white] (1,-6) circle [radius=0.1] node[below] {\scriptsize 2N-1};
\filldraw[fill= white] (2,-6) circle [radius=0.1] node[below] {\scriptsize 3N-1};
\filldraw[fill= white] (3,-6) circle [radius=0.1] node[below] {\scriptsize 4N-1};
\filldraw[fill= white] (4,-6) circle [radius=0.1] node[below] {\scriptsize 5N-1};
\filldraw[fill= white] (5,-6) circle [radius=0.1] node[below] {\scriptsize 6N-1};
\filldraw[fill= white] (6,-6) circle [radius=0.1] node[below] {\scriptsize 4N-1};
\filldraw[fill= white] (5,-5) circle [radius=0.1] node[above] {\scriptsize 3N};
\filldraw[fill= white] (7,-6) circle [radius=0.1] node[below] {\scriptsize 2N-1};
\filldraw[fill= white] (4,-5) circle [radius=0.1] node[above] {\scriptsize 1};

\draw [thick] (4.1, -5) -- (4.9,-5) ;
\draw [thick] (0.1, -6) -- (0.9,-6) ;
\draw [thick] (1.1, -6) -- (1.9,-6) ;
\draw [thick] (2.1, -6) -- (2.9,-6) ;
\draw [thick] (3.1, -6) -- (3.9,-6) ;
\draw [thick] (4.1, -6) -- (4.9,-6) ;
\draw [thick] (5.1, -6) -- (5.9,-6) ;
\draw [thick] (5, -5.1) -- (5,-5.9) ;
\draw [thick] (6.1, -6) -- (6.9,-6) ;
\end{tikzpicture}
\ee 
where the \emph{canonical} theory appears as an intermediate step after the first mass deformation. With the second deformation we recover \eqref{E8k3Na8}, which is the theory whose Higgs branch is the $\mathfrak{e}_8$ instanton moduli space. The spectrum of CB operators of this second canonical theory is
$$(4,6,7)\,;\,(10,12,13)\,;\, \dots\,;\,(6N-2, 6N)\,.$$

\subsection{Non-conformal theories}\label{Sec:NonConf}

As we have seen, the theories we get using the procedure described above are conformal in 4d whenever the corresponding magnetic quiver is star-shaped. This however is not always the case and in this section we would like to understand the physics of 4d theories whose magnetic quiver does not have this property. We focus for simplicity on the (families of) theories whose magnetic quiver is obtained by adding a single $U(1)$ node to a star-shaped affine  Dynkin diagram.

In order to see how such quivers arise in our construction, let us start from the theories described in Section \ref{Sec:E8} and turn on the mass term which corresponds to deforming the 7-brane of type E$_8$ to that of type E$_7$ (always wrapped on $\mathbb{C}^2/\mathbb{Z}_2$). We work again at the level of the corresponding magnetic quivers and their FI deformations. We can easily see that there is a mass deformation connecting \eqref{D8NMQ} and \eqref{SU8NMQ}:
\be\label{E8toE7}
\begin{tikzpicture}
\filldraw[fill= white] (0,0) circle [radius=0.1] node[below] {\scriptsize N};
\filldraw[fill= white] (1,0) circle [radius=0.1] node[below] {\scriptsize 2N};
\filldraw[fill= red] (2,0) circle [radius=0.1] node[below] {\scriptsize 3N};
\filldraw[fill= white] (3,0) circle [radius=0.1] node[below] {\scriptsize 4N};
\filldraw[fill= white] (4,0) circle [radius=0.1] node[below] {\scriptsize 5N};
\filldraw[fill= white] (5,0) circle [radius=0.1] node[below] {\scriptsize 6N};
\filldraw[fill= white] (6,0) circle [radius=0.1] node[below] {\scriptsize 4N};
\filldraw[fill= red] (5,1) circle [radius=0.1] node[above] {\scriptsize 3N};
\filldraw[fill= white] (7,0) circle [radius=0.1] node[below] {\scriptsize 2N};
\filldraw[fill= white] (8,0) circle [radius=0.1] node[below] {\scriptsize 1};
\draw [thick] (0.1, 0) -- (0.9,0) ;
\draw [thick] (1.1, 0) -- (1.9,0) ;
\draw [thick] (2.1, 0) -- (2.9,0) ;
\draw [thick] (3.1, 0) -- (3.9,0) ;
\draw [thick] (4.1, 0) -- (4.9,0) ;
\draw [thick] (5.1, 0) -- (5.9,0) ;
\draw [thick] (5, 0.1) -- (5,0.9) ;
\draw [thick] (6.1, 0) -- (6.9,0) ;
\draw [thick] (7.1, 0) -- (7.9,0) ; 

\draw [->] (2,-0.7) -- (2,-1.7) ;

\filldraw[fill= white] (0,-2) circle [radius=0.1] node[below] {\scriptsize N};
\filldraw[fill= white] (1,-2) circle [radius=0.1] node[below] {\scriptsize 2N};
\filldraw[fill= white] (2,-2) circle [radius=0.1] node[below] {\scriptsize 3N};
\filldraw[fill= white] (3,-2) circle [radius=0.1] node[below] {\scriptsize 4N};
\filldraw[fill= white] (4,-2) circle [radius=0.1] node[below] {\scriptsize 3N};
\filldraw[fill= white] (5,-2) circle [radius=0.1] node[below] {\scriptsize 2N};
\filldraw[fill= white] (6,-2) circle [radius=0.1] node[below] {\scriptsize N};
\filldraw[fill= white] (3,-1) circle [radius=0.1] node[above] {\scriptsize 2N};
\filldraw[fill= white] (4,-1) circle [radius=0.1] node[above] {\scriptsize 1}; 
\draw [thick] (0.1, -2) -- (0.9,-2) ;
\draw [thick] (1.1, -2) -- (1.9,-2) ;
\draw [thick] (2.1, -2) -- (2.9,-2) ;
\draw [thick] (3.1, -2) -- (3.9,-2) ;
\draw [thick] (4.1, -2) -- (4.9,-2) ;
\draw [thick] (5.1, -2) -- (5.9,-2) ;
\draw [thick] (3, -1.1) -- (3,-1.9) ;
\draw [thick] (3.1, -1) -- (3.9,-1) ;
\end{tikzpicture} \ee
We therefore conclude that upon mass deformation the $SO(16)$-preserving holonomy of $E_8$ is mapped to the $SU(8)$-preserving holonomy for $E_7$. In this case we end up with a star-shaped quiver which corresponds to a SCFT in 4d. The conclusion is different if we instead consider the other two holonomy choices, as we will now see. The analysis is slightly more involved since we need to activate a FI deformation at the abelian node as well. If we consider the $E_7\times SU(2)$-preserving holonomy for $E_8$ we find the following FI deformation 
\be\label{E8toE7II}
\begin{tikzpicture}
\filldraw[fill= white] (0,0) circle [radius=0.1] node[below] {\scriptsize N-1};
\filldraw[fill= white] (1,0) circle [radius=0.1] node[below] {\scriptsize 2N-2};
\filldraw[fill= red] (2,0) circle [radius=0.1] node[below] {\scriptsize 3N-2};
\filldraw[fill= white] (3,0) circle [radius=0.1] node[below] {\scriptsize 4N-2};
\filldraw[fill= white] (4,0) circle [radius=0.1] node[below] {\scriptsize 5N-2};
\filldraw[fill= white] (5,0) circle [radius=0.1] node[below] {\scriptsize 6N-2};
\filldraw[fill= white] (6,0) circle [radius=0.1] node[below] {\scriptsize 4N-1};
\filldraw[fill= red] (5,1) circle [radius=0.1] node[above] {\scriptsize 3N-1};
\filldraw[fill= white] (7,0) circle [radius=0.1] node[below] {\scriptsize 2N};
\filldraw[fill= red] (8,0) circle [radius=0.1] node[below] {\scriptsize 1};
\draw [thick] (0.1, 0) -- (0.9,0) ;
\draw [thick] (1.1, 0) -- (1.9,0) ;
\draw [thick] (2.1, 0) -- (2.9,0) ;
\draw [thick] (3.1, 0) -- (3.9,0) ;
\draw [thick] (4.1, 0) -- (4.9,0) ;
\draw [thick] (5.1, 0) -- (5.9,0) ;
\draw [thick] (5, 0.1) -- (5,0.9) ;
\draw [thick] (6.1, 0) -- (6.9,0) ;
\draw [thick] (7.1, 0) -- (7.9,0) ; 

\draw [->] (2,-0.7) -- (2,-1.7) ;

\filldraw[fill= white] (0,-2) circle [radius=0.1] node[below] {\scriptsize N-1};
\filldraw[fill= white] (1,-2) circle [radius=0.1] node[below] {\scriptsize 2N-2};
\filldraw[fill= white] (2,-2) circle [radius=0.1] node[below] {\scriptsize 3N-2};
\filldraw[fill= white] (3,-2) circle [radius=0.1] node[below] {\scriptsize 4N-2};
\filldraw[fill= white] (4,-2) circle [radius=0.1] node[below] {\scriptsize 3N-1};
\filldraw[fill= white] (5,-2) circle [radius=0.1] node[below] {\scriptsize 2N};
\filldraw[fill= white] (6,-2) circle [radius=0.1] node[below] {\scriptsize N};
\filldraw[fill= white] (3,-1) circle [radius=0.1] node[above] {\scriptsize 2N-1};
\filldraw[fill= white] (5,-1) circle [radius=0.1] node[above] {\scriptsize 1}; 
\draw [thick] (0.1, -2) -- (0.9,-2) ;
\draw [thick] (1.1, -2) -- (1.9,-2) ;
\draw [thick] (2.1, -2) -- (2.9,-2) ;
\draw [thick] (3.1, -2) -- (3.9,-2) ;
\draw [thick] (4.1, -2) -- (4.9,-2) ;
\draw [thick] (5.1, -2) -- (5.9,-2) ;
\draw [thick] (3, -1.1) -- (3,-1.9) ;
\draw [thick] (5, -1.1) -- (5,-1.9) ;
\end{tikzpicture} \ee
which leads to the $SO(12)\times SU(2)$-preserving holonomy for $E_7$. Finally, the trivial holonomy for $E_8$ is mapped to the trivial holonomy for $E_7$: 
\be\label{E8toE7III}
\begin{tikzpicture}
\filldraw[fill= white] (0,0) circle [radius=0.1] node[below] {\scriptsize N-2};
\filldraw[fill= white] (1,0) circle [radius=0.1] node[below] {\scriptsize 2N-2};
\filldraw[fill= red] (2,0) circle [radius=0.1] node[below] {\scriptsize 3N-2};
\filldraw[fill= white] (3,0) circle [radius=0.1] node[below] {\scriptsize 4N-2};
\filldraw[fill= white] (4,0) circle [radius=0.1] node[below] {\scriptsize 5N-2};
\filldraw[fill= white] (5,0) circle [radius=0.1] node[below] {\scriptsize 6N-2};
\filldraw[fill= white] (6,0) circle [radius=0.1] node[below] {\scriptsize 4N-1};
\filldraw[fill= red] (5,1) circle [radius=0.1] node[above] {\scriptsize 3N-1};
\filldraw[fill= white] (7,0) circle [radius=0.1] node[below] {\scriptsize 2N};
\filldraw[fill= red] (8,0) circle [radius=0.1] node[below] {\scriptsize 1};
\draw [thick] (0.1, 0) -- (0.9,0) ;
\draw [thick] (1.1, 0) -- (1.9,0) ;
\draw [thick] (2.1, 0) -- (2.9,0) ;
\draw [thick] (3.1, 0) -- (3.9,0) ;
\draw [thick] (4.1, 0) -- (4.9,0) ;
\draw [thick] (5.1, 0) -- (5.9,0) ;
\draw [thick] (5, 0.1) -- (5,0.9) ;
\draw [thick] (6.1, 0) -- (6.9,0) ;
\draw [thick] (7.1, 0) -- (7.9,0) ; 

\draw [->] (2,-0.7) -- (2,-1.7) ;

\filldraw[fill= white] (0,-2) circle [radius=0.1] node[below] {\scriptsize N-2};
\filldraw[fill= white] (1,-2) circle [radius=0.1] node[below] {\scriptsize 2N-2};
\filldraw[fill= white] (2,-2) circle [radius=0.1] node[below] {\scriptsize 3N-2};
\filldraw[fill= white] (3,-2) circle [radius=0.1] node[below] {\scriptsize 4N-2};
\filldraw[fill= white] (4,-2) circle [radius=0.1] node[below] {\scriptsize 3N-1};
\filldraw[fill= white] (5,-2) circle [radius=0.1] node[below] {\scriptsize 2N};
\filldraw[fill= white] (6,-2) circle [radius=0.1] node[below] {\scriptsize N};
\filldraw[fill= white] (3,-1) circle [radius=0.1] node[above] {\scriptsize 2N-1};
\filldraw[fill= white] (5,-1) circle [radius=0.1] node[above] {\scriptsize 1}; 
\draw [thick] (0.1, -2) -- (0.9,-2) ;
\draw [thick] (1.1, -2) -- (1.9,-2) ;
\draw [thick] (2.1, -2) -- (2.9,-2) ;
\draw [thick] (3.1, -2) -- (3.9,-2) ;
\draw [thick] (4.1, -2) -- (4.9,-2) ;
\draw [thick] (5.1, -2) -- (5.9,-2) ;
\draw [thick] (3, -1.1) -- (3,-1.9) ;
\draw [thick] (5, -1.1) -- (5,-1.9) ;
\end{tikzpicture} \ee
We have exhibited in \eqref{E8toE7}-\eqref{E8toE7III} a correspondence between $E_8$ and $E_7$ $\mathbb{Z}_2$ holonomies, the only outlier being the $E_6\times U(1)$-preserving holonomy (described by the quiver \eqref{E6NMQ}) for $E_7$ which has no $E_8$ origin. Indeed we see that the $E_7$ quivers in \eqref{E8toE7II} and \eqref{E8toE7III} are not star-shaped and therefore we do not expect them to describe conformal models. We can also notice that, if we proceed from \eqref{E8toE7} and further deform to the E$_6$ 7-brane, we land on another non star-shaped quiver 
\be\label{E7toE6}
\begin{tikzpicture}
\filldraw[fill= white] (0,0) circle [radius=0.1] node[below] {\scriptsize N};
\filldraw[fill= white] (1,0) circle [radius=0.1] node[below] {\scriptsize 2N};
\filldraw[fill= red] (2,0) circle [radius=0.1] node[below] {\scriptsize 3N};
\filldraw[fill= white] (3,0) circle [radius=0.1] node[below] {\scriptsize 4N};
\filldraw[fill= red] (4,0) circle [radius=0.1] node[below] {\scriptsize 3N};
\filldraw[fill= white] (5,0) circle [radius=0.1] node[below] {\scriptsize 2N};
\filldraw[fill= white] (6,0) circle [radius=0.1] node[below] {\scriptsize N};
\filldraw[fill= white] (3,1) circle [radius=0.1] node[above] {\scriptsize 2N};
\filldraw[fill= white] (4,1) circle [radius=0.1] node[above] {\scriptsize 1}; 
\draw [thick] (0.1, 0) -- (0.9,0) ;
\draw [thick] (1.1, 0) -- (1.9,0) ;
\draw [thick] (2.1, 0) -- (2.9,0) ;
\draw [thick] (3.1, 0) -- (3.9,0) ;
\draw [thick] (4.1, 0) -- (4.9,0) ;
\draw [thick] (5.1, 0) -- (5.9,0) ;
\draw [thick] (3, 0.1) -- (3,0.9) ;
\draw [thick] (3.1, 1) -- (3.9,1) ;

\draw [->] (2,-0.5) -- (2,-1.5) ;

\filldraw[fill= white] (0,-2) circle [radius=0.1] node[above] {\scriptsize N};
\filldraw[fill= white] (1,-2) circle [radius=0.1] node[above] {\scriptsize 2N};
\filldraw[fill= white] (2,-2) circle [radius=0.1] node[above] {\scriptsize 3N};
\filldraw[fill= white] (3,-2) circle [radius=0.1] node[above] {\scriptsize 2N};
\filldraw[fill= white] (4,-2) circle [radius=0.1] node[above] {\scriptsize N};
\filldraw[fill= white] (2,-3) circle [radius=0.1] node[left] {\scriptsize 2N};
\filldraw[fill= white] (2,-4) circle [radius=0.1] node[left] {\scriptsize N};
\filldraw[fill= white] (3,-3) circle [radius=0.1] node[below] {\scriptsize 1}; 
\draw [thick] (0.1, -2) -- (0.9,-2) ;
\draw [thick] (1.1, -2) -- (1.9,-2) ;
\draw [thick] (2.1, -2) -- (2.9,-2) ;
\draw [thick] (3.1, -2) -- (3.9,-2) ;
\draw [thick] (2, -2.1) -- (2,-2.9) ;
\draw [thick] (2, -3.1) -- (2,-3.9) ;
\draw [thick] (3, -2.1) -- (3,-2.9) ;
\end{tikzpicture} \ee
associated with the $SU(6)\times SU(2)$-preserving holonomy for $E_6$. 

The above discussion clearly shows that magnetic quivers which are not star-shaped are rather common in our construction and it is therefore important to understand the underlying four-dimensional theories. As we will now see, these turn out to describe infrared-free vector multiplets coupled to (generically) non lagrangian matter.

\subsubsection*{E$_6$-type 7-brane on $\mathbb{C}^2/\mathbb{Z}_2$}

Let us start from the $E_6$ theory described by the second quiver in \eqref{E7toE6}:
\be\label{nostarE6}
\begin{tikzpicture}
\filldraw[fill= white] (0,0) circle [radius=0.1] node[below] {\scriptsize N};
\filldraw[fill= white] (1,0) circle [radius=0.1] node[below] {\scriptsize 2N};
\filldraw[fill= white] (2,0) circle [radius=0.1] node[below] {\scriptsize 3N};
\filldraw[fill= white] (2,2) circle [radius=0.1] node[right] {\scriptsize N};
\filldraw[fill= white] (3,0) circle [radius=0.1] node[below] {\scriptsize 2N};
\filldraw[fill= white] (4,0) circle [radius=0.1] node[below] {\scriptsize N};
\filldraw[fill= white] (2,1) circle [radius=0.1] node[right] {\scriptsize 2N};
\filldraw[fill= white] (3,1) circle [radius=0.1] node[right] {\scriptsize 1}; 
\draw [thick] (0.1, 0) -- (0.9,0) ;
\draw [thick] (1.1, 0) -- (1.9,0) ;
\draw [thick] (2.1, 0) -- (2.9,0) ;
\draw [thick] (2, 1.1) -- (2,1.9) ;

\draw [thick] (2, 0.1) -- (2,0.9) ;
\draw [thick] (3.1, 0) -- (3.9,0) ;
\draw [thick] (3, 0.1) -- (3,0.9) ;
\end{tikzpicture}\ee
We can interpret the above as the magnetic quiver of an infrared free $SU(2)$ vector multiplet coupled to a SCFT (which we call $\mathcal{T}$) whose magnetic quiver is given by:
\be\label{nostarE6mo}
\begin{tikzpicture}
\filldraw[fill= white] (0,0) circle [radius=0.1] node[below] {\scriptsize N};
\filldraw[fill= white] (1,0) circle [radius=0.1] node[below] {\scriptsize 2N};
\filldraw[fill= white] (2,0) circle [radius=0.1] node[below] {\scriptsize 3N};
\filldraw[fill= white] (2,2) circle [radius=0.1] node[right] {\scriptsize N};
\filldraw[fill= white] (3,0) circle [radius=0.1] node[below] {\scriptsize 2N};
\filldraw[fill= white] (4,0) circle [radius=0.1] node[below] {\scriptsize N+1};
\filldraw[fill= white] (2,1) circle [radius=0.1] node[right] {\scriptsize 2N};
\filldraw[fill= white] (5,0) circle [radius=0.1] node[below] {\scriptsize 2};
\filldraw[fill= white] (6,0) circle [radius=0.1] node[below] {\scriptsize 1};

\draw [thick] (0.1, 0) -- (0.9,0) ;
\draw [thick] (1.1, 0) -- (1.9,0) ;
\draw [thick] (2.1, 0) -- (2.9,0) ;
\draw [thick] (2, 0.1) -- (2,0.9) ;
\draw [thick] (3.1, 0) -- (3.9,0) ;
\draw [thick] (2, 1.1) -- (2,1.9) ;
\draw [thick] (4.1, 0) -- (4.9,0) ;
\draw [thick] (5.1, 0) -- (5.9,0) ;
\end{tikzpicture}\ee
The global symmetry of the theory $\mathcal{T}$ is $SU(6)\times SU(2)^2\times U(1)$ and its Coulomb branch has dimension $2N-1$. 

Our claim can be proven with the following procedure: We first couple the theory $\mathcal{T}$ to a $T_2$ (i.e. two doublets of $SU(2)$) via a $SU(2)$ gauging. At the level of magnetic quivers this is implemented by fusing together the $T(SU(2))$ tails of the two magnetic quivers (of $\mathcal{T}$ and of $T_2$)
 \be
\begin{tikzpicture}
\filldraw[fill= white] (0,0) circle [radius=0.1] node[below] {\scriptsize N};
\filldraw[fill= white] (1,0) circle [radius=0.1] node[below] {\scriptsize 2N};
\filldraw[fill= white] (2,2) circle [radius=0.1] node[right] {\scriptsize N};
\filldraw[fill= white] (2,1) circle [radius=0.1] node[right] {\scriptsize 2N};
\filldraw[fill= white] (2,0) circle [radius=0.1] node[below] {\scriptsize 3N};
\filldraw[fill= white] (3,0) circle [radius=0.1] node[below] {\scriptsize 2N};
\filldraw[fill= white] (4,0) circle [radius=0.1] node[below] {\scriptsize N+1};
\filldraw[fill= white] (5,0) circle [radius=0.1] node[below] {\scriptsize 2};
\filldraw[fill= white] (6,0) circle [radius=0.1] node[below] {\scriptsize 1};
\filldraw[fill= white] (10,0) circle [radius=0.1] node[below] {\scriptsize 1};
\filldraw[fill= white] (11,0) circle [radius=0.1] node[below] {\scriptsize 2};
\filldraw[fill= white] (11,1) circle [radius=0.1] node[above] {\scriptsize 1};
\filldraw[fill= white] (12,0) circle [radius=0.1] node[below] {\scriptsize 1};
\node[] at (8,0) {SU(2)};

\draw [->] (7.5,0) -- (6.5,0) ;
\draw [->] (8.5,0) -- (9.5,0) ;

\draw [thick] (2,0.1) -- (2,0.9) ;
\draw [thick] (2,1.1) -- (2,1.9) ;
\draw [thick] (0.1, 0) -- (0.9,0) ;
\draw [thick] (1.1, 0) -- (1.9,0) ;
\draw [thick] (2.1, 0) -- (2.9,0) ;
\draw [thick] (3.1, 0) -- (3.9,0) ;
\draw [thick] (4.1, 0) -- (4.9,0) ;
\draw [thick] (5.1, 0) -- (5.9,0) ;
\draw [thick] (11, 0.1) -- (11,0.9) ;
\draw [thick] (10.1, 0) -- (10.9,0) ;
\draw [thick] (11.1, 0) -- (11.9,0) ;
\end{tikzpicture}\ee
or equivalently by adding to \eqref{nostarE6mo} an abelian node attached to the $U(2)$ node, leading to the quiver 
\be\label{nostarE6mo2}
\begin{tikzpicture}
\filldraw[fill= white] (0,0) circle [radius=0.1] node[below] {\scriptsize N};
\filldraw[fill= white] (1,0) circle [radius=0.1] node[below] {\scriptsize 2N};
\filldraw[fill= white] (2,0) circle [radius=0.1] node[below] {\scriptsize 3N};
\filldraw[fill= white] (2,2) circle [radius=0.1] node[right] {\scriptsize N};
\filldraw[fill= white] (3,0) circle [radius=0.1] node[below] {\scriptsize 2N};
\filldraw[fill= white] (4,0) circle [radius=0.1] node[below] {\scriptsize N+1};
\filldraw[fill= white] (2,1) circle [radius=0.1] node[right] {\scriptsize 2N};
\filldraw[fill= white] (5,0) circle [radius=0.1] node[below] {\scriptsize 2};
\filldraw[fill= white] (6,0) circle [radius=0.1] node[below] {\scriptsize 1};
\filldraw[fill= white] (5,1) circle [radius=0.1] node[right] {\scriptsize 1};

\draw [thick] (0.1, 0) -- (0.9,0) ;
\draw [thick] (1.1, 0) -- (1.9,0) ;
\draw [thick] (2.1, 0) -- (2.9,0) ;
\draw [thick] (2, 0.1) -- (2,0.9) ;
\draw [thick] (3.1, 0) -- (3.9,0) ;
\draw [thick] (2, 1.1) -- (2,1.9) ;
\draw [thick] (4.1, 0) -- (4.9,0) ;
\draw [thick] (5.1, 0) -- (5.9,0) ;
\draw [thick] (5, 0.1) -- (5,0.9) ;
\end{tikzpicture}\ee
At this stage we have the theory $$\mathcal{T}-SU(2)-\boxed{2}$$ and all is left to do is to make the two $SU(2)$ doublets massive. This is implemented at the quiver level by turning on FI terms at the two abelian nodes, both in \eqref{nostarE6mo2} and also in the resulting quiver. This leads precisely to \eqref{nostarE6}: 
\be\label{nostarE6mo3}
\begin{tikzpicture}
\filldraw[fill= white] (0,0) circle [radius=0.1] node[below] {\scriptsize N};
\filldraw[fill= white] (1,0) circle [radius=0.1] node[below] {\scriptsize 2N};
\filldraw[fill= white] (2,0) circle [radius=0.1] node[below] {\scriptsize 3N};
\filldraw[fill= white] (2,2) circle [radius=0.1] node[right] {\scriptsize N};
\filldraw[fill= white] (3,0) circle [radius=0.1] node[below] {\scriptsize 2N};
\filldraw[fill= white] (4,0) circle [radius=0.1] node[below] {\scriptsize N+1};
\filldraw[fill= white] (2,1) circle [radius=0.1] node[right] {\scriptsize 2N};
\filldraw[fill= white] (5,0) circle [radius=0.1] node[below] {\scriptsize 2};
\filldraw[fill= red] (6,0) circle [radius=0.1] node[below] {\scriptsize 1};
\filldraw[fill= red] (5,1) circle [radius=0.1] node[right] {\scriptsize 1};

\draw [thick] (0.1, 0) -- (0.9,0) ;
\draw [thick] (1.1, 0) -- (1.9,0) ;
\draw [thick] (2.1, 0) -- (2.9,0) ;
\draw [thick] (2, 0.1) -- (2,0.9) ;
\draw [thick] (3.1, 0) -- (3.9,0) ;
\draw [thick] (2, 1.1) -- (2,1.9) ;
\draw [thick] (4.1, 0) -- (4.9,0) ;
\draw [thick] (5.1, 0) -- (5.9,0) ;
\draw [thick] (5, 0.1) -- (5,0.9) ;

\draw [->] (3,-0.5) -- (3,-1.5);

\filldraw[fill= white] (0,-3) circle [radius=0.1] node[below] {\scriptsize N};
\filldraw[fill= white] (1,-3) circle [radius=0.1] node[below] {\scriptsize 2N};
\filldraw[fill= white] (2,-3) circle [radius=0.1] node[below] {\scriptsize 3N};
\filldraw[fill= white] (2,-1) circle [radius=0.1] node[right] {\scriptsize N};
\filldraw[fill= white] (3,-3) circle [radius=0.1] node[below] {\scriptsize 2N};
\filldraw[fill= white] (4,-3) circle [radius=0.1] node[below] {\scriptsize N+1};
\filldraw[fill= white] (2,-2) circle [radius=0.1] node[right] {\scriptsize 2N};
\filldraw[fill= red] (5,-3) circle [radius=0.1] node[below] {\scriptsize 1};
\filldraw[fill= red] (4,-2) circle [radius=0.1] node[right] {\scriptsize 1};

\draw [thick] (0.1, -3) -- (0.9,-3) ;
\draw [thick] (1.1, -3) -- (1.9,-3) ;
\draw [thick] (2.1, -3) -- (2.9,-3) ;
\draw [thick] (2, -2.1) -- (2,-2.9) ;
\draw [thick] (3.1, -3) -- (3.9,-3) ;
\draw [thick] (2, -1.1) -- (2,-1.9) ;
\draw [thick] (4.1, -3) -- (4.9,-3) ;
\draw [thick] (4, -2.1) -- (4,-2.9) ;

\draw [->] (3,-3.5) -- (3,-4.5); 

\filldraw[fill= white] (0,-6) circle [radius=0.1] node[below] {\scriptsize N};
\filldraw[fill= white] (1,-6) circle [radius=0.1] node[below] {\scriptsize 2N};
\filldraw[fill= white] (2,-6) circle [radius=0.1] node[below] {\scriptsize 3N};
\filldraw[fill= white] (2,-4) circle [radius=0.1] node[right] {\scriptsize N};
\filldraw[fill= white] (3,-6) circle [radius=0.1] node[below] {\scriptsize 2N};
\filldraw[fill= white] (4,-6) circle [radius=0.1] node[below] {\scriptsize N};
\filldraw[fill= white] (2,-5) circle [radius=0.1] node[right] {\scriptsize 2N};
\filldraw[fill= white] (3,-5) circle [radius=0.1] node[right] {\scriptsize 1}; 
\draw [thick] (0.1, -6) -- (0.9,-6) ;
\draw [thick] (1.1, -6) -- (1.9,-6) ;
\draw [thick] (2.1, -6) -- (2.9,-6) ;
\draw [thick] (2, -4.1) -- (2,-4.9) ;
\draw [thick] (2, -5.1) -- (2,-5.9) ;
\draw [thick] (3.1, -6) -- (3.9,-6) ;
\draw [thick] (3, -5.1) -- (3,-5.9) ;
\end{tikzpicture}\ee
Including the CB operator of dimension 2 coming from the $SU(2)$ vector multiplet we find that the CB spectrum of the theory \eqref{nostarE6} is 
\be \{2,3,5,6,\dots, 3N-1,3N\}.\ee

\subsubsection*{E$_7$-type 7-brane on $\mathbb{C}^2/\mathbb{Z}_2$}

The same argument can be applied to the $E_7$ quivers appearing in \eqref{E8toE7II} and  \eqref{E8toE7III}. Let us focus on \eqref{E8toE7II} for simplicity:
 \be\label{E7k2non}
\begin{tikzpicture}
\filldraw[fill= white] (0,0) circle [radius=0.1] node[below] {\scriptsize N-1};
\filldraw[fill= white] (1,0) circle [radius=0.1] node[below] {\scriptsize 2N-2};
\filldraw[fill= white] (2,0) circle [radius=0.1] node[below] {\scriptsize 3N-2};
\filldraw[fill= white] (3,0) circle [radius=0.1] node[below] {\scriptsize 4N-2};
\filldraw[fill= white] (4,0) circle [radius=0.1] node[below] {\scriptsize 3N-1};
\filldraw[fill= white] (5,0) circle [radius=0.1] node[below] {\scriptsize 2N};
\filldraw[fill= white] (6,0) circle [radius=0.1] node[below] {\scriptsize N};
\filldraw[fill= white] (3,1) circle [radius=0.1] node[above] {\scriptsize 2N-1};
\filldraw[fill= white] (5,1) circle [radius=0.1] node[above] {\scriptsize 1}; 
\draw [thick] (0.1, 0) -- (0.9,0) ;
\draw [thick] (1.1, 0) -- (1.9,0) ;
\draw [thick] (2.1, 0) -- (2.9,0) ;
\draw [thick] (3.1, 0) -- (3.9,0) ;
\draw [thick] (4.1, 0) -- (4.9,0) ;
\draw [thick] (5.1, 0) -- (5.9,0) ;
\draw [thick] (3, 0.1) -- (3,0.9) ;
\draw [thick] (5, 0.1) -- (5,0.9) ;
\end{tikzpicture}\ee
Also in this case we can interpret the  quiver as describing the $SU(2)$ gauging of a strongly coupled SCFT:
 \be
\begin{tikzpicture}
\filldraw[fill= white] (-2,0) circle [radius=0.1] node[below] {\scriptsize N-1};
\filldraw[fill= white] (-1,0) circle [radius=0.1] node[below] {\scriptsize 2N-2};
\filldraw[fill= white] (0,0) circle [radius=0.1] node[below] {\scriptsize 3N-2};
\filldraw[fill= white] (1,0) circle [radius=0.1] node[below] {\scriptsize 4N-2};
\filldraw[fill= white] (2,0) circle [radius=0.1] node[below] {\scriptsize 3N-1};
\filldraw[fill= white] (3,0) circle [radius=0.1] node[below] {\scriptsize 2N};
\filldraw[fill= white] (4,0) circle [radius=0.1] node[below] {\scriptsize N+1};
\filldraw[fill= white] (5,0) circle [radius=0.1] node[below] {\scriptsize 2};
\filldraw[fill= white] (6,0) circle [radius=0.1] node[below] {\scriptsize 1};
\filldraw[fill= white] (1,1) circle [radius=0.1] node[above] {\scriptsize 2N-1};
\filldraw[fill= white] (10,0) circle [radius=0.1] node[below] {\scriptsize 1};
\filldraw[fill= white] (11,0) circle [radius=0.1] node[below] {\scriptsize 2};
\filldraw[fill= white] (11,1) circle [radius=0.1] node[above] {\scriptsize 1};
\filldraw[fill= white] (12,0) circle [radius=0.1] node[below] {\scriptsize 1};
\node[] at (8,0) {SU(2)};

\draw [->] (7.5,0) -- (6.5,0) ;
\draw [->] (8.5,0) -- (9.5,0) ;

\draw [thick] (-0.1, 0) -- (-0.9,0) ;
\draw [thick] (-1.1, 0) -- (-1.9,0) ;
\draw [thick] (0.1, 0) -- (0.9,0) ;
\draw [thick] (1.1, 0) -- (1.9,0) ;
\draw [thick] (2.1, 0) -- (2.9,0) ;
\draw [thick] (3.1, 0) -- (3.9,0) ;
\draw [thick] (4.1, 0) -- (4.9,0) ;
\draw [thick] (5.1, 0) -- (5.9,0) ;
\draw [thick] (1, 0.1) -- (1,0.9) ;
\draw [thick] (11, 0.1) -- (11,0.9) ;
\draw [thick] (10.1, 0) -- (10.9,0) ;
\draw [thick] (11.1, 0) -- (11.9,0) ;
\end{tikzpicture}\ee
the SCFT (described by the quiver on the left) has global symmetry $SO(12)\times SU(2)^2\times U(1)$. \ref{E7k2non} is obtained again by fusing the $T(SU(2))$ tails and then by turning on FI deformations which give mass to the $SU(2)$ doublets.

\subsubsection*{$\bf\mathbb{C}^2/\mathbb{Z}_3$}
 
For an E$_7$ stack of 7-branes, we can identify another non-conformal quiver configuration, corresponding to wrapping the orbifold $\mathbb{C}^2/\mathbb{Z}_3$:
  \be\label{E7k3non}
\begin{tikzpicture}
\filldraw[fill= white] (0,0) circle [radius=0.1] node[below] {\scriptsize N};
\filldraw[fill= white] (1,0) circle [radius=0.1] node[below] {\scriptsize 2N};
\filldraw[fill= white] (2,0) circle [radius=0.1] node[below] {\scriptsize 3N};
\filldraw[fill= white] (3,0) circle [radius=0.1] node[below] {\scriptsize 4N};
\filldraw[fill= white] (4,0) circle [radius=0.1] node[below] {\scriptsize 3N};
\filldraw[fill= white] (5,0) circle [radius=0.1] node[below] {\scriptsize 2N};
\filldraw[fill= white] (6,0) circle [radius=0.1] node[below] {\scriptsize N};
\filldraw[fill= white] (3,1) circle [radius=0.1] node[above] {\scriptsize 2N};
\filldraw[fill= white] (4,1) circle [radius=0.1] node[above] {\scriptsize 1}; 
\draw [thick] (0.1, 0) -- (0.9,0) ;
\draw [thick] (1.1, 0) -- (1.9,0) ;
\draw [thick] (2.1, 0) -- (2.9,0) ;
\draw [thick] (3.1, 0) -- (3.9,0) ;
\draw [thick] (4.1, 0) -- (4.9,0) ;
\draw [thick] (5.1, 0) -- (5.9,0) ;
\draw [thick] (3, 0.1) -- (3,0.9) ;
\draw [thick] (4, 0.1) -- (4,0.9) ;
\end{tikzpicture}\ee
This arises by deforming the last quiver in \eqref{SU9def}. Again we can interpret \eqref{E7k3non} as an infrared-free gauge theory, this time with $SU(3)$ gauge group: 
 \be\label{E7Z3}
\begin{tikzpicture}
\filldraw[fill= white] (0,0) circle [radius=0.1] node[below] {\scriptsize N};
\filldraw[fill= white] (1,0) circle [radius=0.1] node[below] {\scriptsize 2N};
\filldraw[fill= white] (2,0) circle [radius=0.1] node[below] {\scriptsize 3N};
\filldraw[fill= white] (3,0) circle [radius=0.1] node[below] {\scriptsize 4N};
\filldraw[fill= white] (4,0) circle [radius=0.1] node[below] {\scriptsize 3N};
\filldraw[fill= white] (5,0) circle [radius=0.1] node[below] {\scriptsize 2N+1};
\filldraw[fill= white] (6,0) circle [radius=0.1] node[below] {\scriptsize N+2};
\filldraw[fill= white] (8,0) circle [radius=0.1] node[below] {\scriptsize 2};
\filldraw[fill= white] (7,0) circle [radius=0.1] node[below] {\scriptsize 3};
\filldraw[fill= white] (9,0) circle [radius=0.1] node[below] {\scriptsize 1};
\filldraw[fill= white] (3,1) circle [radius=0.1] node[above] {\scriptsize 2N};
\filldraw[fill= white] (13,0) circle [radius=0.1] node[below] {\scriptsize 1};
\filldraw[fill= white] (14,0) circle [radius=0.1] node[below] {\scriptsize 2};
\filldraw[fill= white] (15,0) circle [radius=0.1] node[below] {\scriptsize 3};
\filldraw[fill= white] (15,1) circle [radius=0.1] node[above] {\scriptsize 1};
\filldraw[fill= white] (16,0) circle [radius=0.1] node[below] {\scriptsize 2};
\filldraw[fill= white] (17,0) circle [radius=0.1] node[below] {\scriptsize 1};
\node[] at (11,0) {SU(3)};

\draw [->] (10.5,0) -- (9.5,0) ;
\draw [->] (11.5,0) -- (12.5,0) ;

\draw [thick] (0.1, 0) -- (0.9,0) ;
\draw [thick] (1.1, 0) -- (1.9,0) ;
\draw [thick] (2.1, 0) -- (2.9,0) ;
\draw [thick] (3.1, 0) -- (3.9,0) ;
\draw [thick] (4.1, 0) -- (4.9,0) ;
\draw [thick] (5.1, 0) -- (5.9,0) ;
\draw [thick] (3, 0.1) -- (3,0.9) ;
\draw [thick] (6.1, 0) -- (6.9,0) ;
\draw [thick] (7.1, 0) -- (7.9,0) ;
\draw [thick] (8.1, 0) -- (8.9,0) ;
\draw [thick] (15, 0.1) -- (15,0.9) ;
\draw [thick] (13.1, 0) -- (13.9,0) ;
\draw [thick] (14.1, 0) -- (14.9,0) ;
\draw [thick] (15.1, 0) -- (15.9,0) ;
\draw [thick] (16.1, 0) -- (16.9,0) ;
\end{tikzpicture}\nonumber \ee
The SCFT involved is described by the magnetic quiver above on the left, and it has global symmetry $SU(6)\times SU(3)^2\times U(1)$. After the gauging we find the magnetic quiver 
\be
\begin{tikzpicture}
\filldraw[fill= white] (0,0) circle [radius=0.1] node[below] {\scriptsize N};
\filldraw[fill= white] (1,0) circle [radius=0.1] node[below] {\scriptsize 2N};
\filldraw[fill= white] (2,0) circle [radius=0.1] node[below] {\scriptsize 3N};
\filldraw[fill= white] (3,0) circle [radius=0.1] node[below] {\scriptsize 4N};
\filldraw[fill= white] (4,0) circle [radius=0.1] node[below] {\scriptsize 3N};
\filldraw[fill= white] (5,0) circle [radius=0.1] node[below] {\scriptsize 2N+1};
\filldraw[fill= white] (6,0) circle [radius=0.1] node[below] {\scriptsize N+2};
\filldraw[fill= white] (8,0) circle [radius=0.1] node[below] {\scriptsize 2};
\filldraw[fill= white] (7,0) circle [radius=0.1] node[below] {\scriptsize 3};
\filldraw[fill= white] (9,0) circle [radius=0.1] node[below] {\scriptsize 1};
\filldraw[fill= white] (3,1) circle [radius=0.1] node[above] {\scriptsize 2N};
\filldraw[fill= white] (7,1) circle [radius=0.1] node[above] {\scriptsize 1};

\draw [thick] (0.1, 0) -- (0.9,0) ;
\draw [thick] (1.1, 0) -- (1.9,0) ;
\draw [thick] (2.1, 0) -- (2.9,0) ;
\draw [thick] (3.1, 0) -- (3.9,0) ;
\draw [thick] (4.1, 0) -- (4.9,0) ;
\draw [thick] (5.1, 0) -- (5.9,0) ;
\draw [thick] (3, 0.1) -- (3,0.9) ;
\draw [thick] (6.1, 0) -- (6.9,0) ;
\draw [thick] (7.1, 0) -- (7.9,0) ;
\draw [thick] (8.1, 0) -- (8.9,0) ;
\draw [thick] (7, 0.1) -- (7,0.9) ;
\end{tikzpicture}\ee
The theory in 4d includes, besides the $SU(3)$ vector multiplet, three hypermultiplets in the fundamental of $SU(3)$. Upon activating FI deformations at the abelian nodes to give them a mass, we end up with \eqref{E7k3non}.

\subsubsection*{The general rule} 

Building on the examples we have seen so far, we can now extrapolate the general rule for quivers obtained by adding an abelian node attached to a single node of a star-shaped quiver.  Without loss of generality we can assume that the abelian node is connected to a node along a tail, say the k-th node from the end of the tail. Then the gauge group is $SU(k)$ and the magnetic quiver of the matter sector can be constructed with the following procedure: 
\begin{itemize}
\item Remove the abelian node from the quiver; 
\item Attach at the end of the tail the sequence of nodes $(k)-(k-1)-\dots-(2)-(1)$;
\item Modify the rank of the nodes in the tail as follows: increase by $k-1$ the rank of the last node, by $k-2$ the rank of the node next to it and so on. Overall the rank of the $j$-th node from the end of the tail is increased by $k-j$ for $j<k$. The other nodes are not affected.
\end{itemize} 
Let us illustrate this procedure with an example. One of the possible worldvolume theories associated to $N$ D3 branes probing a 7-brane of type E$_8$ wrapped on $\mathbb{C}^2/\mathbb{Z}_4$ is described, for a suitable choice of holonomies, by the magnetic quiver 
\be\label{E8Z4ex}
\begin{tikzpicture}
\filldraw[fill= white] (0,0) circle [radius=0.1] node[below] {\scriptsize N};
\filldraw[fill= white] (1,0) circle [radius=0.1] node[below] {\scriptsize 2N};
\filldraw[fill= white] (2,0) circle [radius=0.1] node[below] {\scriptsize 3N};
\filldraw[fill= white] (3,0) circle [radius=0.1] node[below] {\scriptsize 4N};
\filldraw[fill= white] (4,0) circle [radius=0.1] node[below] {\scriptsize 5N};
\filldraw[fill= white] (5,0) circle [radius=0.1] node[below] {\scriptsize 6N};
\filldraw[fill= white] (6,0) circle [radius=0.1] node[below] {\scriptsize 4N};
\filldraw[fill= white] (7,0) circle [radius=0.1] node[below] {\scriptsize 2N};
\filldraw[fill= white] (5,1) circle [radius=0.1] node[above] {\scriptsize 3N};
\filldraw[fill= white] (3,1) circle [radius=0.1] node[above] {\scriptsize 1};

\draw [thick] (0.1, 0) -- (0.9,0) ;
\draw [thick] (1.1, 0) -- (1.9,0) ;
\draw [thick] (2.1, 0) -- (2.9,0) ;
\draw [thick] (3.1, 0) -- (3.9,0) ;
\draw [thick] (4.1, 0) -- (4.9,0) ;
\draw [thick] (5.1, 0) -- (5.9,0) ;
\draw [thick] (5, 0.1) -- (5,0.9) ;
\draw [thick] (6.1, 0) -- (6.9,0) ;
\draw [thick] (3, 0.1) -- (3,0.9) ;
\end{tikzpicture}\ee
In this case the abelian node is connected to the tail of length six, at the fourth node from the end of the tail. According to our procedure we therefore conclude that the four-dimensional theory involves a $SU(4)$ vector multiplet, coupled to a matter sector described by the magnetic quiver 
\be\label{E8Z42}
\begin{tikzpicture}
\filldraw[fill= white] (-1,0) circle [radius=0.1] node[below] {\scriptsize 4};
\filldraw[fill= white] (-2,0) circle [radius=0.1] node[below] {\scriptsize 3};
\filldraw[fill= white] (-3,0) circle [radius=0.1] node[below] {\scriptsize 2};
\filldraw[fill= white] (-4,0) circle [radius=0.1] node[below] {\scriptsize 1};
\filldraw[fill= white] (0,0) circle [radius=0.1] node[below] {\scriptsize N+3};
\filldraw[fill= white] (1,0) circle [radius=0.1] node[below] {\scriptsize 2N+2};
\filldraw[fill= white] (2,0) circle [radius=0.1] node[below] {\scriptsize 3N+1};
\filldraw[fill= white] (3,0) circle [radius=0.1] node[below] {\scriptsize 4N};
\filldraw[fill= white] (4,0) circle [radius=0.1] node[below] {\scriptsize 5N};
\filldraw[fill= white] (5,0) circle [radius=0.1] node[below] {\scriptsize 6N};
\filldraw[fill= white] (6,0) circle [radius=0.1] node[below] {\scriptsize 4N};
\filldraw[fill= white] (7,0) circle [radius=0.1] node[below] {\scriptsize 2N};
\filldraw[fill= white] (5,1) circle [radius=0.1] node[above] {\scriptsize 3N};

\draw [thick] (-0.1, 0) -- (-0.9,0) ;
\draw [thick] (-1.1, 0) -- (-1.9,0) ;
\draw [thick] (-2.1, 0) -- (-2.9,0) ;
\draw [thick] (-3.1, 0) -- (-3.9,0) ;
\draw [thick] (0.1, 0) -- (0.9,0) ;
\draw [thick] (1.1, 0) -- (1.9,0) ;
\draw [thick] (2.1, 0) -- (2.9,0) ;
\draw [thick] (3.1, 0) -- (3.9,0) ;
\draw [thick] (4.1, 0) -- (4.9,0) ;
\draw [thick] (5.1, 0) -- (5.9,0) ;
\draw [thick] (5, 0.1) -- (5,0.9) ;
\draw [thick] (6.1, 0) -- (6.9,0) ;
\end{tikzpicture}\ee
where we have increased by three the rank of the last node of the length-six tail, by two the rank of the second-last node and by one the rank of the third last. If we instead start from the quiver 
\be
\begin{tikzpicture}
\filldraw[fill= white] (0,0) circle [radius=0.1] node[below] {\scriptsize N};
\filldraw[fill= white] (1,0) circle [radius=0.1] node[below] {\scriptsize 2N};
\filldraw[fill= white] (2,0) circle [radius=0.1] node[below] {\scriptsize 3N};
\filldraw[fill= white] (3,0) circle [radius=0.1] node[below] {\scriptsize 4N};
\filldraw[fill= white] (4,0) circle [radius=0.1] node[below] {\scriptsize 5N};
\filldraw[fill= white] (5,0) circle [radius=0.1] node[below] {\scriptsize 6N};
\filldraw[fill= white] (6,0) circle [radius=0.1] node[below] {\scriptsize 4N};
\filldraw[fill= white] (7,0) circle [radius=0.1] node[below] {\scriptsize 2N};
\filldraw[fill= white] (5,1) circle [radius=0.1] node[above] {\scriptsize 3N};
\filldraw[fill= white] (6,1) circle [radius=0.1] node[above] {\scriptsize 1};

\draw [thick] (0.1, 0) -- (0.9,0) ;
\draw [thick] (1.1, 0) -- (1.9,0) ;
\draw [thick] (2.1, 0) -- (2.9,0) ;
\draw [thick] (3.1, 0) -- (3.9,0) ;
\draw [thick] (4.1, 0) -- (4.9,0) ;
\draw [thick] (5.1, 0) -- (5.9,0) ;
\draw [thick] (5, 0.1) -- (5,0.9) ;
\draw [thick] (6.1, 0) -- (6.9,0) ;
\draw [thick] (6, 0.1) -- (6,0.9) ;
\end{tikzpicture}\ee 
we conclude that the gauge group is $SU(2)$, since the abelian node is attached to the second-last node in the tail, and the magnetic quiver of the matter sector is 
\be\label{E8Z4ex2}
\begin{tikzpicture}
\filldraw[fill= white] (0,0) circle [radius=0.1] node[below] {\scriptsize N};
\filldraw[fill= white] (1,0) circle [radius=0.1] node[below] {\scriptsize 2N};
\filldraw[fill= white] (2,0) circle [radius=0.1] node[below] {\scriptsize 3N};
\filldraw[fill= white] (3,0) circle [radius=0.1] node[below] {\scriptsize 4N};
\filldraw[fill= white] (4,0) circle [radius=0.1] node[below] {\scriptsize 5N};
\filldraw[fill= white] (5,0) circle [radius=0.1] node[below] {\scriptsize 6N};
\filldraw[fill= white] (6,0) circle [radius=0.1] node[below] {\scriptsize 4N};
\filldraw[fill= white] (7,0) circle [radius=0.1] node[below] {\scriptsize 2N+1};
\filldraw[fill= white] (5,1) circle [radius=0.1] node[above] {\scriptsize 3N};
\filldraw[fill= white] (8,0) circle [radius=0.1] node[below] {\scriptsize 2};
\filldraw[fill= white] (9,0) circle [radius=0.1] node[below] {\scriptsize 1};

\draw [thick] (0.1, 0) -- (0.9,0) ;
\draw [thick] (1.1, 0) -- (1.9,0) ;
\draw [thick] (2.1, 0) -- (2.9,0) ;
\draw [thick] (3.1, 0) -- (3.9,0) ;
\draw [thick] (4.1, 0) -- (4.9,0) ;
\draw [thick] (5.1, 0) -- (5.9,0) ;
\draw [thick] (5, 0.1) -- (5,0.9) ;
\draw [thick] (6.1, 0) -- (6.9,0) ;
\draw [thick] (7.1, 0) -- (7.9,0) ;
\draw [thick] (8.1, 0) -- (8.9,0) ;
\end{tikzpicture}\ee

\section{Conclusions}\label{Sec:Concl}

In this paper we addressed the problem of determining what is the field theory seen at low energy by D3 branes probing parallel stacks of non-perturbative 7-branes extended along abelian orbifolds. For the sake of simplicity, we mainly focused on superconformal theories, and gave a straightforward algorithm to characterize them. In particular, we specified their global symmetry, spectrum of Coulomb-branch operators, and pattern of Higgs-branch flows. Analogously to their perturbative counterparts, these orbifolds admit either a single or a pair of  inequivalent realizations. For any such realization there is always a family of SCFTs standing out, which we dubbed the ``canonical'' family, that corresponds to a homogeneous distribution of the fractional D3 charge among the various irreducible representations of the orbifold group. Other families, associated to different choices of holonomy at infinity and at the origin for the 7-brane gauge field, emanate from the canonical one through Higgs-branch deformations.

In order to reach the above conclusions, we relied on the technique of magnetic quivers. This method has proven very effective due to the relation of our stringy setup with the M-theory realization of 6d SCFTs. Our analysis further elucidates the correspondence noticed in \cite{Giacomelli:2020gee} and suggests that the six-dimensional origin of four-dimensional SCFTs might be a very general statement. Considering the fact that we have a classification of 6d SCFTs \cite{Heckman:2013pva, Heckman:2015bfa}, this might result in a very simple and natural organizing principle for SCFTs in four dimensions, which is worth further investigations (see \cite{Argyres:2022mnu} for a recent overview of superconformal theories). 

By comparing with the partial classification results at rank two, in particular \cite{Martone:2021ixp, Martone:2021drm}, we notice that in the present work all the models in the $\mathfrak{e}_8-\mathfrak{so}(20)$ appear. The other theories, at least most of them, are expected to arise by including $\mathcal{S}$-fold quotients in our F-theory setup. It will be important to analyze them systematically in order to improve our understanding of the superconformal landscape. This should also connect the present work to the recent analysis of SW compactification of 6d SCFTs performed in \cite{Heckman:2022suy}. 

The systematics we highlighted in this paper points towards the existence of a deeper mathematical structure underlying the brane systems we studied. We proposed that the main properties of the probe SCFT for a given choice of orbifold and 7-brane stack be inferable from a specific extension by the orbifold group of the cyclic group acting on the 7-brane transverse space. In analogy with the perturbative case, this extension would play the role of symmetry group of the fractional-D3-brane lattice and, as such, should in particular encode in some manner the conformal dimensions of the various Coulomb-branch operators. It would be important to pursue this idea further.

\subsection*{Acknowledgements}

We would like to thank Massimo Bianchi and Noppadol Mekareeya for useful discussions, and Yuji Tachikawa for carefully reading our manuscript. S.G.~would like to thank the physics Department of Tor Vergata University for hospitality during the completion of this project. The work of S.G.~is supported by the INFN grant ``Per attivit\`a di formazione per sostenere progetti di ricerca'' (GRANT 73/STRONGQFT).

\begin{appendix}

\section{Mass defomations via FI parameters} \label{FIdef}

In this appendix we review the technique developed in \cite{vanBeest:2021xyt} to implement mass deformations at the level of magnetic quivers, by activating suitable FI parameters.

\subsection{FI deformations of Unitary Quivers}

Let us start by reviewing how FI deformations affect the equations of motion at a given gauge node in the quiver. We turn on a  complex FI parameter $\lambda$ in a $U(N)$ gauge theory with $F$ fundamental hypermultiplets $\widetilde{Q}_f$ and $Q_f$ (in four supercharges notation) and we denote with $\Phi$ the adjoint chiral in the $\mathcal{N}=4$ vector multiplet.
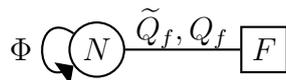
\begin{center}
\begin{tikzpicture}[->,thick, scale=0.4]
\node[](L3) at (-5,0) {$\Phi$};
\node[] (L2) at (0.3,0.7) {$\widetilde{Q}_f,Q_f$};
\node[circle, draw, inner sep=2.5](L4) at (-2.5,0){$N$};
\node[rectangle, draw, inner sep=1.7,minimum height=.6cm,minimum width=.6cm](L5) at (3,0){$F$};

  \path[every node/.style={font=\sffamily\small,
  		fill=white,inner sep=1pt}]
(L4) edge [loop, out=145, in=215, looseness=4] (L4);
\draw[-] (L5) -- (L4);
\end{tikzpicture}
\end{center}
After the FI deformation we end up with the superpotential 
\begin{equation} 
\mathcal{W}= \widetilde{Q}_f\Phi Q^f + \lambda\Tr\Phi\,,
\end{equation}
and consequently we should solve the following F-term and D-term equations: 
\be\begin{array}{l} 
Q^f\widetilde{Q}_f = \lambda I_N\,,\\ 
Q^fQ_f^{\dagger}-\widetilde{Q}^{\dagger f} \widetilde{Q}_f=0\,,
\end{array}\ee
where the summation over flavor indices implies a summation over all the bifundamental hypermultiplets charged under the $U(N)$ gauge group, when this is embedded as a node inside a bigger quiver. 
From F-terms we can deduce that
\be 
\label{eq:traceFterm}
\Tr(Q^f\widetilde{Q}_f)=N\lambda\,.
\ee
Considering each node in the quiver in turn, and combining this with the identity $\Tr(Q^f\widetilde{Q}_f)=\Tr(\widetilde{Q}_fQ^f)$, we conclude that for any quiver with unitary gauge groups and bifundamental hypermultiplets the FI parameters should obey the relation 
\be\label{constr} \sum_i N_i\lambda^i=0\,,\ee
where $N_i$ denotes the rank of the $i$'th node. As a result we need to turn on FI parameters at multiple nodes and we will mainly focus on the minimal option with only two FI parameters, although sometimes we will find it more convenient to turn on three or more FI parameters. We will see some examples of this below.

Consider the case of FI parameters turned on at two abelian nodes. Because of (\ref{constr}), the two FI parameters are $\lambda$ and $-\lambda$. The deformation induces a nontrivial expectation value for a chain of bifundamentals connecting the two nodes. This follows from the constraint on the trace of bifundamental bilinears in \eqref{eq:traceFterm}. We can easily find an explicit solution to the F- and D-term equations in this case. First, we choose a subquiver beginning and ending at the nodes at which we have turned on the FI parameter. All the bifundamentals along the subquiver, which we denote $B_i$ and $\widetilde{B}_i$, acquire a nontrivial vev and modulo gauge transformations we can set
\be\label{solab} \langle B_i\rangle=\sqrt{\lambda}(v_1,0,\dots,0); \quad \langle \widetilde{B}_i\rangle=\langle B_i\rangle^T\,, \ee 
where $v_1$ is the unit vector whose entries are all trivial except for the first. Of course the size of the matrix $\langle B_i\rangle$ is dictated by the rank of the gauge groups under which the bifundamental is charged.
All the unitary gauge groups along the subquiver are spontaneously broken as $U(n_i)\rightarrow U(n_i-1)$, whereas the other nodes in the quiver are unaffected. Among all the broken $U(1)$ factors, the diagonal combination survives and gives rise to a new $U(1)$ node, which is coupled to all the nodes of the quiver connected to those of the subquiver.  Overall, this is equivalent to subtracting \cite{Cabrera:2018ann} an abelian quiver, isomorphic to the subquiver described above, and the new $U(1)$ is identified with the rebalancing node. 

Generalizing to the case of FI parameters turned on at two nodes of the same rank $k \geq 1$ is straightforward. The vev for the bifundamentals is obtained from (\ref{solab}) by picking the tensor product with the $k\times k$ identity matrix $I_k$. All the groups along the subquiver are broken as $U(n_i)\rightarrow U(n_i-k)$ and finally we need to add a $U(k)$ node associated with the surviving gauge symmetry.  This operation corresponds to a modified quiver subtraction, where we subtract a quiver with $U(k)$ nodes only and rebalance with a $U(k)$ node. 

Let us now consider a more elaborate variant which involves three nodes. Say we turn on FI parameters at the nodes $U(n)$, $U(m)$ and $U(n+m)$. The FI parameters satisfy the relation (\ref{constr}) and we further impose the constraint $\lambda_m=\lambda_n$, so that we still have only one independent parameter. The equations of motion can be solved as follows: We set the vev of all the bifundamentals in the subquiver connecting the nodes $U(m)$ and $U(n+m)$ to be 
\be\label{solab1} \langle B_i\rangle=\sqrt{\lambda_m}(I_m,0,\dots,0); \quad \langle \widetilde{B}_i\rangle=\langle B_i\rangle^T\,, \ee 
and the vev of the bifundamentals in the subquiver connecting the nodes $U(n+m)$ and $U(n)$ to be 
\be\label{solab2} \langle B_i\rangle=\sqrt{\lambda_m}(I_n,0,\dots,0); \quad \langle \widetilde{B}_i\rangle=\langle B_i\rangle^T\,. \ee 
Here we are assuming that the two subquivers meet at the $U(n+m)$ node only. 

The higgsing of the theory can be described in terms of a sequence of two modified quiver subtractions: We first subtract a quiver of $U(n)$ nodes going from $U(n)$ to $U(n+m)$ and we rebalance with a $U(n)$ node. Then we subtract from the resulting quiver a quiver of $U(m)$ nodes going from $U(m)$ to $U(n+m)$ and rebalance with a $U(m)$ node. A careful analysis of the higgsing reveals that the $U(n)$ node we introduced at the first step should not be rebalanced at the second step. 
Let us illustrate the procedure for $m=1$ and $n=2$ in the case of the $E_8$ quiver: 
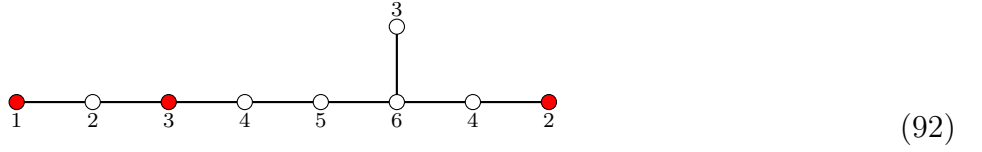
\be\label{E8MQ}
\begin{tikzpicture}
\filldraw[fill= red] (0,0) circle [radius=0.1] node[below] {\scriptsize 1};
\filldraw[fill= white] (1,0) circle [radius=0.1] node[below] {\scriptsize 2};
\filldraw[fill= red] (2,0) circle [radius=0.1] node[below] {\scriptsize 3};
\filldraw[fill= white] (3,0) circle [radius=0.1] node[below] {\scriptsize 4};
\filldraw[fill= white] (4,0) circle [radius=0.1] node[below] {\scriptsize 5};
\filldraw[fill= white] (5,0) circle [radius=0.1] node[below] {\scriptsize 6};
\filldraw[fill= white] (6,0) circle [radius=0.1] node[below] {\scriptsize 4};
\filldraw[fill= white] (5,1) circle [radius=0.1] node[above] {\scriptsize 3};
\filldraw[fill= red] (7,0) circle [radius=0.1] node[below] {\scriptsize 2};
\draw [thick] (0.1, 0) -- (0.9,0) ;
\draw [thick] (1.1, 0) -- (1.9,0) ;
\draw [thick] (2.1, 0) -- (2.9,0) ;
\draw [thick] (3.1, 0) -- (3.9,0) ;
\draw [thick] (4.1, 0) -- (4.9,0) ;
\draw [thick] (5.1, 0) -- (5.9,0) ;
\draw [thick] (5, 0.1) -- (5,0.9) ;
\draw [thick] (6.1, 0) -- (6.9,0) ;
\end{tikzpicture} 
\ee 
We turn on FI parameters at the nodes in red. We first subtract an $A_6$ quiver with $U(2)$ nodes, getting
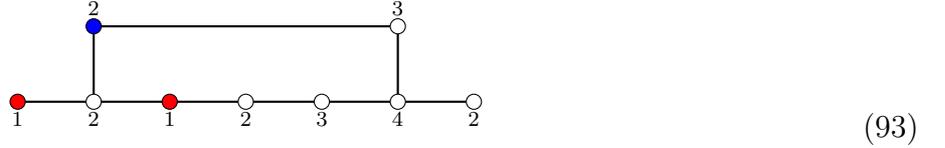
\be\label{E8MQ2}
\begin{tikzpicture}
\filldraw[fill= red] (0,0) circle [radius=0.1] node[below] {\scriptsize 1};
\filldraw[fill= white] (1,0) circle [radius=0.1] node[below] {\scriptsize 2};
\filldraw[fill= red] (2,0) circle [radius=0.1] node[below] {\scriptsize 1};
\filldraw[fill= white] (3,0) circle [radius=0.1] node[below] {\scriptsize 2};
\filldraw[fill= white] (4,0) circle [radius=0.1] node[below] {\scriptsize 3};
\filldraw[fill= white] (5,0) circle [radius=0.1] node[below] {\scriptsize 4};
\filldraw[fill= white] (6,0) circle [radius=0.1] node[below] {\scriptsize 2};
\filldraw[fill= white] (5,1) circle [radius=0.1] node[above] {\scriptsize 3};
\filldraw[fill= blue] (1,1) circle [radius=0.1] node[above] {\scriptsize 2};
\draw [thick] (0.1, 0) -- (0.9,0) ;
\draw [thick] (1.1, 0) -- (1.9,0) ;
\draw [thick] (2.1, 0) -- (2.9,0) ;
\draw [thick] (3.1, 0) -- (3.9,0) ;
\draw [thick] (4.1, 0) -- (4.9,0) ;
\draw [thick] (5.1, 0) -- (5.9,0) ;
\draw [thick] (5, 0.1) -- (5,0.9) ;
\draw [thick] (1.1, 1) -- (4.9,1) ;
\draw [thick] (1, 0.1) -- (1,0.9) ;
\end{tikzpicture} 
\ee 
The node in blue is used to rebalance. Then we subtract an $A_3$ abelian quiver, obtaining 
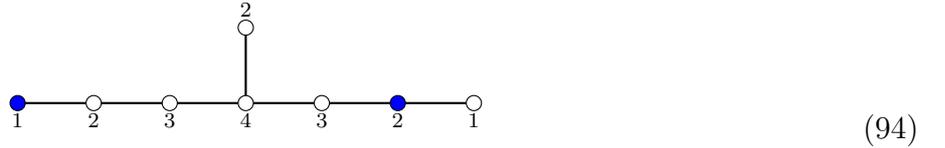
\be\label{E8MQ3}
\begin{tikzpicture}
\filldraw[fill= blue] (0,0) circle [radius=0.1] node[below] {\scriptsize 1};
\filldraw[fill= white] (1,0) circle [radius=0.1] node[below] {\scriptsize 2};
\filldraw[fill= white] (2,0) circle [radius=0.1] node[below] {\scriptsize 3};
\filldraw[fill= white] (3,0) circle [radius=0.1] node[below] {\scriptsize 4};
\filldraw[fill= white] (4,0) circle [radius=0.1] node[below] {\scriptsize 3};
\filldraw[fill= blue] (5,0) circle [radius=0.1] node[below] {\scriptsize 2};
\filldraw[fill= white] (6,0) circle [radius=0.1] node[below] {\scriptsize 1};
\filldraw[fill= white] (3,1) circle [radius=0.1] node[above] {\scriptsize 2};
\draw [thick] (0.1, 0) -- (0.9,0) ;
\draw [thick] (1.1, 0) -- (1.9,0) ;
\draw [thick] (2.1, 0) -- (2.9,0) ;
\draw [thick] (3.1, 0) -- (3.9,0) ;
\draw [thick] (4.1, 0) -- (4.9,0) ;
\draw [thick] (5.1, 0) -- (5.9,0) ;
\draw [thick] (3, 0.1) -- (3,0.9) ;
\end{tikzpicture} 
\ee 
The $U(1)$ node in blue is introduced to rebalance in the second subtraction, and, as we have explained before, is not connected to the blue $U(2)$ node. Overall, this FI deformation describes the transition from the E$_8$ to the E$_7$ Minahan-Nemeschansky theory.

\subsection{Other FI deformations}\label{sec:genFI}

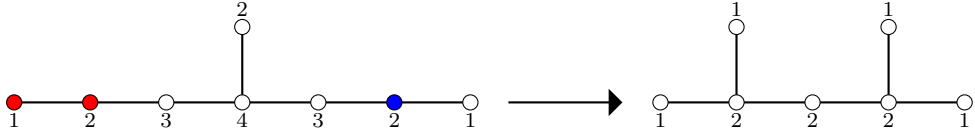
\begin{figure}[ht]
\begin{center}
\begin{tikzpicture}
\filldraw[fill= red] (0,0) circle [radius=0.1] node[below] {\scriptsize 1};
\filldraw[fill= red] (1,0) circle [radius=0.1] node[below] {\scriptsize 2};
\filldraw[fill= white] (2,0) circle [radius=0.1] node[below] {\scriptsize 3};
\filldraw[fill= white] (3,0) circle [radius=0.1] node[below] {\scriptsize 4};
\filldraw[fill= white] (4,0) circle [radius=0.1] node[below] {\scriptsize 3};
\filldraw[fill= blue] (5,0) circle [radius=0.1] node[below] {\scriptsize 2};
\filldraw[fill= white] (6,0) circle [radius=0.1] node[below] {\scriptsize 1};
\filldraw[fill= white] (3,1) circle [radius=0.1] node[above] {\scriptsize 2};
\draw [thick] (0.1, 0) -- (0.9,0) ;
\draw [thick] (1.1, 0) -- (1.9,0) ;
\draw [thick] (2.1, 0) -- (2.9,0) ;
\draw [thick] (3.1, 0) -- (3.9,0) ;
\draw [thick] (4.1, 0) -- (4.9,0) ;
\draw [thick] (3, 0.1) -- (3,0.9) ;
\draw [thick] (5.1, 0) -- (5.9,0) ;

\draw[->, thick] (6.5,0)--(8,0);

\filldraw[fill= white] (8.5,0) circle [radius=0.1] node[below] {\scriptsize 1};
\filldraw[fill= white] (9.5,0) circle [radius=0.1] node[below] {\scriptsize 2};
\filldraw[fill= white] (10.5,0) circle [radius=0.1] node[below] {\scriptsize 2};
\filldraw[fill= white] (11.5,0) circle [radius=0.1] node[below] {\scriptsize 2};
\filldraw[fill= white] (9.5,1) circle [radius=0.1] node[above] {\scriptsize 1};
\filldraw[fill= white] (12.5,0) circle [radius=0.1] node[below] {\scriptsize 1};
\filldraw[fill= white] (11.5,1) circle [radius=0.1] node[above] {\scriptsize 1};
\draw [thick] (8.6,0) -- (9.4,0) ;
\draw [thick] (9.6,0) -- (10.4,0) ;
\draw [thick] (10.6,0) -- (11.4,0) ;
\draw [thick] (11.6,0) -- (12.4,0) ;
\draw [thick] (9.5,0.1) -- (9.5,0.9) ;
\draw [thick] (11.5,0.1) -- (11.5,0.9) ;

\end{tikzpicture} 
\end{center}
\caption{The $SO(12)$-preserving FI deformation of the rank 1 $E_7$ theory. We turn on FI parameters at the red nodes, and we indicate in blue the node of the $E_7$ quiver which is not higgsed but is connected to a higgsed node. The extra $U(1)$ node is coupled to the blue node only.}\label{e7gaugedef2}
\end{figure}

When the quiver contains a tail of the form $U(1)-U(2)-\cdots$ we can turn on FI parameters at the $U(1)$ and $U(2)$ nodes only. As before, the generalization to the case $U(k)-U(2k)-\cdots$ is straightforward. Because of (\ref{constr}), we set $\lambda_1=-2\lambda_2\equiv2\lambda$. If we denote the $U(1)\times U(2)$ bifundamentals as $\widetilde{Q}$, $Q$ and the other $U(2)$ fundamentals as $\widetilde{P}_f$, $P^f$, the relevant F-terms are 
\be\label{soldef2} \langle\widetilde{Q}Q\rangle=2\lambda;\quad \langle\widetilde{P}_fP^f\rangle-\langle Q\widetilde{Q}\rangle=\lambda I_2\,,\ee 
where we are summing over flavor indices. These equations are solved by 
\be\label{u1u2}\langle\widetilde{Q}\rangle=\sqrt{\lambda}(1,1);\; \langle Q\rangle=\langle\widetilde{Q}\rangle^T;\; \langle\widetilde{P}\rangle=\sqrt{\lambda}\left(\begin{array}{cccc}1 & 0 & 0 & \dots \\ 0 & 1 & 0 & \dots\\\end{array}\right);\; \langle P\rangle^T=\sqrt{\lambda}\left(\begin{array}{cccc}0 & 1 & 0 & \dots \\ 1 & 0 & 0 & \dots\\\end{array}\right).\ee 
This vev spontaneously breaks $U(1)\times U(2)$ to a diagonal $U(1)$ subgroup. It is easy to check that D-terms are satisfied as well. The vev for the $U(2)$ fundamentals is
\be 
\langle\widetilde{P}_fP^f\rangle=\left(\begin{array}{cc}0 & \lambda \\ \lambda & 0 \\\end{array}\right)=\left(\begin{array}{cc}0 & \lambda \\ 0 & 0 \\\end{array}\right)+\left(\begin{array}{cc}0 & 0 \\ \lambda & 0 \\\end{array}\right)\,.
\ee 
This propagates along the quiver, breaking all the groups as $U(n)\rightarrow U(n-2)$, until we find a junction where we can ``decompose'' the vev as above.
The two nodes connected to the junction are higgsed as $U(n)\rightarrow U(n-1)$ and the vev does not propagate any further. All the nodes connected to the subquiver of nodes which are (partially) higgsed are now coupled to a new $U(1)$ node, which is left unbroken by the vev. 
We give an example of this process in Figure \ref{e7gaugedef2}.

\section{The Coulomb-branch spectrum and central charges of class $\mathcal{S}$ theories of type $A$}\label{Sec:CBSpectrum} 

The CB spectrum of class $\mathcal{S}$ theories is encoded in the puncture data on the Riemann surface. We will concentrate on the case of the sphere for a $\mathcal{N}=(2,0)$ theory of type $A_{N-1}$, which is enough for our purposes. The corresponding magnetic quiver was determined in \cite{Benini:2010uu} and is given by a unitary star-shaped quiver with a $U(N)$ group at the center. The tails are in one-to-one correspondence with the punctures and are determined as follows: As is well known, $A_{N-1}$ punctures are classified by Young diagrams with $N$ boxes (or partitions of $N$), where the diagram with a row of length $N$ corresponds to the full puncture (partition $(1^N)$) and the Young diagram with a single column (partition $N$) corresponds to the trivial puncture (i.e. a generic point on the Riemann surface). Each part of the partition corresponds to the height of a column in the corresponding Young diagram. The rank of the first unitary group along a tail starting from the center is given by the number of boxes of the corresponding Young diagram with the first column removed; the rank of the second is given by the number of boxes upon removal of the first two columns and so on. With this rule we see that the trivial puncture corresponds to no tail at all whereas the tail associated with the full puncture is given by $T(SU(N))$: 
$$ U(1)-U(2)-\dots-U(N-1)-(N)\,,$$ 
whre $(N)$ is identified with the central node of the quiver. 

The algorithm for determining the CB spectrum of the theory is described in \cite{Chacaltana:2010ks} and works as follows: We start by numbering the boxes of each Young diagram. The number associated with a box in the i-th row and j-th column\footnote{Both rows and columns of the Young diagram are ordered according to their size, so the first row is the longest and the first column is the tallest.} is given by $\sum_{k<i}\ell_k+j$, where $\ell_k$ denotes the length of the k-th row. The last box of the last row of course corresponds to the number $N$. We also label the n-th box with an integer $p_n=n-i$, where again we assume that the box belongs to the i-th row. With this notation at hand, the number of CB operators of dimension $n>1$ is given by 
\be\label{spectrumCB} \sum_k p_{n,k}-2n+1\,,\ee 
where the sum runs over punctures (or equivalently Young diagrams) and $p_{n,k}$ denotes the label of the n-th box of the k-th Young diagram. 

The punctures also encode the global symmetry of the theory since each puncture contributes a factor \cite{Gaiotto:2009we, Chacaltana:2010ks} 
\be G_p=S\left( \prod_h U(n_h)\right),\ee 
where $n_h$ is the number of columns of height $h$ in the corresponding Young diagram. Generically, the global symmetry is the product of all $G_p$ factors, although symmetry enhancements are possible. The corresponding flavor central charges are encoded in the length of the rows of the Young diagram: The flavor central charge of $SU(\ell_{j+1}-\ell_j)$ is given by the expression \cite{Chacaltana:2011ze}
\be k=2\sum_{i\leq j}\ell_i\,.\ee
As is well known, if we gauge the flavor symmetry the contribution to the beta function of the gauge coupling is half the flavor central charge (see e.g.~\cite{Benini:2009mz}).
Similarly, we can compute the a and c central charges from the class $\mathcal{S}$ data as follows: The combination $2a-c$ is given by the Tachikawa-Shapere formula \cite{Shapere:2008zf}
\be 8a-4c=\sum_i(2D_i-1)\,,\ee 
where the sum runs over CB operators and $D_i$ denotes their scaling dimension. The combination $24(c-a)$ gives the quaternionic dimension of the Higgs branch and is equal to the rank of the corresponding magnetic quiver (each $U(n)$ gauge group in the quiver contributes $n$ to the rank).

\section{Higher-order abelian orbifolds}\label{Sec:MoreEx}

In this appendix we discuss in detail some more examples of theories arising from non-perturbative 7-branes wrapped on abelian orbifolds. When possible, we provide a uniform description in terms of their six-dimensional avatars.

\subsection{E$_6$-type 7-brane}
\subsubsection*{$\bf\mathbb{C}^2/\mathbb{Z}_3$}
The allowed $\mathbb{Z}_3$ non-trivial holonomies break the $E_6$ to: $SO(8)\times U(1)^2$, $SO(10)\times U(1)$, $SU(6)\times U(1)$, $SU(5)\times SU(2)\times U(1)$, $SU(3)^3$. However, not all of them will be visible with the technique of magnetic quivers.

According to the rules of Section \ref{Sec:GenDisc}, there are two canonical families of theories associated to the E$_6$ 7-brane on the $\mathbb{Z}_3$ orbifold.

\paragraph{$\bullet$} The first family has the following magnetic quiver
\be\label{E6Z3NMQ}
\begin{tikzpicture}
\filldraw[fill= white] (3,0) circle [radius=0.1] node[below] {\scriptsize N};
\filldraw[fill= white] (4,0) circle [radius=0.1] node[below] {\scriptsize 2N};
\filldraw[fill= white] (5,0) circle [radius=0.1] node[below] {\scriptsize 3N};
\filldraw[fill= white] (6,0) circle [radius=0.1] node[below] {\scriptsize 2N};
\filldraw[fill= white] (5,1) circle [radius=0.1] node[above] {\scriptsize 2N};
\filldraw[fill= white] (5,-1) circle [radius=0.1] node[below] {\scriptsize 1};
\filldraw[fill= white] (7,0) circle [radius=0.1] node[below] {\scriptsize N};
\filldraw[fill= white] (6,1) circle [radius=0.1] node[above] {\scriptsize N};

\draw [thick] (5.1, 1) -- (5.9,1) ;
\draw [thick] (3.1, 0) -- (3.9,0) ;
\draw [thick] (4.1, 0) -- (4.9,0) ;
\draw [thick] (5.1, 0) -- (5.9,0) ;
\draw [thick] (5, 0.1) -- (5,0.9) ;
\draw [thick] (5, -0.1) -- (5,-0.9) ;
\draw [thick] (6.1, 0) -- (6.9,0) ;
\end{tikzpicture} \ee 
This family has $SU(3)^3$ non-abelian global symmetry and spectrum of CB operators
$$(2,3,3)\,;\,(5,6,6)\,;\,\dots\,;\, (3N-1,3N,3N)\,.$$

We can obtain \ref{E6Z3NMQ} starting from the first M-theory quiver of those listed in \eqref{Table:SU(3)}, where we first turn on FI's for the leftmost $U(k)$ nodes ($k=1,2,3$):\\
\be
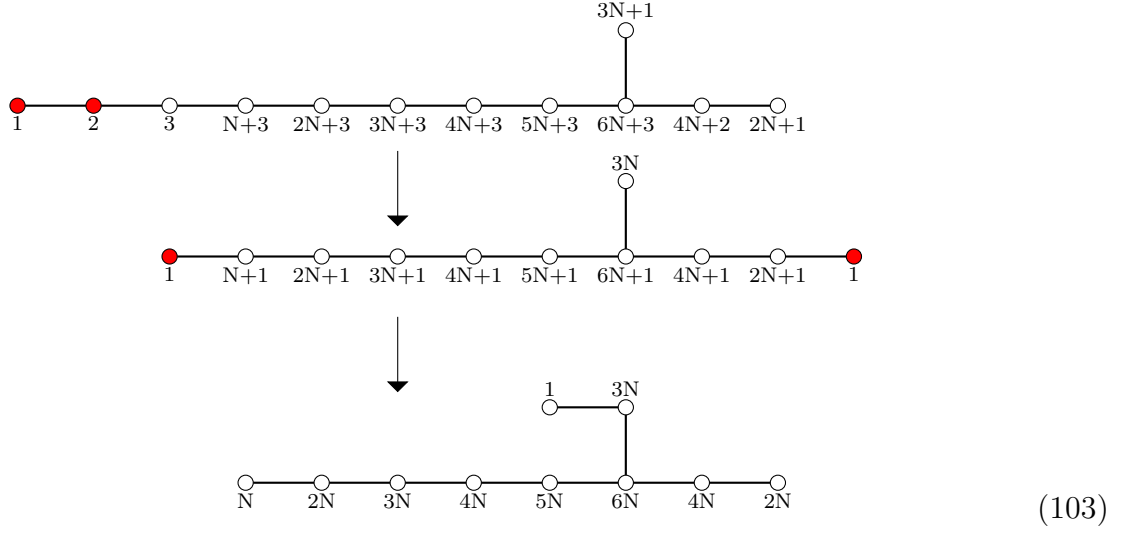
 \begin{tikzpicture}
\filldraw[fill= red] (-3,0) circle [radius=0.1] node[below] {\scriptsize 1};
\filldraw[fill= red] (-2,0) circle [radius=0.1] node[below] {\scriptsize 2};
\filldraw[fill= white] (-1,0) circle [radius=0.1] node[below] {\scriptsize 3};
\filldraw[fill= white] (0,0) circle [radius=0.1] node[below] {\scriptsize N+3};
\filldraw[fill= white] (1,0) circle [radius=0.1] node[below] {\scriptsize 2N+3};
\filldraw[fill= white] (2,0) circle [radius=0.1] node[below] {\scriptsize 3N+3};
\filldraw[fill= white] (3,0) circle [radius=0.1] node[below] {\scriptsize 4N+3};
\filldraw[fill= white] (4,0) circle [radius=0.1] node[below] {\scriptsize 5N+3};
\filldraw[fill= white] (5,0) circle [radius=0.1] node[below] {\scriptsize 6N+3};
\filldraw[fill= white] (6,0) circle [radius=0.1] node[below] {\scriptsize 4N+2};
\filldraw[fill= white] (5,1) circle [radius=0.1] node[above] {\scriptsize 3N+1};
\filldraw[fill= white] (7,0) circle [radius=0.1] node[below] {\scriptsize 2N+1};

\draw [thick] (-0.1, 0) -- (-0.9,0) ;
\draw [thick] (-1.1, 0) -- (-1.9,0) ;
\draw [thick] (-2.1, 0) -- (-2.9,0) ;
\draw [thick] (0.1, 0) -- (0.9,0) ;
\draw [thick] (1.1, 0) -- (1.9,0) ;
\draw [thick] (2.1, 0) -- (2.9,0) ;
\draw [thick] (3.1, 0) -- (3.9,0) ;
\draw [thick] (4.1, 0) -- (4.9,0) ;
\draw [thick] (5.1, 0) -- (5.9,0) ;
\draw [thick] (5, 0.1) -- (5,0.9) ;
\draw [thick] (6.1, 0) -- (6.9,0) ;
\draw [->] (2,-0.6) -- (2,-1.6) ;

\filldraw[fill= red] (-1,-2) circle [radius=0.1] node[below] {\scriptsize 1};
\filldraw[fill= white] (0,-2) circle [radius=0.1] node[below] {\scriptsize N+1};
\filldraw[fill= white] (1,-2) circle [radius=0.1] node[below] {\scriptsize 2N+1};
\filldraw[fill= white] (2,-2) circle [radius=0.1] node[below] {\scriptsize 3N+1};
\filldraw[fill= white] (3,-2) circle [radius=0.1] node[below] {\scriptsize 4N+1};
\filldraw[fill= white] (4,-2) circle [radius=0.1] node[below] {\scriptsize 5N+1};
\filldraw[fill= white] (5,-2) circle [radius=0.1] node[below] {\scriptsize 6N+1};
\filldraw[fill= white] (6,-2) circle [radius=0.1] node[below] {\scriptsize 4N+1};
\filldraw[fill= white] (5,-1) circle [radius=0.1] node[above] {\scriptsize 3N};
\filldraw[fill= white] (7,-2) circle [radius=0.1] node[below] {\scriptsize 2N+1};
\filldraw[fill= red] (8,-2) circle [radius=0.1] node[below] {\scriptsize 1};

\draw [thick] (-0.1,-2) -- (-0.9,-2) ;
\draw [thick] (0.1, -2) -- (0.9,-2) ;
\draw [thick] (1.1, -2) -- (1.9,-2) ;
\draw [thick] (2.1, -2) -- (2.9,-2) ;
\draw [thick] (3.1, -2) -- (3.9,-2) ;
\draw [thick] (4.1, -2) -- (4.9,-2) ;
\draw [thick] (5.1, -2) -- (5.9,-2) ;
\draw [thick] (5, -1.1) -- (5,-1.9) ;
\draw [thick] (6.1, -2) -- (6.9,-2) ;
\draw [thick] (7.1, -2) -- (7.9,-2) ;

\draw [->] (2,-2.8) -- (2,-3.8) ;

\filldraw[fill= white] (4,-4) circle [radius=0.1] node[above] {\scriptsize 1};
\filldraw[fill= white] (0,-5) circle [radius=0.1] node[below] {\scriptsize N};
\filldraw[fill= white] (1,-5) circle [radius=0.1] node[below] {\scriptsize 2N};
\filldraw[fill= white] (2,-5) circle [radius=0.1] node[below] {\scriptsize 3N};
\filldraw[fill= white] (3,-5) circle [radius=0.1] node[below] {\scriptsize 4N};
\filldraw[fill= white] (4,-5) circle [radius=0.1] node[below] {\scriptsize 5N};
\filldraw[fill= white] (5,-5) circle [radius=0.1] node[below] {\scriptsize 6N};
\filldraw[fill= white] (6,-5) circle [radius=0.1] node[below] {\scriptsize 4N};
\filldraw[fill= white] (5,-4) circle [radius=0.1] node[above] {\scriptsize 3N};
\filldraw[fill= white] (7,-5) circle [radius=0.1] node[below] {\scriptsize 2N};

\draw [thick] (0.1, -5) -- (0.9,-5) ;
\draw [thick] (1.1, -5) -- (1.9,-5) ;
\draw [thick] (2.1, -5) -- (2.9,-5) ;
\draw [thick] (3.1, -5) -- (3.9,-5) ;
\draw [thick] (4.1, -5) -- (4.9,-5) ;
\draw [thick] (5.1, -5) -- (5.9,-5) ;
\draw [thick] (5, -4.1) -- (5,-4.9) ;
\draw [thick] (6.1, -5) -- (6.9,-5) ;
\draw [thick] (4.1, -4) -- (4.9,-4) ;

\end{tikzpicture}\ee

\noindent Now we turn on FI parameters for the $3N$ nodes twice, which corresponds to mass deformations of the original theory involving the global symmetry carried by the 7-brane stack. This will break $SU(9)$ down to $SU(3)^3$:

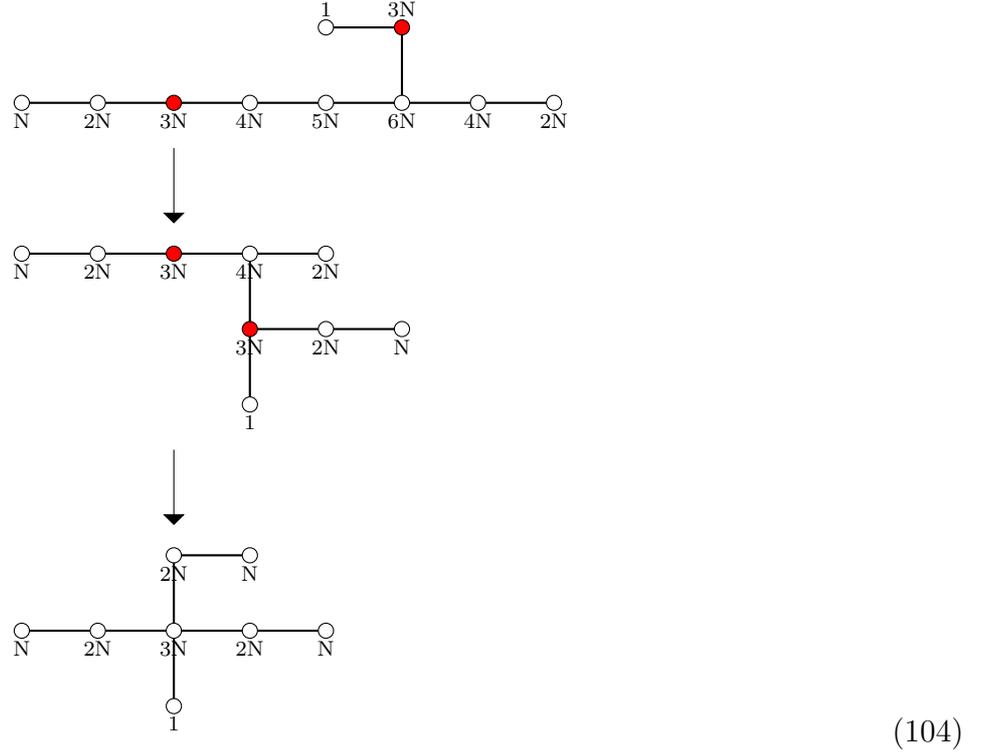
\be
 \begin{tikzpicture}
 \filldraw[fill= white] (4,1) circle [radius=0.1] node[above] {\scriptsize 1};
\filldraw[fill= white] (0,0) circle [radius=0.1] node[below] {\scriptsize N};
\filldraw[fill= white] (1,0) circle [radius=0.1] node[below] {\scriptsize 2N};
\filldraw[fill= red] (2,0) circle [radius=0.1] node[below] {\scriptsize 3N};
\filldraw[fill= white] (3,0) circle [radius=0.1] node[below] {\scriptsize 4N};
\filldraw[fill= white] (4,0) circle [radius=0.1] node[below] {\scriptsize 5N};
\filldraw[fill= white] (5,0) circle [radius=0.1] node[below] {\scriptsize 6N};
\filldraw[fill= white] (6,0) circle [radius=0.1] node[below] {\scriptsize 4N};
\filldraw[fill= red] (5,1) circle [radius=0.1] node[above] {\scriptsize 3N};
\filldraw[fill= white] (7,0) circle [radius=0.1] node[below] {\scriptsize 2N};

\draw [thick] (0.1, 0) -- (0.9,0) ;
\draw [thick] (1.1, 0) -- (1.9,0) ;
\draw [thick] (2.1, 0) -- (2.9,0) ;
\draw [thick] (3.1, 0) -- (3.9,0) ;
\draw [thick] (4.1, 0) -- (4.9,0) ;
\draw [thick] (5.1, 0) -- (5.9,0) ;
\draw [thick] (5, 0.1) -- (5,0.9) ;
\draw [thick] (6.1, 0) -- (6.9,0) ;
\draw [thick] (4.1, 1) -- (4.9,1) ;

\draw [->] (2,-0.6) -- (2,-1.6) ;

\filldraw[fill= white] (0,-2) circle [radius=0.1] node[below] {\scriptsize N};
\filldraw[fill= white] (1,-2) circle [radius=0.1] node[below] {\scriptsize 2N};
\filldraw[fill= red] (2,-2) circle [radius=0.1] node[below] {\scriptsize 3N};
\filldraw[fill= white] (3,-2) circle [radius=0.1] node[below] {\scriptsize 4N};
\filldraw[fill= white] (4,-2) circle [radius=0.1] node[below] {\scriptsize 2N};
\filldraw[fill= red] (3,-3) circle [radius=0.1] node[below] {\scriptsize 3N};
\filldraw[fill= white] (4,-3) circle [radius=0.1] node[below] {\scriptsize 2N};
\filldraw[fill= white] (5,-3) circle [radius=0.1] node[below] {\scriptsize N};
\filldraw[fill= white] (3,-4) circle [radius=0.1] node[below] {\scriptsize 1};

\draw [thick] (0.1, -2) -- (0.9,-2) ;
\draw [thick] (1.1, -2) -- (1.9,-2) ;
\draw [thick] (2.1, -2) -- (2.9,-2) ;
\draw [thick] (3.1, -2) -- (3.9,-2) ;
\draw [thick] (3, -2.1) -- (3,-2.9) ;
\draw [thick] (3.1, -3) -- (3.9,-3) ;
\draw [thick] (4.1, -3) -- (4.9,-3) ;
\draw [thick] (3, -3.1) -- (3,-3.9) ;
\draw [->] (2,-4.6) -- (2,-5.6) ;

\filldraw[fill= white] (0,-7) circle [radius=0.1] node[below] {\scriptsize N};
\filldraw[fill= white] (1,-7) circle [radius=0.1] node[below] {\scriptsize 2N};
\filldraw[fill= white] (2,-7) circle [radius=0.1] node[below] {\scriptsize 3N};
\filldraw[fill= white] (3,-7) circle [radius=0.1] node[below] {\scriptsize 2N};
\filldraw[fill= white] (4,-7) circle [radius=0.1] node[below] {\scriptsize N};
\filldraw[fill= white] (2,-6) circle [radius=0.1] node[below] {\scriptsize 2N};
\filldraw[fill= white] (3,-6) circle [radius=0.1] node[below] {\scriptsize N};

\filldraw[fill= white] (2,-8) circle [radius=0.1] node[below] {\scriptsize 1};

\draw [thick] (0.1, -7) -- (0.9,-7) ;
\draw [thick] (1.1, -7) -- (1.9,-7) ;
\draw [thick] (2.1, -7) -- (2.9,-7) ;
\draw [thick] (3.1, -7) -- (3.9,-7) ;
\draw [thick] (2, -6.1) -- (2,-6.9) ;
\draw [thick] (2, -7.1) -- (2,-7.9) ;
\draw [thick] (2.1, -6) -- (2.9,-6) ;
\end{tikzpicture}\ee

A chain of higgsing processes allows to realize other holonomy choices. This is summarized in the following Hasse diagram:
\be\label{HDE6z3} 
\begin{tikzpicture} 
\node[] at (-0.9,0) {$\cdots$};
\node[] (A) at (0,0) {\scriptsize $SU(3)^3$}; 
\node[] at (0,0.3) {\scriptsize (N)};
\node[] (B) at (3,0) {\scriptsize $SU(6)\times U(1)$}; 
\node[] at (3,0.3) {\scriptsize (N)};
\node[] (C) at (6,0) {\scriptsize $SU(3)^3$}; 
\node[] at (6,0.3) {\scriptsize (N-1)}; 
\node[] (D) at (9,0) {\scriptsize $SU(6)\times U(1)$}; 
\node[] at (9,0.3) {\scriptsize (N-1)};
\node[] at (1.2,0.2) {\scriptsize $\mathfrak{a}_2$};
\node[] at (4.5,0.2) {\scriptsize $\mathfrak{a}_5$};
\node[] at (7.5,0.2) {\scriptsize $\mathfrak{a}_2$};
\node[] at (10.5,0) {$\cdots$};
\draw[->] (A) -- (B); 
\draw[->] (B) -- (C); 
\draw[->] (C) -- (D); 
\end{tikzpicture}
\ee
\begin{tikzpicture} 
\node[] at (-0.9,0) {$\cdots$};
\node[] (A) at (0,0) {\scriptsize $SU(3)^3$}; 
\node[] at (0,0.3) {\scriptsize (2)};
\node[] (B) at (3,0) {\scriptsize $SU(6)\times U(1)$}; 
\node[] at (3,0.3) {\scriptsize (2)};
\node[] (C) at (6,0) {\scriptsize $SO(8)$}; 
\node[] at (6,0.3) {\scriptsize (2)}; 
\node[] (D) at (9,0) {\scriptsize $SU(3)^3$}; 
\node[] at (9,0.3) {\scriptsize (1)};
\node[] (E) at (12,0) {\scriptsize $SU(6)\times U(1)$}; 
\node[] at (12,0.3) {\scriptsize (1)};

\node[] (F) at (15,0) {\scriptsize $SO(8)$}; 
\node[] at (15,0.3) {\scriptsize (1)};
\node[] at (1.2,0.2) {\scriptsize $\mathfrak{a}_2$};
\node[] at (4.5,0.2) {\scriptsize $\mathfrak{a}_5$};
\node[] at (7.5,0.2) {\scriptsize $\mathfrak{d}_4$};
\node[] at (10.2,0.2) {\scriptsize $\mathfrak{a}_2$};
\node[] at (13.5,0.2) {\scriptsize $\mathfrak{a}_5$};
\draw[->] (A) -- (B); 
\draw[->] (B) -- (C); 
\draw[->] (C) -- (D); 
\draw[->] (D) -- (E);
\draw[->] (E) -- (F);  
\end{tikzpicture}

\paragraph{$\bullet$} The second canonical family of SCFTs for this case has the following magnetic quiver ($N>1$)
\be\label{E6Z3bisNMQ}
\begin{tikzpicture}
\filldraw[fill= white] (3,0) circle [radius=0.1] node[below] {\scriptsize N};
\filldraw[fill= white] (4,0) circle [radius=0.1] node[below] {\scriptsize 2N};
\filldraw[fill= white] (5,0) circle [radius=0.1] node[below] {\scriptsize 3N};
\filldraw[fill= white] (6,0) circle [radius=0.1] node[below] {\scriptsize 2N};
\filldraw[fill= white] (5,1) circle [radius=0.1] node[above] {\scriptsize 2N};
\filldraw[fill= white] (2,0) circle [radius=0.1] node[below] {\scriptsize 1};
\filldraw[fill= white] (8,0) circle [radius=0.1] node[below] {\scriptsize 1};
\filldraw[fill= white] (7,0) circle [radius=0.1] node[below] {\scriptsize N};
\filldraw[fill= white] (6,1) circle [radius=0.1] node[above] {\scriptsize N};
\filldraw[fill= white] (7,1) circle [radius=0.1] node[above] {\scriptsize 1};
\draw [thick] (5.1, 1) -- (5.9,1) ;
\draw [thick] (6.1, 1) -- (6.9,1) ;
\draw [thick] (3.1, 0) -- (3.9,0) ;
\draw [thick] (4.1, 0) -- (4.9,0) ;
\draw [thick] (5.1, 0) -- (5.9,0) ;
\draw [thick] (5, 0.1) -- (5,0.9) ;
\draw [thick] (2.1, 0) -- (2.9,0) ;
\draw [thick] (6.1, 0) -- (6.9,0) ;
\draw [thick] (7.1, 0) -- (7.9,0) ;
\end{tikzpicture} \ee 
This family has global symmetry $SO(8)\times U(1)^2$ times extra abelian factors\footnote{These factors enhance to $SU(2)$'s when $N=2$.}, and spectrum of CB operators
$$(3,4,4)\,;\,(6,7,7)\,;\,\dots\,;\, (3N)\,.$$
\\

\subsection{E$_7$-type 7-brane}
\subsubsection*{$\bf\mathbb{C}^2/\mathbb{Z}_3$}
According to the rules of Section \ref{Sec:GenDisc}, there is only one realization of the $\mathbb{Z}_3$ orbifold for a 7-brane of type E$_7$. The corresponding canonical family of superconformal field theories has the following magnetic quiver ($N>1$)
\be\label{E7Z3NMQ}
\begin{tikzpicture}
\filldraw[fill= white] (0,0) circle [radius=0.1] node[below] {\scriptsize N};
\filldraw[fill= white] (1,0) circle [radius=0.1] node[below] {\scriptsize 2N};
\filldraw[fill= white] (2,0) circle [radius=0.1] node[below] {\scriptsize 3N};
\filldraw[fill= white] (3,0) circle [radius=0.1] node[below] {\scriptsize 4N};
\filldraw[fill= white] (4,0) circle [radius=0.1] node[below] {\scriptsize 3N};
\filldraw[fill= white] (5,0) circle [radius=0.1] node[below] {\scriptsize 2N};
\filldraw[fill= white] (6,0) circle [radius=0.1] node[below] {\scriptsize N};
\filldraw[fill= white] (3,1) circle [radius=0.1] node[above] {\scriptsize 2N};
\filldraw[fill= white] (4,1) circle [radius=0.1] node[above] {\scriptsize 1}; 
\filldraw[fill= white] (-1,0) circle [radius=0.1] node[below] {\scriptsize 1}; 
\draw [thick] (0.1, 0) -- (0.9,0) ;
\draw [thick] (1.1, 0) -- (1.9,0) ;
\draw [thick] (2.1, 0) -- (2.9,0) ;
\draw [thick] (3.1, 0) -- (3.9,0) ;
\draw [thick] (4.1, 0) -- (4.9,0) ;
\draw [thick] (5.1, 0) -- (5.9,0) ;
\draw [thick] (3, 0.1) -- (3,0.9) ;
\draw [thick] (3.1, 1) -- (3.9,1) ;
\draw [thick] (-0.9, 0) -- (-0.1,0) ;
\end{tikzpicture}\ee
and flavor symmetry\footnote{There is an enhancement to $SU(7)\times SU(2)$ for the minimal choice $N=2$. Higgsing the $SU(2)$, one binds two of the three fractional D3 stacks together ending up with the quiver \eqref{SU8NMQ}: The probe is no longer able to see the full singularity, but only part of it.} $SU(7)\times U(1)$ times extra abelian factors. The associated CB operators have dimension:
$$(3,4,5)\,;\,(7,8,9)\,;\,\dots\,;\,(4N-1,4N)\,.$$
Following the systematics of the previous case we can derive \ref{E7Z3NMQ} from the M-theory quiver in \eqref{Table:SU(3)} that corresponds to the global symmetry $SO(14)\times U(1)$:
\be
\begin{tikzpicture}
\filldraw[fill= red] (-3,0) circle [radius=0.1] node[below] {\scriptsize 1};
\filldraw[fill= red] (-2,0) circle [radius=0.1] node[below] {\scriptsize 2};
\filldraw[fill= white] (-1,0) circle [radius=0.1] node[below] {\scriptsize 3};
\filldraw[fill= white] (0,0) circle [radius=0.1] node[below] {\scriptsize N+2};
\filldraw[fill= white] (1,0) circle [radius=0.1] node[below] {\scriptsize 2N+2};
\filldraw[fill= white] (2,0) circle [radius=0.1] node[below] {\scriptsize 3N+2};
\filldraw[fill= white] (3,0) circle [radius=0.1] node[below] {\scriptsize 4N+2};
\filldraw[fill= white] (4,0) circle [radius=0.1] node[below] {\scriptsize 5N+2};
\filldraw[fill= white] (5,0) circle [radius=0.1] node[below] {\scriptsize 6N+2};
\filldraw[fill= white] (6,0) circle [radius=0.1] node[below] {\scriptsize 4N+1};
\filldraw[fill= white] (5,1) circle [radius=0.1] node[above] {\scriptsize 3N+1};
\filldraw[fill= white] (7,0) circle [radius=0.1] node[below] {\scriptsize 2N};

\draw [thick] (-0.1, 0) -- (-0.9,0) ;
\draw [thick] (-1.1, 0) -- (-1.9,0) ;
\draw [thick] (-2.1, 0) -- (-2.9,0) ;
\draw [thick] (0.1, 0) -- (0.9,0) ;
\draw [thick] (1.1, 0) -- (1.9,0) ;
\draw [thick] (2.1, 0) -- (2.9,0) ;
\draw [thick] (3.1, 0) -- (3.9,0) ;
\draw [thick] (4.1, 0) -- (4.9,0) ;
\draw [thick] (5.1, 0) -- (5.9,0) ;
\draw [thick] (5, 0.1) -- (5,0.9) ;
\draw [thick] (6.1, 0) -- (6.9,0) ;
\draw [->] (2,-0.6) -- (2,-1.6) ;
\filldraw[fill= white] (-1,-2) circle [radius=0.1] node[below] {\scriptsize 1};
\filldraw[fill= white] (0,-2) circle [radius=0.1] node[below] {\scriptsize N};
\filldraw[fill= white] (1,-2) circle [radius=0.1] node[below] {\scriptsize 2N};
\filldraw[fill= red] (2,-2) circle [radius=0.1] node[below] {\scriptsize 3N};
\filldraw[fill= white] (3,-2) circle [radius=0.1] node[below] {\scriptsize 4N};
\filldraw[fill= white] (4,-2) circle [radius=0.1] node[below] {\scriptsize 5N};
\filldraw[fill= white] (5,-2) circle [radius=0.1] node[below] {\scriptsize 6N};
\filldraw[fill= white] (6,-2) circle [radius=0.1] node[below] {\scriptsize 4N};
\filldraw[fill= red] (5,-1) circle [radius=0.1] node[above] {\scriptsize 3N};
\filldraw[fill= white] (7,-2) circle [radius=0.1] node[below] {\scriptsize 2N};
\filldraw[fill= white] (8,-2) circle [radius=0.1] node[below] {\scriptsize 1};

\draw [thick] (-0.1,-2) -- (-0.9,-2) ;
\draw [thick] (0.1, -2) -- (0.9,-2) ;
\draw [thick] (1.1, -2) -- (1.9,-2) ;
\draw [thick] (2.1, -2) -- (2.9,-2) ;
\draw [thick] (3.1, -2) -- (3.9,-2) ;
\draw [thick] (4.1, -2) -- (4.9,-2) ;
\draw [thick] (5.1, -2) -- (5.9,-2) ;
\draw [thick] (5, -1.1) -- (5,-1.9) ;
\draw [thick] (6.1, -2) -- (6.9,-2) ;
\draw [thick] (7.1, -2) -- (7.9,-2) ;

\draw [->] (2,-2.8) -- (2,-3.8) ;

\filldraw[fill= white] (0,-5) circle [radius=0.1] node[below] {\scriptsize N};
\filldraw[fill= white] (1,-5) circle [radius=0.1] node[below] {\scriptsize 2N};
\filldraw[fill= white] (2,-5) circle [radius=0.1] node[below] {\scriptsize 3N};
\filldraw[fill= white] (3,-5) circle [radius=0.1] node[below] {\scriptsize 4N};
\filldraw[fill= white] (4,-5) circle [radius=0.1] node[below] {\scriptsize 2N};
\filldraw[fill= white] (5,-5) circle [radius=0.1] node[below] {\scriptsize 1};
\filldraw[fill= white] (3,-6) circle [radius=0.1] node[below] {\scriptsize 3N};
\filldraw[fill= white] (2,-6) circle [radius=0.1] node[below] {\scriptsize 2N};
\filldraw[fill= white] (1,-6) circle [radius=0.1] node[below] {\scriptsize N};
\filldraw[fill= white] (0,-6) circle [radius=0.1] node[below] {\scriptsize 1};

\draw [thick] (0.1, -5) -- (0.9,-5) ;
\draw [thick] (1.1, -5) -- (1.9,-5) ;
\draw [thick] (2.1, -5) -- (2.9,-5) ;
\draw [thick] (3.1, -5) -- (3.9,-5) ;
\draw [thick] (4.1, -5) -- (4.9,-5) ;
\draw [thick] (3, -5.1) -- (3,-5.9) ;
\draw [thick] (2.1, -6) -- (2.9,-6) ;
\draw [thick] (1.1, -6) -- (1.9,-6) ;
\draw [thick] (0.1, -6) -- (0.9,-6) ;
\end{tikzpicture}
\ee

By iteratively higgsing the flavor symmetries we derive the following Hasse Diagram:
\be\label{E7k3hasse} 
\begin{tikzpicture} 
\node[] at (-12.8,0) {$\cdots$};
\node[] (H) at (-11.5,0) {\scriptsize $SU(7)\times U(1)$}; 
\node[] at (-11.5,0.3) {\scriptsize (N+1)};
\node[] (A) at (1,0) {\scriptsize $SU(2)\times SO(10)\times U(1)$}; 
\node[] at (1,0.3) {\scriptsize (N)};

\node[] (D) at (-2.3,0) {\scriptsize $SU(7)\times U(1)$}; 
\node[] at (-2.3,0.3) {\scriptsize (N)};
\node[] at (-5,0.3) {\scriptsize (N+1)};
\node[] (E) at (-5,0) {\scriptsize $E_6\times U(1)$}; 
\node[] (F) at (-8,0) {\scriptsize $SU(2)\times SO(10)\times U(1)$}; 
\node[] at (-8,0.3) {\scriptsize (N+1)};
\node[] (I) at (-3.5,-1) {\scriptsize $SO(12)$}; 
\node[] at (-3.5,-1.3) {\scriptsize (N+1)};

\node[] at (-1.1,0.2) {\scriptsize $\mathfrak{a}_6$};
\node[] at (-1.5,-0.7) {\scriptsize $\mathfrak{d}_6$};
\node[] at (-5.5,-0.7) {\scriptsize $\mathfrak{d}_5$};
\node[] at (-10.2,0.2) {\scriptsize $\mathfrak{a}_6$};
\node[] at (-6.1,0.2) {\scriptsize $\mathfrak{a}_1$};

\node[] at (-4,0.2) {\scriptsize $\mathfrak{e}_6$};

\node[] at (3,0) {$\cdots$};

\draw[->] (D) -- (A); 
\draw[->] (F) -- (I);
\draw[->] (I) -- (A);  
\draw[->] (F) -- (E); 
\draw[->] (E) -- (D); 
\draw[->] (H) -- (F);

\end{tikzpicture}
\ee

where all the non-trivial $\mathbb{Z}_3$ holonomy choices are realized, except for $SU(6)\times SU(3)$.

\subsubsection*{$\bf\mathbb{C}^2/\mathbb{Z}_4$}

The allowed $\mathbb{Z}_4$ (non-trivial) holonomies for $E_7$ preserve the following subgroups: $SU(4)^2\times SU(2)$, $SU(2)\times SU(6) \times U(1)$, $SU(7)\times U(1)$, $SO(12)\times U(1)$, $SO(10)\times SU(2)\times U(1)$, $SO(12)\times SU(2)$, $E_6\times U(1)$, $SO(8)\times SU(2)^2\times U(1)$, $SU(8)$,  $SU(6)\times U(1)^2$, $SU(5)\times SU(3)$. We will see that all of these possibilities but $SU(5)\times SU(3)$ are realized (or at least visible to magnetic quivers) on the worldvolume of D3 branes probing the singularity of a $\mathbb{C}^2/\mathbb{Z}_4$ orbifold wrapped by $E_7$ 7-branes. In this case there are however two inequivalent realizations of the orbifold, and hence two canonical families of theories.

\paragraph{$\bullet$} Let us start with the rank $4N$ theory with non-abelian global symmetry $SU(4)^2\times SU(2)$. This is described by the following magnetic quiver:
\be\label{$SU(4)^2xSU(2)N$}
\begin{tikzpicture}
\filldraw[fill= white] (0,0) circle [radius=0.1] node[below] {\scriptsize N};
\filldraw[fill= white] (1,0) circle [radius=0.1] node[below] {\scriptsize 2N};
\filldraw[fill= white] (2,0) circle [radius=0.1] node[below] {\scriptsize 3N};
\filldraw[fill= white] (3,0) circle [radius=0.1] node[below] {\scriptsize 4N};
\filldraw[fill= white] (4,0) circle [radius=0.1] node[below] {\scriptsize 3N};
\filldraw[fill= white] (5,0) circle [radius=0.1] node[below] {\scriptsize 2N};
\filldraw[fill= white] (6,0) circle [radius=0.1] node[below] {\scriptsize N};
\filldraw[fill= white] (3,1) circle [radius=0.1] node[above] {\scriptsize 2N};
\filldraw[fill= white] (3,-1) circle [radius=0.1] node[below] {\scriptsize 1}; 
\draw [thick] (0.1, 0) -- (0.9,0) ;
\draw [thick] (1.1, 0) -- (1.9,0) ;
\draw [thick] (2.1, 0) -- (2.9,0) ;
\draw [thick] (3.1, 0) -- (3.9,0) ;
\draw [thick] (4.1, 0) -- (4.9,0) ;
\draw [thick] (5.1, 0) -- (5.9,0) ;
\draw [thick] (3, 0.1) -- (3,0.9) ;
\draw [thick] (3, -0.1) -- (3,-0.9) ;
\end{tikzpicture}\ee
The associated CB operators have dimension:
$$(2,3,4,4)\,;\, (6, 7, 8, 8)\,;\,\dots\,;\,(4N-2, 4N-1, 4N, 4N)\,.$$
The central charges match those of \eqref{formulacc} provided that $\alpha=\beta=0$ and $\epsilon=\frac{3}{32}$. The quiver \ref{$SU(4)^2xSU(2)N$} can be reached by introducing mass deformations for the $SU(4)$ global symmetry and for the $SU(8)$ subgroup of $E_8$ of a 6d theory living on $N$ M5 branes, which lie on top of a M9 wall wrapping a $\mathbb{C}^2/\mathbb{Z}_4$ orbifold. The corresponding quiver is given by:
\be
 \begin{tikzpicture}

\filldraw[fill= white] (-4,0) circle [radius=0.1] node[below] {\scriptsize 1};
\filldraw[fill= white] (-3,0) circle [radius=0.1] node[below] {\scriptsize 2};

\filldraw[fill= white] (-2,0) circle [radius=0.1] node[below] {\scriptsize 3};
\filldraw[fill= white] (-1,0) circle [radius=0.1] node[below] {\scriptsize 4};
\filldraw[fill= white] (0,0) circle [radius=0.1] node[below] {\scriptsize N+4};
\filldraw[fill= white] (1,0) circle [radius=0.1] node[below] {\scriptsize 2N+4};
\filldraw[fill= white] (2,0) circle [radius=0.1] node[below] {\scriptsize 3N+4};
\filldraw[fill= white] (3,0) circle [radius=0.1] node[below] {\scriptsize 4N+4};
\filldraw[fill= white] (4,0) circle [radius=0.1] node[below] {\scriptsize 5N+4};
\filldraw[fill= white] (5,0) circle [radius=0.1] node[below] {\scriptsize 6N+4};
\filldraw[fill= white] (6,0) circle [radius=0.1] node[below] {\scriptsize 4N+2};
\filldraw[fill= white] (5,1) circle [radius=0.1] node[above] {\scriptsize 3N+2};
\filldraw[fill= white] (7,0) circle [radius=0.1] node[below] {\scriptsize 2N+1};

\draw [thick] (-3.1, 0) -- (-3.9,0) ;
\draw [thick] (-2.1, 0) -- (-2.9,0) ;

\draw [thick] (-0.1, 0) -- (-0.9,0) ;
\draw [thick] (-1.1, 0) -- (-1.9,0) ;
\draw [thick] (0.1, 0) -- (0.9,0) ;
\draw [thick] (1.1, 0) -- (1.9,0) ;
\draw [thick] (2.1, 0) -- (2.9,0) ;
\draw [thick] (3.1, 0) -- (3.9,0) ;
\draw [thick] (4.1, 0) -- (4.9,0) ;
\draw [thick] (5.1, 0) -- (5.9,0) ;
\draw [thick] (5, 0.1) -- (5,0.9) ;
\draw [thick] (6.1, 0) -- (6.9,0) ;
\end{tikzpicture}\ee
A complete classification of the 6d theories that emerge in the $\mathbb{Z}_4$ orbifold case can be found in \cite{Mekareeya:2017jgc}. 
We first turn on FI terms for the four leftmost $U(k)$ ($k=1,2,3,4$) nodes, following the same procedure we have applied for $E_6$ on $\mathbb{Z}_3$:
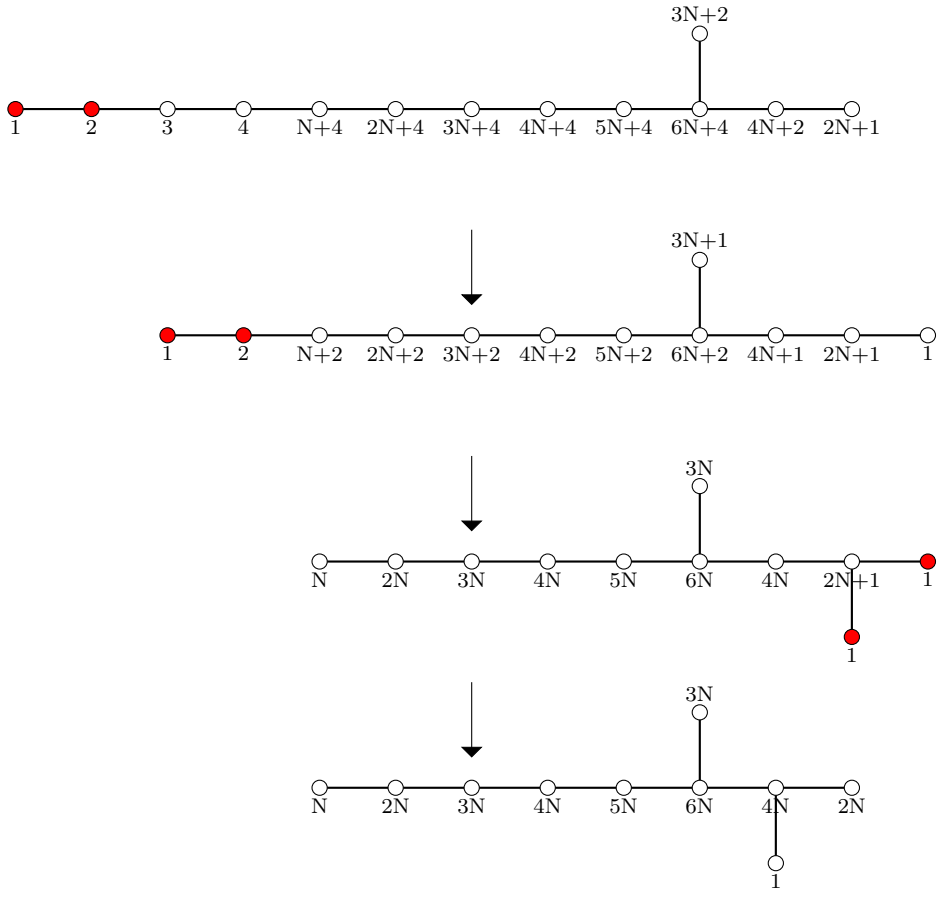
\be
 \begin{tikzpicture}

\filldraw[fill= red] (-4,0) circle [radius=0.1] node[below] {\scriptsize 1};
\filldraw[fill= red] (-3,0) circle [radius=0.1] node[below] {\scriptsize 2};

\filldraw[fill= white] (-2,0) circle [radius=0.1] node[below] {\scriptsize 3};
\filldraw[fill= white] (-1,0) circle [radius=0.1] node[below] {\scriptsize 4};
\filldraw[fill= white] (0,0) circle [radius=0.1] node[below] {\scriptsize N+4};
\filldraw[fill= white] (1,0) circle [radius=0.1] node[below] {\scriptsize 2N+4};
\filldraw[fill= white] (2,0) circle [radius=0.1] node[below] {\scriptsize 3N+4};
\filldraw[fill= white] (3,0) circle [radius=0.1] node[below] {\scriptsize 4N+4};
\filldraw[fill= white] (4,0) circle [radius=0.1] node[below] {\scriptsize 5N+4};
\filldraw[fill= white] (5,0) circle [radius=0.1] node[below] {\scriptsize 6N+4};
\filldraw[fill= white] (6,0) circle [radius=0.1] node[below] {\scriptsize 4N+2};
\filldraw[fill= white] (5,1) circle [radius=0.1] node[above] {\scriptsize 3N+2};
\filldraw[fill= white] (7,0) circle [radius=0.1] node[below] {\scriptsize 2N+1};

\draw [thick] (-3.1, 0) -- (-3.9,0) ;
\draw [thick] (-2.1, 0) -- (-2.9,0) ;

\draw [thick] (-0.1, 0) -- (-0.9,0) ;
\draw [thick] (-1.1, 0) -- (-1.9,0) ;
\draw [thick] (0.1, 0) -- (0.9,0) ;
\draw [thick] (1.1, 0) -- (1.9,0) ;
\draw [thick] (2.1, 0) -- (2.9,0) ;
\draw [thick] (3.1, 0) -- (3.9,0) ;
\draw [thick] (4.1, 0) -- (4.9,0) ;
\draw [thick] (5.1, 0) -- (5.9,0) ;
\draw [thick] (5, 0.1) -- (5,0.9) ;
\draw [thick] (6.1, 0) -- (6.9,0) ;
\draw [->] (2,-1.6) -- (2,-2.6) ;

\filldraw[fill= red] (-2,-3) circle [radius=0.1] node[below] {\scriptsize 1};
\filldraw[fill= red] (-1,-3) circle [radius=0.1] node[below] {\scriptsize 2};
\filldraw[fill= white] (0,-3) circle [radius=0.1] node[below] {\scriptsize N+2};
\filldraw[fill= white] (1,-3) circle [radius=0.1] node[below] {\scriptsize 2N+2};
\filldraw[fill= white] (2,-3) circle [radius=0.1] node[below] {\scriptsize 3N+2};
\filldraw[fill= white] (3,-3) circle [radius=0.1] node[below] {\scriptsize 4N+2};
\filldraw[fill= white] (4,-3) circle [radius=0.1] node[below] {\scriptsize 5N+2};
\filldraw[fill= white] (5,-3) circle [radius=0.1] node[below] {\scriptsize 6N+2};
\filldraw[fill= white] (6,-3) circle [radius=0.1] node[below] {\scriptsize 4N+1};
\filldraw[fill= white] (5,-2) circle [radius=0.1] node[above] {\scriptsize 3N+1};
\filldraw[fill= white] (7,-3) circle [radius=0.1] node[below] {\scriptsize 2N+1};
\filldraw[fill= white] (8,-3) circle [radius=0.1] node[below] {\scriptsize 1};

\draw [thick] (-0.1, -3) -- (-0.9,-3) ;
\draw [thick] (-1.1, -3) -- (-1.9,-3) ;
\draw [thick] (0.1, -3) -- (0.9,-3) ;
\draw [thick] (1.1, -3) -- (1.9,-3) ;
\draw [thick] (2.1, -3) -- (2.9,-3) ;
\draw [thick] (3.1, -3) -- (3.9,-3) ;
\draw [thick] (4.1, -3) -- (4.9,-3) ;
\draw [thick] (5.1, -3) -- (5.9,-3) ;
\draw [thick] (5, -2.1) -- (5,-2.9) ;
\draw [thick] (6.1,- 3) -- (6.9,-3) ;
\draw [thick] (7.1,- 3) -- (7.9,-3) ;
\draw [->] (2,-4.6) -- (2,-5.6) ;

\filldraw[fill= white] (0,-6) circle [radius=0.1] node[below] {\scriptsize N};
\filldraw[fill= white] (1,-6) circle [radius=0.1] node[below] {\scriptsize 2N};
\filldraw[fill= white] (2,-6) circle [radius=0.1] node[below] {\scriptsize 3N};
\filldraw[fill= white] (3,-6) circle [radius=0.1] node[below] {\scriptsize 4N};
\filldraw[fill= white] (4,-6) circle [radius=0.1] node[below] {\scriptsize 5N};
\filldraw[fill= white] (5,-6) circle [radius=0.1] node[below] {\scriptsize 6N};
\filldraw[fill= white] (6,-6) circle [radius=0.1] node[below] {\scriptsize 4N};
\filldraw[fill= white] (5,-5) circle [radius=0.1] node[above] {\scriptsize 3N};
\filldraw[fill= white] (7,-6) circle [radius=0.1] node[below] {\scriptsize 2N+1};
\filldraw[fill= red] (8,-6) circle [radius=0.1] node[below] {\scriptsize 1};
\filldraw[fill= red] (7,-7) circle [radius=0.1] node[below] {\scriptsize 1};

\draw [thick] (0.1, -6) -- (0.9,-6) ;
\draw [thick] (1.1, -6) -- (1.9,-6) ;
\draw [thick] (2.1, -6) -- (2.9,-6) ;
\draw [thick] (3.1, -6) -- (3.9,-6) ;
\draw [thick] (4.1, -6) -- (4.9,-6) ;
\draw [thick] (5.1, -6) -- (5.9,-6) ;
\draw [thick] (5, -5.1) -- (5,-5.9) ;
\draw [thick] (6.1,- 6) -- (6.9,-6) ;
\draw [thick] (7.1,- 6) -- (7.9,-6) ;
\draw [thick] (7,- 6.1) -- (7,-6.9) ;
\draw [->] (2,-7.6) -- (2,-8.6) ;

\filldraw[fill= white] (0,-9) circle [radius=0.1] node[below] {\scriptsize N};
\filldraw[fill= white] (1,-9) circle [radius=0.1] node[below] {\scriptsize 2N};
\filldraw[fill= white] (2,-9) circle [radius=0.1] node[below] {\scriptsize 3N};
\filldraw[fill= white] (3,-9) circle [radius=0.1] node[below] {\scriptsize 4N};
\filldraw[fill= white] (4,-9) circle [radius=0.1] node[below] {\scriptsize 5N};
\filldraw[fill= white] (5,-9) circle [radius=0.1] node[below] {\scriptsize 6N};
\filldraw[fill= white] (6,-9) circle [radius=0.1] node[below] {\scriptsize 4N};
\filldraw[fill= white] (5,-8) circle [radius=0.1] node[above] {\scriptsize 3N};
\filldraw[fill= white] (7,-9) circle [radius=0.1] node[below] {\scriptsize 2N};

\filldraw[fill= white] (6,-10) circle [radius=0.1] node[below] {\scriptsize 1};

\draw [thick] (0.1, -9) -- (0.9,-9) ;
\draw [thick] (1.1, -9) -- (1.9,-9) ;
\draw [thick] (2.1, -9) -- (2.9,-9) ;
\draw [thick] (3.1, -9) -- (3.9,-9) ;
\draw [thick] (4.1, -9) -- (4.9,-9) ;
\draw [thick] (5.1, -9) -- (5.9,-9) ;
\draw [thick] (5, -8.1) -- (5,-8.9) ;
\draw [thick] (6.1,- 9) -- (6.9,-9) ;

\draw [thick] (6,- 9.1) -- (6,-9.9) ;

\end{tikzpicture}\ee
At this point we turn on FI parameters for the $U(3N)$ nodes to break $SU(8)$ down to $SU(4)^2\times SU(2)$:
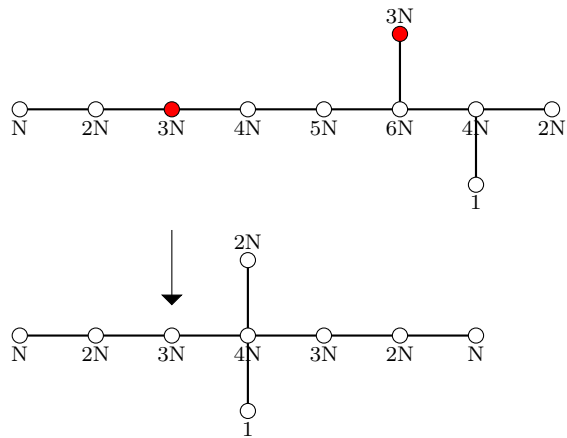
\be
 \begin{tikzpicture}
\filldraw[fill= white] (0,0) circle [radius=0.1] node[below] {\scriptsize N};
\filldraw[fill= white] (1,0) circle [radius=0.1] node[below] {\scriptsize 2N};
\filldraw[fill= red] (2,0) circle [radius=0.1] node[below] {\scriptsize 3N};
\filldraw[fill= white] (3,0) circle [radius=0.1] node[below] {\scriptsize 4N};
\filldraw[fill= white] (4,0) circle [radius=0.1] node[below] {\scriptsize 5N};
\filldraw[fill= white] (5,0) circle [radius=0.1] node[below] {\scriptsize 6N};
\filldraw[fill= white] (6,0) circle [radius=0.1] node[below] {\scriptsize 4N};
\filldraw[fill= red] (5,1) circle [radius=0.1] node[above] {\scriptsize 3N};
\filldraw[fill= white] (7,0) circle [radius=0.1] node[below] {\scriptsize 2N};

\filldraw[fill= white] (6,-1) circle [radius=0.1] node[below] {\scriptsize 1};

\draw [thick] (0.1, 0) -- (0.9,0) ;
\draw [thick] (1.1, 0) -- (1.9,0) ;
\draw [thick] (2.1, 0) -- (2.9,0) ;
\draw [thick] (3.1, 0) -- (3.9,0) ;
\draw [thick] (4.1, 0) -- (4.9,0) ;
\draw [thick] (5.1, 0) -- (5.9,0) ;
\draw [thick] (5, 0.1) -- (5,0.9) ;
\draw [thick] (6.1,0) -- (6.9,0) ;
\draw [thick] (6,- 0.1) -- (6,-0.9) ;

\draw [->] (2,-1.6) -- (2,-2.6) ;

\filldraw[fill= white] (0,-3) circle [radius=0.1] node[below] {\scriptsize N};
\filldraw[fill= white] (1,-3) circle [radius=0.1] node[below] {\scriptsize 2N};
\filldraw[fill= white] (2,-3) circle [radius=0.1] node[below] {\scriptsize 3N};
\filldraw[fill= white] (3,-3) circle [radius=0.1] node[below] {\scriptsize 4N};
\filldraw[fill= white] (4,-3) circle [radius=0.1] node[below] {\scriptsize 3N};
\filldraw[fill= white] (5,-3) circle [radius=0.1] node[below] {\scriptsize 2N};
\filldraw[fill= white] (6,-3) circle [radius=0.1] node[below] {\scriptsize N};
\filldraw[fill= white] (3,-2) circle [radius=0.1] node[above] {\scriptsize 2N};
\filldraw[fill= white] (3,-4) circle [radius=0.1] node[below] {\scriptsize 1}; 
\draw [thick] (0.1, -3) -- (0.9,-3) ;
\draw [thick] (1.1, -3) -- (1.9,-3) ;
\draw [thick] (2.1, -3) -- (2.9,-3) ;
\draw [thick] (3.1, -3) -- (3.9,-3) ;
\draw [thick] (4.1, -3) -- (4.9,-3) ;
\draw [thick] (5.1, -3) -- (5.9,-3) ;
\draw [thick] (3, -2.1) -- (3,-2.9) ;
\draw [thick] (3, -3.1) -- (3,-3.9) ;

\end{tikzpicture}\ee

Different holonomy possibilities are reached through a chain of higgsing processes, with rank decreasing by one unit after each higgsing. The related diagrams can be obtained by means of the quiver-subtraction rule. 

The resulting series of RG flows is depicted in the following Hasse diagram:
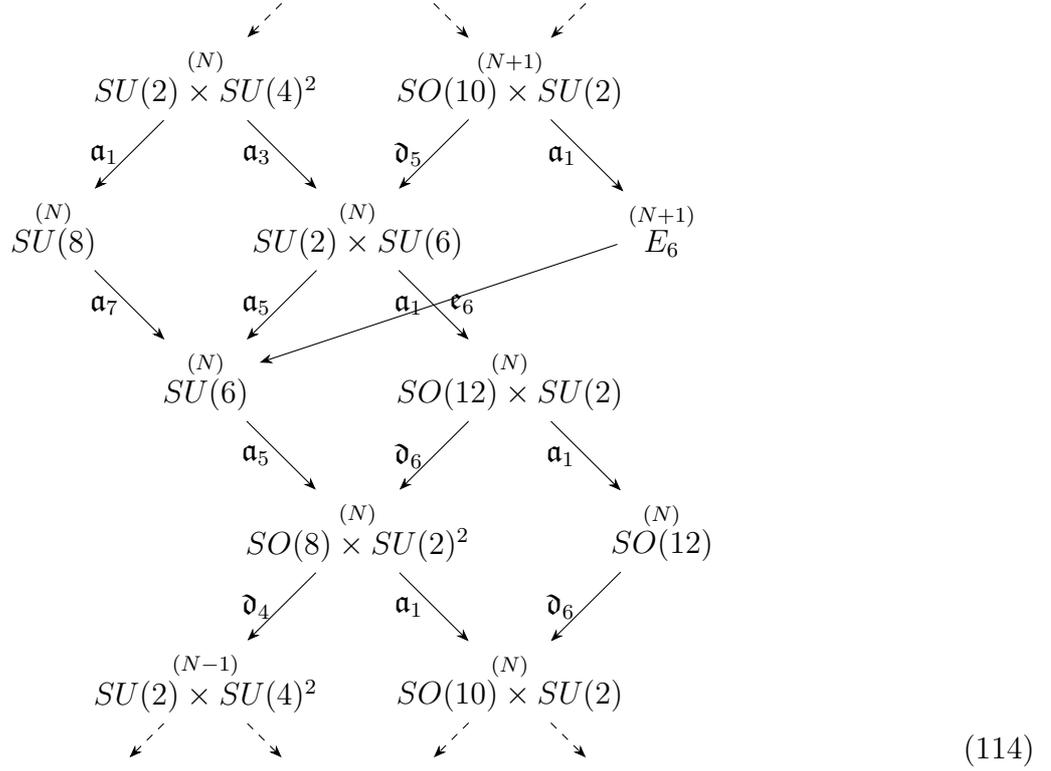
\be
\begin{tikzpicture} 
  \node at ( -2,0) (0') {$\overset{(N)}{SU(2)\times SU(4)^2}$};
   \node at ( 2,0) (0) {$\overset{(N+1)}{SO(10)\times SU(2)}$};
   \node at (-4,-2) (child10') {$\overset{(N)}{SU(8)}$};
    \node at (-2,-4) (cc10') {$\overset{(N)}{SU(6)}$};
       \node at (4,-2) (child10) {$\overset{(N+1)}{E_6}$};
         \node at (0,-2) (child30') {$\overset{(N)}{SU(2)\times SU(6)}$};
        \node at (2,-4) (cc30') {$\overset{(N)}{SO(12)\times SU(2)}$};
         \node at (4,-6) (ccc130') {$\overset{(N)}{SO(12)}$};
          \node at (2,-8) (cccc130') {$\overset{(N)}{SO(10)\times SU(2)}$};
         \node at (0,-6) (ccc230') {$\overset{(N)}{SO(8)\times SU(2)^2}$};
          \node at (-2,-8) (cccc230') {$\overset{(N-1)}{SU(2)\times SU(4)^2}$};
         \draw  [dashed,arrows = {-Stealth[]}](-1,1)--(0');
           \draw  [dashed,arrows = {-Stealth[]}](3,1)--(0);
             \draw  [dashed,arrows = {-Stealth[]}] (1,1)--(0);
         \draw [arrows = {-Stealth[]}]  (0')--(child10')node[midway, left] {$\mathfrak{a}_1$};
  \draw[arrows = {-Stealth[]}]   (0')--(child30')node[midway, left] {$\mathfrak{a}_3$};
     \draw[arrows = {-Stealth[]}]   (0)--(child30')node[midway, left] {$\mathfrak{d}_5$};
      \draw[arrows = {-Stealth[]}]   (0)--(child10)node[midway, left] {$\mathfrak{a}_1$};
           \draw[arrows = {-Stealth[]}]   (child10')--(cc10')node[midway, left] {$\mathfrak{a}_7$};         
            \draw [arrows = {-Stealth[]}]  (child30')--(cc10')node[midway, left] {$\mathfrak{a}_5$};        
             \draw [arrows = {-Stealth[]}]  (child30')--(cc30')node[midway, left] {$\mathfrak{a}_1$};     
              \draw[arrows = {-Stealth[]}]   (child10)--(cc10')node[midway, right] {$\mathfrak{e}_6$};
                  \draw[arrows = {-Stealth[]}]   (cc10')--(ccc230')node[midway, left] {$\mathfrak{a}_5$};
                   \draw[arrows = {-Stealth[]}]   (cc30')--(ccc230')node[midway, left] {$\mathfrak{d}_6$};
                     \draw [arrows = {-Stealth[]}]  (cc30')--(ccc130')node[midway, left] {$\mathfrak{a}_1$};
                       \draw [arrows = {-Stealth[]}]  (ccc230')--(cccc130')node[midway, left] {$\mathfrak{a}_1$};
                         \draw [arrows = {-Stealth[]}]  (ccc130')--(cccc130')node[midway, left] {$\mathfrak{d}_6$};
                           \draw [arrows = {-Stealth[]}]  (ccc230')--(cccc230')node[midway, left] {$\mathfrak{d}_4$};
                          \draw [dashed,arrows = {-Stealth[]}] (cccc230')--(-3,-9); 
                            \draw [dashed,arrows = {-Stealth[]}](cccc230')--(-1,-9); 
                              \draw  [dashed,arrows = {-Stealth[]}] (cccc130')--(3,-9); 
                                \draw  [dashed,arrows = {-Stealth[]}] (cccc130')--(1,-9); 
                                 \end{tikzpicture}
                                 \ee

  
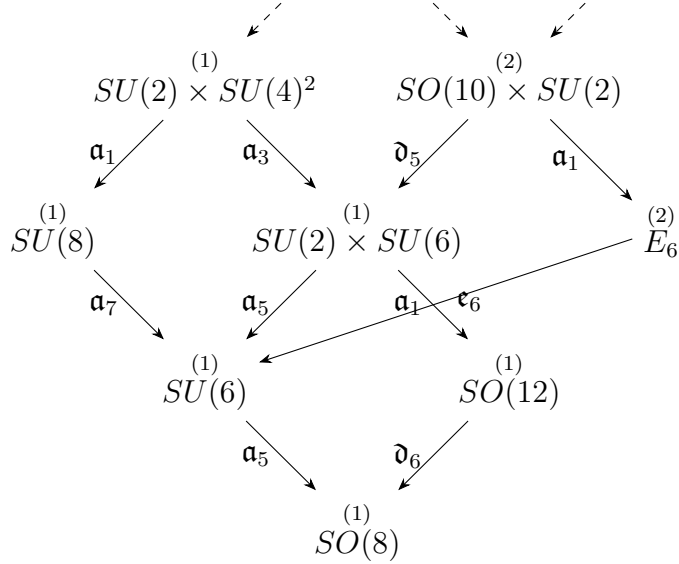
\begin{eqnarray*}
  \begin{tikzpicture} 
  \node at ( -2,0) (0') {$\overset{(1)}{SU(2)\times SU(4)^2}$};
   \node at ( 2,0) (0) {$\overset{(2)}{SO(10)\times SU(2)}$};
   \node at (-4,-2) (child10') {$\overset{(1)}{SU(8)}$};
    \node at (-2,-4) (cc10') {$\overset{(1)}{SU(6)}$};
       \node at (4,-2) (child10) {$\overset{(2)}{E_6}$};
         \node at (0,-2) (child30') {$\overset{(1)}{SU(2)\times SU(6)}$};
        \node at (2,-4) (cc30') {$\overset{(1)}{SO(12)}$};
         \node at (0,-6) (ccc30') {$\overset{(1)}{SO(8)}$};
         \draw  [dashed,arrows = {-Stealth[]}]  (-1,1)--(0');
           \draw  [dashed,arrows = {-Stealth[]}] (3,1)--(0);
             \draw [dashed,arrows = {-Stealth[]}]  (1,1)--(0);
         \draw[arrows = {-Stealth[]}]  (0')--(child10')node[midway, left] {$\mathfrak{a}_1$};
  \draw[arrows = {-Stealth[]}]  (0')--(child30')node[midway, left] {$\mathfrak{a}_3$};
     \draw[arrows = {-Stealth[]}]  (0)--(child30')node[midway, left] {$\mathfrak{d}_5$};
      \draw[arrows = {-Stealth[]}]  (0)--(child10)node[midway, left] {$\mathfrak{a}_1$};
           \draw[arrows = {-Stealth[]}]  (child10')--(cc10')node[midway, left] {$\mathfrak{a}_7$};         
            \draw[arrows = {-Stealth[]}]  (child30')--(cc10')node[midway, left] {$\mathfrak{a}_5$};        
             \draw [arrows = {-Stealth[]}] (child30')--(cc30')node[midway, left] {$\mathfrak{a}_1$};     
              \draw[arrows = {-Stealth[]}]  (child10)--(cc10')node[midway, right] {$\mathfrak{e}_6$};
                  \draw [arrows = {-Stealth[]}] (cc10')--(ccc30')node[midway, left] {$\mathfrak{a}_5$};
                   \draw[arrows = {-Stealth[]}]  (cc30')--(ccc30')node[midway, left] {$\mathfrak{d}_6$};
  \end{tikzpicture}
\end{eqnarray*}

\paragraph{$\bullet$} Let us come now to the other canonical family of superconformal theories, which is described by the magnetic quiver ($N>1$)
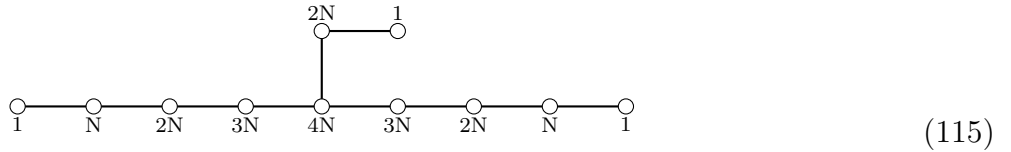
\be\label{E7Z4bisNMQ}
\begin{tikzpicture}
\filldraw[fill= white] (0,0) circle [radius=0.1] node[below] {\scriptsize N};
\filldraw[fill= white] (1,0) circle [radius=0.1] node[below] {\scriptsize 2N};
\filldraw[fill= white] (2,0) circle [radius=0.1] node[below] {\scriptsize 3N};
\filldraw[fill= white] (3,0) circle [radius=0.1] node[below] {\scriptsize 4N};
\filldraw[fill= white] (4,0) circle [radius=0.1] node[below] {\scriptsize 3N};
\filldraw[fill= white] (5,0) circle [radius=0.1] node[below] {\scriptsize 2N};
\filldraw[fill= white] (6,0) circle [radius=0.1] node[below] {\scriptsize N};
\filldraw[fill= white] (3,1) circle [radius=0.1] node[above] {\scriptsize 2N};
\filldraw[fill= white] (4,1) circle [radius=0.1] node[above] {\scriptsize 1}; 
\filldraw[fill= white] (-1,0) circle [radius=0.1] node[below] {\scriptsize 1}; 
\filldraw[fill= white] (7,0) circle [radius=0.1] node[below] {\scriptsize 1}; 
\draw [thick] (0.1, 0) -- (0.9,0) ;
\draw [thick] (1.1, 0) -- (1.9,0) ;
\draw [thick] (2.1, 0) -- (2.9,0) ;
\draw [thick] (3.1, 0) -- (3.9,0) ;
\draw [thick] (4.1, 0) -- (4.9,0) ;
\draw [thick] (5.1, 0) -- (5.9,0) ;
\draw [thick] (3, 0.1) -- (3,0.9) ;
\draw [thick] (3.1, 1) -- (3.9,1) ;
\draw [thick] (-0.9, 0) -- (-0.1,0) ;
\draw [thick] (6.1, 0) -- (6.9,0) ;
\end{tikzpicture}\ee
and has flavor symmetry $SU(6)\times U(1)^2$ (modulo extra abelian factors), enhancing to $SU(6)\times SU(2)^2$ for $N=2$. The associated CB operators have dimension:
$$(3,4,5,5)\,;\,(7,8,9,9)\,;\,\dots\,;\,(4N-1,4N)\,.$$

\subsection{E$_8$-type 7-brane}

\subsubsection*{$\bf\mathbb{C}^2/\mathbb{Z}_6$}

For the $\mathbb{Z}_6$ orbifold we have the following $E_8$ non-trivial holonomies: $SU(6)\times SU(3)\times SU(2)$, $SU(3)\times SU(6)\times U(1)^2$, $SU(7)\times U(1)^2$, $SO(12)\times SU(2)\times U(1)$, $SO(14)\times U(1)$, $SU(6)\times SU(2)^2\times U(1)$, $SO(10)\times SU(2)^2\times U(1)$, $SU(8)\times U(1)$, $SU(4)\times SO(8)\times U(1)$, $SU(5)\times SU(4)\times U(1)$, $SU(2)\times SU(7)\times U(1)$. We will see that all of these possibilities are covered by the series of SCFTs living on the D3-brane probes. Again we have two inequivalent realizations here. Let us discuss them in turn.

\paragraph{$\bullet$} We start by the one associated to the rank-$6N$ theory with non-abelian global symmetry $SU(6)\times SU(3)\times SU(2)$, described by the magnetic quiver:
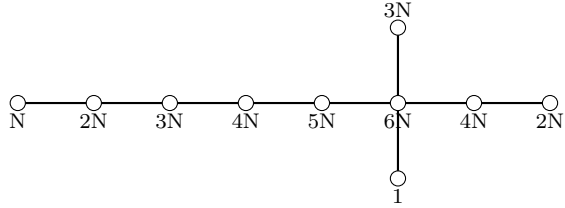
\be\label{E8k6N}
\begin{tikzpicture}
\filldraw[fill= white] (0,0) circle [radius=0.1] node[below] {\scriptsize N};
\filldraw[fill= white] (1,0) circle [radius=0.1] node[below] {\scriptsize 2N};
\filldraw[fill= white] (2,0) circle [radius=0.1] node[below] {\scriptsize 3N};
\filldraw[fill= white] (3,0) circle [radius=0.1] node[below] {\scriptsize 4N};
\filldraw[fill= white] (4,0) circle [radius=0.1] node[below] {\scriptsize 5N};
\filldraw[fill= white] (5,0) circle [radius=0.1] node[below] {\scriptsize 6N};
\filldraw[fill= white] (6,0) circle [radius=0.1] node[below] {\scriptsize 4N};
\filldraw[fill= white] (5,1) circle [radius=0.1] node[above] {\scriptsize 3N};
\filldraw[fill= white] (5,-1) circle [radius=0.1] node[below] {\scriptsize 1};
\filldraw[fill= white] (7,0) circle [radius=0.1] node[below] {\scriptsize 2N};

\draw [thick] (0.1, 0) -- (0.9,0) ;
\draw [thick] (1.1, 0) -- (1.9,0) ;
\draw [thick] (2.1, 0) -- (2.9,0) ;
\draw [thick] (3.1, 0) -- (3.9,0) ;
\draw [thick] (4.1, 0) -- (4.9,0) ;
\draw [thick] (5.1, 0) -- (5.9,0) ;
\draw [thick] (5, 0.1) -- (5,0.9) ;
\draw [thick] (5, -0.1) -- (5,-0.9) ;
\draw [thick] (6.1, 0) -- (6.9,0) ;
\end{tikzpicture} \ee 
The CB operators have dimension:
$$ (2,3,4,5,6,6)\,;\, (8,9,10,11,12,12)\,;\,\dots\,;\, (6N-4,6N-3,6N-2,6N-1,6N,6N)\,.$$
The central charges match the holographic result \eqref{formulacc} provided that $\alpha=\beta=0$ and $\epsilon=5/72$. 
Now, following the systematics of Section \ref{Sec:M-theory}, it is easy to obtain \ref{E8k6N} by turning on mass deformations for the $SU(6)$ symmetry of the 6d theory engineered by the M5/M9-brane setup. Starting from a magnetic quiver of the form:
\\
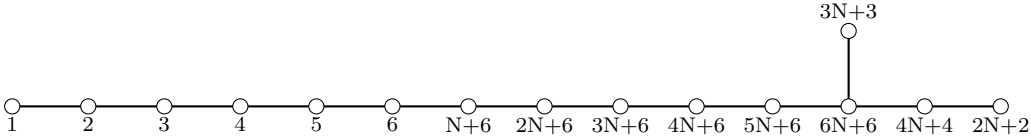
 \begin{tikzpicture}
\filldraw[fill= white] (-6,0) circle [radius=0.1] node[below] {\scriptsize 1};
\filldraw[fill= white] (-5,0) circle [radius=0.1] node[below] {\scriptsize 2};
\filldraw[fill= white] (-4,0) circle [radius=0.1] node[below] {\scriptsize 3};
\filldraw[fill= white] (-3,0) circle [radius=0.1] node[below] {\scriptsize 4};

\filldraw[fill= white] (-2,0) circle [radius=0.1] node[below] {\scriptsize 5};
\filldraw[fill= white] (-1,0) circle [radius=0.1] node[below] {\scriptsize 6};
\filldraw[fill= white] (0,0) circle [radius=0.1] node[below] {\scriptsize N+6};
\filldraw[fill= white] (1,0) circle [radius=0.1] node[below] {\scriptsize 2N+6};
\filldraw[fill= white] (2,0) circle [radius=0.1] node[below] {\scriptsize 3N+6};
\filldraw[fill= white] (3,0) circle [radius=0.1] node[below] {\scriptsize 4N+6};
\filldraw[fill= white] (4,0) circle [radius=0.1] node[below] {\scriptsize 5N+6};
\filldraw[fill= white] (5,0) circle [radius=0.1] node[below] {\scriptsize 6N+6};
\filldraw[fill= white] (6,0) circle [radius=0.1] node[below] {\scriptsize 4N+4};
\filldraw[fill= white] (5,1) circle [radius=0.1] node[above] {\scriptsize 3N+3};
\filldraw[fill= white] (7,0) circle [radius=0.1] node[below] {\scriptsize 2N+2};
\draw [thick] (-5.1, 0) -- (-5.9,0) ;

\draw [thick] (-4.1, 0) -- (-4.9,0) ;
\draw [thick] (-3.1, 0) -- (-3.9,0) ;
\draw [thick] (-2.1, 0) -- (-2.9,0) ;

\draw [thick] (-0.1, 0) -- (-0.9,0) ;
\draw [thick] (-1.1, 0) -- (-1.9,0) ;
\draw [thick] (0.1, 0) -- (0.9,0) ;
\draw [thick] (1.1, 0) -- (1.9,0) ;
\draw [thick] (2.1, 0) -- (2.9,0) ;
\draw [thick] (3.1, 0) -- (3.9,0) ;
\draw [thick] (4.1, 0) -- (4.9,0) ;
\draw [thick] (5.1, 0) -- (5.9,0) ;
\draw [thick] (5, 0.1) -- (5,0.9) ;
\draw [thick] (6.1, 0) -- (6.9,0) ;
\end{tikzpicture}\\
we apply the algorithm illustrated in Section \ref{massdef} to obtain:
\\
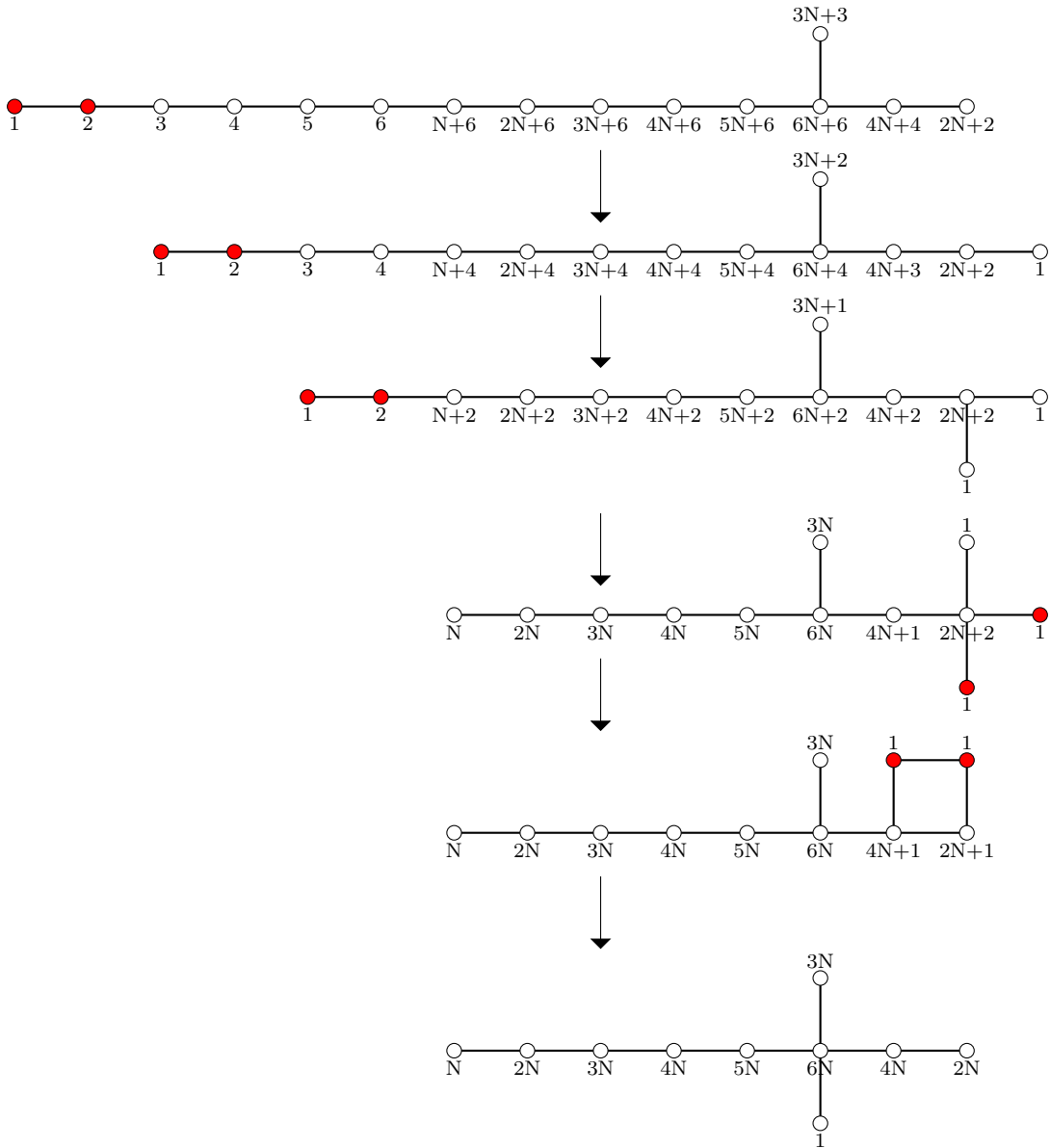
 \begin{tikzpicture}
\filldraw[fill= red] (-6,0) circle [radius=0.1] node[below] {\scriptsize 1};
\filldraw[fill= red] (-5,0) circle [radius=0.1] node[below] {\scriptsize 2};
\filldraw[fill= white] (-4,0) circle [radius=0.1] node[below] {\scriptsize 3};
\filldraw[fill= white] (-3,0) circle [radius=0.1] node[below] {\scriptsize 4};

\filldraw[fill= white] (-2,0) circle [radius=0.1] node[below] {\scriptsize 5};
\filldraw[fill= white] (-1,0) circle [radius=0.1] node[below] {\scriptsize 6};
\filldraw[fill= white] (0,0) circle [radius=0.1] node[below] {\scriptsize N+6};
\filldraw[fill= white] (1,0) circle [radius=0.1] node[below] {\scriptsize 2N+6};
\filldraw[fill= white] (2,0) circle [radius=0.1] node[below] {\scriptsize 3N+6};
\filldraw[fill= white] (3,0) circle [radius=0.1] node[below] {\scriptsize 4N+6};
\filldraw[fill= white] (4,0) circle [radius=0.1] node[below] {\scriptsize 5N+6};
\filldraw[fill= white] (5,0) circle [radius=0.1] node[below] {\scriptsize 6N+6};
\filldraw[fill= white] (6,0) circle [radius=0.1] node[below] {\scriptsize 4N+4};
\filldraw[fill= white] (5,1) circle [radius=0.1] node[above] {\scriptsize 3N+3};
\filldraw[fill= white] (7,0) circle [radius=0.1] node[below] {\scriptsize 2N+2};
\draw [thick] (-5.1, 0) -- (-5.9,0) ;

\draw [thick] (-4.1, 0) -- (-4.9,0) ;
\draw [thick] (-3.1, 0) -- (-3.9,0) ;
\draw [thick] (-2.1, 0) -- (-2.9,0) ;

\draw [thick] (-0.1, 0) -- (-0.9,0) ;
\draw [thick] (-1.1, 0) -- (-1.9,0) ;
\draw [thick] (0.1, 0) -- (0.9,0) ;
\draw [thick] (1.1, 0) -- (1.9,0) ;
\draw [thick] (2.1, 0) -- (2.9,0) ;
\draw [thick] (3.1, 0) -- (3.9,0) ;
\draw [thick] (4.1, 0) -- (4.9,0) ;
\draw [thick] (5.1, 0) -- (5.9,0) ;
\draw [thick] (5, 0.1) -- (5,0.9) ;
\draw [thick] (6.1, 0) -- (6.9,0) ;
\draw [->] (2,-0.6) -- (2,-1.6) ;

\filldraw[fill= red] (-4,-2) circle [radius=0.1] node[below] {\scriptsize 1};
\filldraw[fill= red] (-3,-2) circle [radius=0.1] node[below] {\scriptsize 2};

\filldraw[fill= white] (-2,-2) circle [radius=0.1] node[below] {\scriptsize 3};
\filldraw[fill= white] (-1,-2) circle [radius=0.1] node[below] {\scriptsize 4};
\filldraw[fill= white] (0,-2) circle [radius=0.1] node[below] {\scriptsize N+4};
\filldraw[fill= white] (1,-2) circle [radius=0.1] node[below] {\scriptsize 2N+4};
\filldraw[fill= white] (2,-2) circle [radius=0.1] node[below] {\scriptsize 3N+4};
\filldraw[fill= white] (3,-2) circle [radius=0.1] node[below] {\scriptsize 4N+4};
\filldraw[fill= white] (4,-2) circle [radius=0.1] node[below] {\scriptsize 5N+4};
\filldraw[fill= white] (5,-2) circle [radius=0.1] node[below] {\scriptsize 6N+4};
\filldraw[fill= white] (6,-2) circle [radius=0.1] node[below] {\scriptsize 4N+3};
\filldraw[fill= white] (5,-1) circle [radius=0.1] node[above] {\scriptsize 3N+2};
\filldraw[fill= white] (7,-2) circle [radius=0.1] node[below] {\scriptsize 2N+2};
\filldraw[fill= white] (8,-2) circle [radius=0.1] node[below] {\scriptsize 1};

\draw [thick] (-3.1, -2) -- (-3.9,-2) ;
\draw [thick] (-2.1, -2) -- (-2.9,-2) ;

\draw [thick] (-0.1, -2) -- (-0.9,-2) ;
\draw [thick] (-1.1, -2) -- (-1.9,-2) ;
\draw [thick] (0.1, -2) -- (0.9,-2) ;
\draw [thick] (1.1, -2) -- (1.9,-2) ;
\draw [thick] (2.1, -2) -- (2.9,-2) ;
\draw [thick] (3.1, -2) -- (3.9,-2) ;
\draw [thick] (4.1, -2) -- (4.9,-2) ;
\draw [thick] (5.1, -2) -- (5.9,-2) ;
\draw [thick] (5, -1.1) -- (5,-1.9) ;
\draw [thick] (6.1, -2) -- (6.9,-2) ;
\draw [thick] (7.1, -2) -- (7.9,-2) ;
\draw [->] (2,-2.6) -- (2,-3.6) ;

\filldraw[fill= red] (-2,-4) circle [radius=0.1] node[below] {\scriptsize 1};
\filldraw[fill= red] (-1,-4) circle [radius=0.1] node[below] {\scriptsize 2};
\filldraw[fill= white] (0,-4) circle [radius=0.1] node[below] {\scriptsize N+2};
\filldraw[fill= white] (1,-4) circle [radius=0.1] node[below] {\scriptsize 2N+2};
\filldraw[fill= white] (2,-4) circle [radius=0.1] node[below] {\scriptsize 3N+2};
\filldraw[fill= white] (3,-4) circle [radius=0.1] node[below] {\scriptsize 4N+2};
\filldraw[fill= white] (4,-4) circle [radius=0.1] node[below] {\scriptsize 5N+2};
\filldraw[fill= white] (5,-4) circle [radius=0.1] node[below] {\scriptsize 6N+2};
\filldraw[fill= white] (6,-4) circle [radius=0.1] node[below] {\scriptsize 4N+2};
\filldraw[fill= white] (5,-3) circle [radius=0.1] node[above] {\scriptsize 3N+1};
\filldraw[fill= white] (7,-4) circle [radius=0.1] node[below] {\scriptsize 2N+2};
\filldraw[fill= white] (7,-5) circle [radius=0.1] node[below] {\scriptsize 1};

\filldraw[fill= white] (8,-4) circle [radius=0.1] node[below] {\scriptsize 1};

\draw [thick] (-0.1, -4) -- (-0.9,-4) ;
\draw [thick] (-1.1, -4) -- (-1.9,-4) ;
\draw [thick] (0.1, -4) -- (0.9,-4) ;
\draw [thick] (1.1, -4) -- (1.9,-4) ;
\draw [thick] (2.1, -4) -- (2.9,-4) ;
\draw [thick] (3.1, -4) -- (3.9,-4) ;
\draw [thick] (4.1, -4) -- (4.9,-4) ;
\draw [thick] (5.1, -4) -- (5.9,-4) ;
\draw [thick] (5, -3.1) -- (5,-3.9) ;
\draw [thick] (6.1, -4) -- (6.9,-4) ;
\draw [thick] (7, -4.1) -- (7,-4.9) ;
\draw [thick] (7.1, -4) -- (7.9,-4) ;
\draw [->] (2,-5.6) -- (2,-6.6) ;

\filldraw[fill= white] (0,-7) circle [radius=0.1] node[below] {\scriptsize N};
\filldraw[fill= white] (1,-7) circle [radius=0.1] node[below] {\scriptsize 2N};
\filldraw[fill= white] (2,-7) circle [radius=0.1] node[below] {\scriptsize 3N};
\filldraw[fill= white] (3,-7) circle [radius=0.1] node[below] {\scriptsize 4N};
\filldraw[fill= white] (4,-7) circle [radius=0.1] node[below] {\scriptsize 5N};
\filldraw[fill= white] (5,-7) circle [radius=0.1] node[below] {\scriptsize 6N};
\filldraw[fill= white] (6,-7) circle [radius=0.1] node[below] {\scriptsize 4N+1};
\filldraw[fill= white] (5,-6) circle [radius=0.1] node[above] {\scriptsize 3N};
\filldraw[fill= white] (7,-7) circle [radius=0.1] node[below] {\scriptsize 2N+2};
\filldraw[fill= white] (7,-6) circle [radius=0.1] node[above] {\scriptsize 1};
\filldraw[fill= red] (7,-8) circle [radius=0.1] node[below] {\scriptsize 1};
\filldraw[fill= red] (8,-7) circle [radius=0.1] node[below] {\scriptsize 1};

\draw [thick] (0.1, -7) -- (0.9,-7) ;
\draw [thick] (1.1, -7) -- (1.9,-7) ;
\draw [thick] (2.1, -7) -- (2.9,-7) ;
\draw [thick] (3.1, -7) -- (3.9,-7) ;
\draw [thick] (4.1, -7) -- (4.9,-7) ;
\draw [thick] (5.1, -7) -- (5.9,-7) ;
\draw [thick] (5, -6.1) -- (5,-6.9) ;
\draw [thick] (6.1, -7) -- (6.9,-7) ;
\draw [thick] (7, -7.1) -- (7,-7.9) ;
\draw [thick] (7.1, -7) -- (7.9,-7) ;
\draw [thick] (7, -6.1) -- (7,-6.9) ;
\draw [->] (2,-7.6) -- (2,-8.6) ;

\filldraw[fill= white] (0,-10) circle [radius=0.1] node[below] {\scriptsize N};
\filldraw[fill= white] (1,-10) circle [radius=0.1] node[below] {\scriptsize 2N};
\filldraw[fill= white] (2,-10) circle [radius=0.1] node[below] {\scriptsize 3N};
\filldraw[fill= white] (3,-10) circle [radius=0.1] node[below] {\scriptsize 4N};
\filldraw[fill= white] (4,-10) circle [radius=0.1] node[below] {\scriptsize 5N};
\filldraw[fill= white] (5,-10) circle [radius=0.1] node[below] {\scriptsize 6N};
\filldraw[fill= white] (6,-10) circle [radius=0.1] node[below] {\scriptsize 4N+1};
\filldraw[fill= white] (5,-9) circle [radius=0.1] node[above] {\scriptsize 3N};
\filldraw[fill= white] (7,-10) circle [radius=0.1] node[below] {\scriptsize 2N+1};

\filldraw[fill= red] (6,-9) circle [radius=0.1] node[above] {\scriptsize 1};
\filldraw[fill= red] (7,-9) circle [radius=0.1] node[above] {\scriptsize 1};

\draw [thick] (0.1, -10) -- (0.9,-10) ;
\draw [thick] (1.1, -10) -- (1.9,-10) ;
\draw [thick] (2.1, -10) -- (2.9,-10) ;
\draw [thick] (3.1, -10) -- (3.9,-10) ;
\draw [thick] (4.1, -10) -- (4.9,-10) ;
\draw [thick] (5.1, -10) -- (5.9,-10) ;
\draw [thick] (5, -9.1) -- (5,-9.9) ;
\draw [thick] (6.1, -10) -- (6.9,-10) ;
\draw [thick] (6.1, -9) -- (6.9,-9) ;
\draw [thick] (7, -9.1) -- (7,-9.9) ;
\draw [thick] (6, -9.1) -- (6,-9.9) ;
\draw [->] (2,-10.6) -- (2,-11.6) ;

\filldraw[fill= white] (0,-13) circle [radius=0.1] node[below] {\scriptsize N};
\filldraw[fill= white] (1,-13) circle [radius=0.1] node[below] {\scriptsize 2N};
\filldraw[fill= white] (2,-13) circle [radius=0.1] node[below] {\scriptsize 3N};
\filldraw[fill= white] (3,-13) circle [radius=0.1] node[below] {\scriptsize 4N};
\filldraw[fill= white] (4,-13) circle [radius=0.1] node[below] {\scriptsize 5N};
\filldraw[fill= white] (5,-13) circle [radius=0.1] node[below] {\scriptsize 6N};
\filldraw[fill= white] (5,-14) circle [radius=0.1] node[below] {\scriptsize 1};
\filldraw[fill= white] (6,-13) circle [radius=0.1] node[below] {\scriptsize 4N};
\filldraw[fill= white] (5,-12) circle [radius=0.1] node[above] {\scriptsize 3N};
\filldraw[fill= white] (7,-13) circle [radius=0.1] node[below] {\scriptsize 2N};

\draw [thick] (0.1, -13) -- (0.9,-13) ;
\draw [thick] (1.1, -13) -- (1.9,-13) ;
\draw [thick] (2.1, -13) -- (2.9,-13) ;
\draw [thick] (3.1, -13) -- (3.9,-13) ;
\draw [thick] (4.1, -13) -- (4.9,-13) ;
\draw [thick] (5.1, -13) -- (5.9,-13) ;
\draw [thick] (5, -12.1) -- (5,-12.9) ;
\draw [thick] (6.1, -13) -- (6.9,-13) ;
\draw [thick] (5, -13.1) -- (5,-13.9) ;

\end{tikzpicture}\\

The series of theories corresponding to all other holonomy choices are derived by multiple higgsing processes, which can be implemented through iterated quiver subtractions. \\
The sequence of RG flows is represented by the following Hasse diagram (the $U(1)$ factors are suppressed for simplicity):
\begin{figure}[H]
\centering
\begin{tikzpicture} 
  \node at ( -2,0) (0') {$\overset{(N)}{SU(2)\times SU(3)\times SU(6)}$};
   \node at ( 2,0) (0) {$\overset{(N+1)}{SO(10)\times SU(2)}$};
   \node at (-4,-2) (child10') {$\overset{(N)}{SU(8)}$};
     \node at (-2,-2) (child20') {$\overset{(N)}{SU(9)}$};
       \node at (0,-2) (child30') {$\overset{(N)}{SU(4)\times SU(5)}$};
         \node at (4,-2) (child10) {$\overset{(N+1)}{SO(12)\times SU(2)}$};
         \node at (6,-4) (cc10) {$\overset{(N+1)}{SO(12)}$};
          \node at (-2,-4) (cc20') {$\overset{(N)}{SU(7)}$};
           \node at ( 2,-4) (cc30') {$\overset{(N)}{SO(10)\times SU(2)^2}$};
            \node at ( 0,-6) (ccc20') {$\overset{(N)}{SU(6)\times SU(2)^2}$};
             \node at ( 4,-6) (ccc30') {$\overset{(N)}{SU(3)\times SO(10)}$};
             \node at (6,-8) (cccc30') {$\overset{(N)}{E_6}$};
               \node at (2,-8) (cccc120') {$\overset{(N)}{SU(2)\times SU(7)}$};
                \node at (4,-10) (ccccc1120') {$\overset{(N)}{SO(14)}$};
                 \node at (0,-10) (ccccc2120') {$\overset{(N)}{SU(7)}$};
                \node at ( -2,-8) (cccc220') {$\overset{(N)}{SU(3)\times SU(6)}$};
                 \node at ( -4,-10) (ccccc220') {$\overset{(N)}{SU(4)\times SO(8)}$};
                  \node at ( -2,-12) (cccccc1220') {$\overset{(N-1)}{SU(2)\times SU(3)\times SU(6)}$};
                  % \node at ( -4,-12) (cccccc1220') {$SU(2)\times SU(3)\times SU(6)$};
                   \node at ( 2,-12) (cccccc2220') {$\overset{(N)}{SO(10)\times SU(2)}$};
                    \node at (-4,-14) (ccccccc11220') {$\overset{(N-1)}{SU(8)}$};
                    \node at (-2,-14) (ccccccc21220') {$\overset{(N-1)}{SU(9)}$};
       \node at (0,-14) (ccccccc31220') {$\overset{(N-1)}{SU(4)\times SU(5)}$};
        \node at (4,-14) (ccccccc2220') {$\overset{(N)}{SO(12)\times SU(2)}$};
   \draw [dashed,arrows = {-Stealth[]}] (-3,1)--(0');
     \draw [dashed,arrows = {-Stealth[]}] (-1,1)--(0);
      \draw [dashed,arrows = {-Stealth[]}] (1,1)--(0);
       \draw [dashed,arrows = {-Stealth[]}] (3,1)--(0);
  \draw [arrows = {-Stealth[]}](0')--(child10')node[midway, left] {$\mathfrak{a}_2$};
   \draw[arrows = {-Stealth[]}] (0')--(child20')node[midway, left] {$\mathfrak{a}_1$};
    \draw[arrows = {-Stealth[]}] (0')--(child30')node[midway, left] {$\mathfrak{a}_5$};
     \draw[arrows = {-Stealth[]}] (0)--(child30')node[midway, left] {$\mathfrak{d}_5$};
      \draw[arrows = {-Stealth[]}] (0)--(child10)node[midway, left] {$\mathfrak{a}_1$};
         \draw [arrows = {-Stealth[]}](child20')--(cc20')node[midway, left] {$\mathfrak{a}_8$};
           \draw[arrows = {-Stealth[]}] (child30')--(cc30')node[midway, left] {$\mathfrak{a}_3$};
             \draw [arrows = {-Stealth[]}](child10)--(cc30')node[midway, left] {$\mathfrak{d}_6$};
               \draw[arrows = {-Stealth[]}] (child10)--(cc10)node[midway, left] {$\mathfrak{a}_1$};
                 \draw [arrows = {-Stealth[]}](cc10)--(ccc30')node[midway, left] {$\mathfrak{d}_6$};
               \draw [arrows = {-Stealth[]}](child10')--(cc20')node[midway, left] {$\mathfrak{a}_7$};
               \draw [arrows = {-Stealth[]}](child30')--(cc20')node[midway, left] {$\mathfrak{a}_4$};
                 \draw[arrows = {-Stealth[]}] (cc30')--(ccc30')node[midway, left] {$\mathfrak{a}_1$};
                   \draw [arrows = {-Stealth[]}](cc30')--(ccc20')node[midway, left] {$\mathfrak{d}_5$};
                     \draw[arrows = {-Stealth[]}] (cc20')--(ccc20')node[midway, left] {$\mathfrak{a}_6$};
                       \draw [arrows = {-Stealth[]}](ccc20')--(cccc220')node[midway, left] {$\mathfrak{a}_5$};
                         \draw[arrows = {-Stealth[]}] (ccc30')--(cccc30')node[midway, left] {$\mathfrak{a}_2$};
                          \draw [arrows = {-Stealth[]}](ccc30')--(cccc120')node[midway, left] {$\mathfrak{d}_5$};
                         \draw[arrows = {-Stealth[]}] (ccc20')--(cccc120')node[midway, left] {$\mathfrak{a}_1$};
                          \draw[arrows = {-Stealth[]}] (cccc30')--(ccccc2120')node[midway, right] {$\mathfrak{e}_6$};
                          \draw [arrows = {-Stealth[]}](cccc220')--(ccccc220')node[midway, left] {$\mathfrak{a}_5$};
                           \draw [arrows = {-Stealth[]}](cccc220')--(ccccc2120')node[midway, left] {$\mathfrak{a}_2$};
                            \draw [arrows = {-Stealth[]}](cccc120')--(ccccc1120')node[midway, left] {$\mathfrak{a}_1$};
                             \draw [arrows = {-Stealth[]}](cccc120')--(ccccc2120')node[midway, left] {$\mathfrak{a}_6$};
                               \draw[arrows = {-Stealth[]}] (ccccc220')--(cccccc1220')node[midway, left] {$\mathfrak{d}_4$};
 \draw[arrows = {-Stealth[]}] (ccccc220')--(cccccc2220')node[midway, left] {$\mathfrak{a}_3$};
  \draw[arrows = {-Stealth[]}] (ccccc2120')--(cccccc2220')node[midway, left] {$\mathfrak{a}_6$};
   \draw [arrows = {-Stealth[]}](ccccc1120')--(cccccc2220')node[midway, left] {$\mathfrak{d}_7$};
    \draw [arrows = {-Stealth[]}](cccccc2220')--(ccccccc2220')node[midway, left] {$\mathfrak{a}_1$};
      \draw[arrows = {-Stealth[]}] (cccccc2220')--(ccccccc31220')node[midway, left] {$\mathfrak{d}_5$};
      \draw [arrows = {-Stealth[]}](cccccc1220')--(ccccccc31220')node[midway, left] {$\mathfrak{a}_5$};       
       \draw [arrows = {-Stealth[]}](cccccc1220')--(ccccccc21220')node[midway, left] {$\mathfrak{a}_1$};    
        \draw[arrows = {-Stealth[]}] (cccccc1220')--(ccccccc11220')node[midway, left] {$\mathfrak{a}_2$};
        	\draw [dashed,arrows = {-Stealth[]}] (ccccccc31220')--(-1,-15);  
		\draw [dashed,arrows = {-Stealth[]}] (ccccccc21220')--(-2,-15);  
			\draw [dashed,arrows = {-Stealth[]}] (ccccccc11220')--(-3,-15);  
		\draw [dashed,arrows = {-Stealth[]}] (ccccccc31220')--(1,-15);  
			\draw [dashed,arrows = {-Stealth[]}] (ccccccc2220')--(3,-15);  
				\draw [dashed,arrows = {-Stealth[]}] (ccccccc2220')--(5,-15);  
  \end{tikzpicture}
%\caption{Hasse diagram: E8, k=6}
\end{figure}
\newpage
which culminates in:
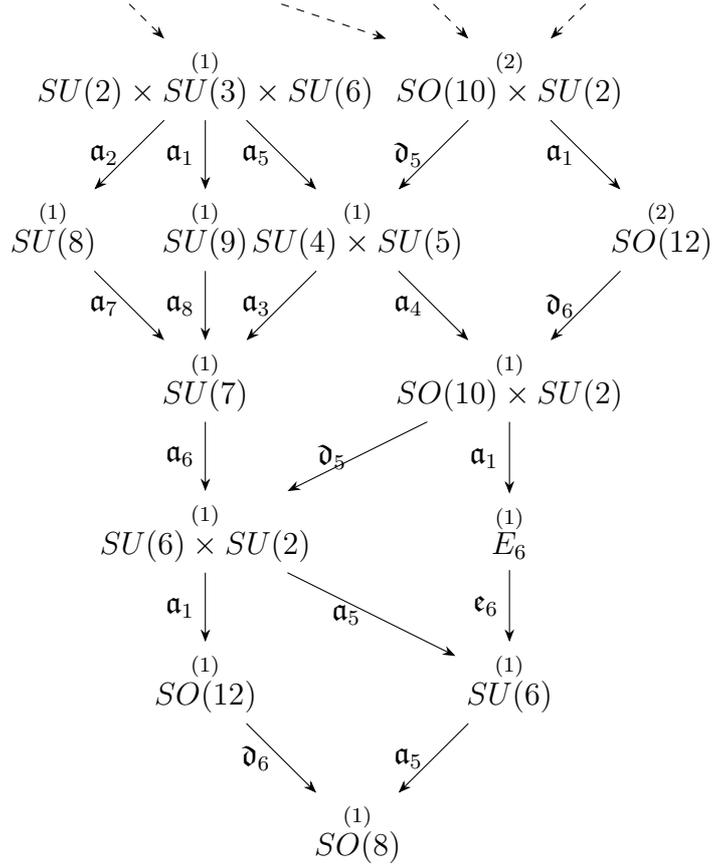
\begin{figure}[H]
\centering
\begin{tikzpicture} 
  \node at ( -2,0) (0') {$\overset{(1)}{SU(2)\times SU(3)\times SU(6)}$};
   \node at ( 2,0) (0) {$\overset{(2)}{SO(10)\times SU(2)}$};
   \node at (-4,-2) (child10') {$\overset{(1)}{SU(8)}$};
     \node at (-2,-2) (child20') {$\overset{(1)}{SU(9)}$};
       \node at (0,-2) (child30') {$\overset{(1)}{SU(4)\times SU(5)}$};
         \node at (4,-2) (child10) {$\overset{(2)}{SO(12)}$};
          \node at (-2,-4) (cc20') {$\overset{(1)}{SU(7)}$};
           \node at ( 2,-4) (cc30') {$\overset{(1)}{SO(10)\times SU(2)}$};
            \node at ( -2,-6) (ccc20') {$\overset{(1)}{SU(6)\times SU(2)}$};
             \node at ( 2,-6) (ccc30') {$\overset{(1)}{E_6}$};
               \node at (2,-8) (cccc30') {$\overset{(1)}{SU(6)}$};
                 \node at ( -2,-8) (cccc20') {$\overset{(1)}{SO(12)}$};
                 \node at ( 0,-10) (ccccc20') {$\overset{(1)}{SO(8)}$};
                  \draw [dashed, arrows = {-Stealth[]}] (-3,1)--(0');
     \draw [dashed, arrows = {-Stealth[]}](-1,1)--(0);
      \draw [dashed, arrows = {-Stealth[]}] (1,1)--(0);
       \draw [dashed, arrows = {-Stealth[]}](3,1)--(0);
  \draw [arrows = {-Stealth[]}] (0')--(child10') node[midway, left] {$\mathfrak{a}_2$};
   \draw [arrows = {-Stealth[]}](0')--(child20')node[midway, left] {$\mathfrak{a}_1$};
    \draw [arrows = {-Stealth[]}](0')--(child30')node[midway, left] {$\mathfrak{a}_5$};
     \draw[arrows = {-Stealth[]}] (0)--(child30')node[midway, left] {$\mathfrak{d}_5$};
      \draw [arrows = {-Stealth[]}](0)--(child10)node[midway, left] {$\mathfrak{a}_1$};
         \draw[arrows = {-Stealth[]}] (child20')--(cc20')node[midway, left] {$\mathfrak{a}_8$};
           \draw [arrows = {-Stealth[]}](child30')--(cc30')node[midway, left] {$\mathfrak{a}_4$};
             \draw [arrows = {-Stealth[]}](child10)--(cc30')node[midway, left] {$\mathfrak{d}_6$};
               \draw [arrows = {-Stealth[]}](child10')--(cc20')node[midway, left] {$\mathfrak{a}_7$};
               \draw [arrows = {-Stealth[]}](child30')--(cc20')node[midway, left] {$\mathfrak{a}_3$};
                 \draw [arrows = {-Stealth[]}](cc30')--(ccc30')node[midway, left] {$\mathfrak{a}_1$};
                   \draw [arrows = {-Stealth[]}](cc30')--(ccc20')node[midway, left] {$\mathfrak{d}_5$};
                     \draw [arrows = {-Stealth[]}](cc20')--(ccc20')node[midway, left] {$\mathfrak{a}_6$};
                       \draw[arrows = {-Stealth[]}] (ccc20')--(cccc20')node[midway, left] {$\mathfrak{a}_1$};
                         \draw [arrows = {-Stealth[]}](ccc30')--(cccc30')node[midway, left] {$\mathfrak{e}_6$};
                         \draw[arrows = {-Stealth[]}] (ccc20')--(cccc30')node[midway, left] {$\mathfrak{a}_5$};
                           \draw [arrows = {-Stealth[]}](cccc20')--(ccccc20')node[midway, left] {$\mathfrak{d}_6$};
                             \draw [arrows = {-Stealth[]}](cccc30')--(ccccc20')node[midway, left] {$\mathfrak{a}_5$};
                              
  \end{tikzpicture}
  \end{figure}
Note that the rank of the theories always decreases by one unit at each step of the chain.\\

\paragraph{$\bullet$} The second canonical family of SCFTs corresponding to the $\mathbb{Z}_6$ orbifold of the E$_8$ 7-brane has magnetic quiver ($N>1$)
\be\label{E8Z6bisNMQ}
\begin{tikzpicture}
\filldraw[fill= white] (0,0) circle [radius=0.1] node[below] {\scriptsize N};
\filldraw[fill= white] (1,0) circle [radius=0.1] node[below] {\scriptsize 2N};
\filldraw[fill= white] (2,0) circle [radius=0.1] node[below] {\scriptsize 3N};
\filldraw[fill= white] (3,0) circle [radius=0.1] node[below] {\scriptsize 4N};
\filldraw[fill= white] (4,0) circle [radius=0.1] node[below] {\scriptsize 5N};
\filldraw[fill= white] (5,0) circle [radius=0.1] node[below] {\scriptsize 6N};
\filldraw[fill= white] (6,0) circle [radius=0.1] node[below] {\scriptsize 4N};
\filldraw[fill= white] (5,1) circle [radius=0.1] node[above] {\scriptsize 3N};
\filldraw[fill= white] (7,0) circle [radius=0.1] node[below] {\scriptsize 2N};
\filldraw[fill= white] (8,0) circle [radius=0.1] node[below] {\scriptsize 1};
\filldraw[fill= white] (-1,0) circle [radius=0.1] node[below] {\scriptsize 1};
\filldraw[fill= white] (6,1) circle [radius=0.1] node[above] {\scriptsize 1};
\draw [thick] (0.1, 0) -- (0.9,0) ;
\draw [thick] (1.1, 0) -- (1.9,0) ;
\draw [thick] (2.1, 0) -- (2.9,0) ;
\draw [thick] (3.1, 0) -- (3.9,0) ;
\draw [thick] (4.1, 0) -- (4.9,0) ;
\draw [thick] (5.1, 0) -- (5.9,0) ;
\draw [thick] (5, 0.1) -- (5,0.9) ;
\draw [thick] (-0.9, 0) -- (-0.1,0) ;
\draw [thick] (7.1, 0) -- (7.9,0) ;
\draw [thick] (6.1, 0) -- (6.9,0) ;
\draw [thick] (5.1, 1) -- (5.9,1) ;

\end{tikzpicture} \ee 
This has flavor symmetry including $SU(7)\times U(1)^2$ (enhancing to $SU(7)\times SU(2)\times U(1)$ for $N=2$), and spectrum of CB operators
$$ (3,4,5,6,7,7)\,;\, (9,10,11,12,13,13)\,;\,\dots\,;\, (6N-3,6N-2,6N-1,6N)\,.$$

\end{appendix}
\bibliographystyle{JHEP}
\bibliography{N2Refs.bib}
\end{document}